\documentclass[aps,prb,reprint,superscriptaddress,amsfonts,amsmath,showpacs,floatfix]{revtex4-1}

\usepackage{graphicx}
\usepackage{epstopdf}

\newcommand{\mrm}[1]{\mathrm{#1}}

\newcommand{\mbb}[1]{\mathbb{#1}}
\newcommand{\mc}[1]{\mathcal{#1}}
\newcommand{\tsf}[1]{\textsf{#1}}
\newcommand{\eref}[1]{(\ref{#1})}
\newcommand{\Eref}[1]{Eq.~(\ref{#1})}
\newcommand{\fref}[1]{Fig.~\ref{#1}}
\newcommand{\tref}[1]{Table~\ref{#1}}
\newcommand{\sref}[1]{Sec.~\ref{#1}}

\newcommand{\pEref}[1]{\protect{Eq.~(\ref{#1})}}
\newcommand{\pfref}[1]{\protect{Fig.~\ref{#1}}}

\newcommand{\rmi}{\mathrm{i}}
\newcommand{\ra}{\rangle}
\newcommand{\la}{\langle}
\newcommand{\p}[1]{\phantom{#1}}
\newcommand{\spinsite}[1]{#1}
\newcommand{\rcite}[1]{Ref.~\onlinecite{#1}}
\newcommand{\olcite}[1]{\onlinecite{#1}}
\newcommand{\Sp}{\mrm{D}}
\newcommand{\T}{\mrm{T}}
\newcommand{\lb}{\left}
\newcommand{\rb}{\right}
\newcommand{\TQFTcite}{\cite{witten1989,blok1990,blok1990a,wen1990a,wen1990d,read1992,bais2009}}
\newcommand{\translop}{\raisebox{-12pt}{\includegraphics{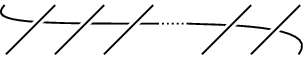}}}
\newcommand{\translopstar}{\raisebox{-12pt}{\includegraphics{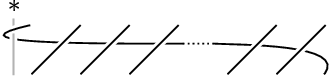}}}
\newcommand{\magtranslop}{\raisebox{-12pt}{\includegraphics{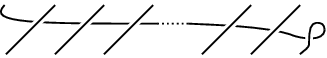}}}
\newcommand{\magtranslopstar}{\raisebox{-12pt}{\includegraphics{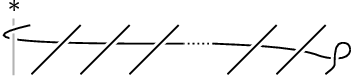}}}
\newcommand{\Fib}{\textsf{Fib}}
\newcommand{\SU}{\mrm{SU}}

\begin{document}

\title{Translation invariance, topology, and protection of criticality\\in chains of interacting anyons}

\author{Robert N. C. Pfeifer}
\email[]{rpfeifer@perimeterinstitute.ca}
\affiliation{School of Mathematics and Physics, The University of Queensland, Brisbane, QLD 4072, Australia}
\affiliation{Perimeter Institute for Theoretical Physics, 31 Caroline St. N, Waterloo ON~~N2L 2Y5, Canada}
\altaffiliation{Current address of R. N. C. Pfeifer and G. Vidal} %
\author{Oliver Buerschaper}
\affiliation{Perimeter Institute for Theoretical Physics, 31 Caroline St. N, Waterloo ON~~N2L 2Y5, Canada}
\author{Simon Trebst}
\affiliation{Microsoft Research, Station Q, University of California, Santa Barbara, CA 93106}
\affiliation{Institute for Theoretical Physics, University of Cologne, 50937 Cologne, Germany}
\author{Andreas W. W. Ludwig}
\affiliation{Department of Physics, University of California, Santa Barbara, CA 93106}
\author{Matthias Troyer}
\affiliation{Theoretische Physik, ETH Zurich, 8093 Zurich, Switzerland}
\author{Guifre Vidal}
\affiliation{School of Mathematics and Physics, The University of Queensland, Brisbane, QLD 4072, Australia}
\affiliation{Perimeter Institute for Theoretical Physics, 31 Caroline St. N, Waterloo ON~~N2L 2Y5, Canada}
\altaffiliation{Current address of R. N. C. Pfeifer and G. Vidal.} %

\date{\today}

\begin{abstract}
Using finite size scaling arguments, the critical properties of a chain of interacting anyons can be extracted from the low energy spectrum of a finite system. In Phys. Rev. Lett. \textbf{98}, 160409 (2007), Feiguin \emph{et al.} showed that an antiferromagnetic (AFM) chain of Fibonacci anyons on a torus is in the same universality class as the tricritical Ising model, and that criticality is protected by a topological symmetry. 
In the present paper we first review the graphical formalism for the study of anyons on the disc and demonstrate how this formalism may be consistently extended to the study of systems on surfaces of higher genus. We then employ this graphical formalism to %
study finite rings of interacting anyons on both the disc and the torus,
and show that %
analysis on the disc necessarily yields an energy spectrum which is a subset of that which is obtained on the torus. For a critical Hamiltonian, one may extract from this subset the scaling dimensions of the local scaling operators which respect the topological symmetry of the system. Related considerations are also shown to apply for open chains.
\end{abstract}

\pacs{05.30.Pr, 73.43.Lp, 03.65.Vf}

\maketitle

\section{Introduction}

The study of collective states of anyonic excitations is an exciting and yet relatively unexplored area of condensed matter physics. The nontrivial exchange behaviour of non-Abelian anyons may be exploited for universal quantum computation,\cite{freedman2002,freedman2002a,freedman2003,kitaev2003,nayak2008} 
with the simplest suitable model being that of Fibonacci anyons. It has been suggested that, as the non-Abelian component of the $k=3$ $Z_k$-parafermion Read-Rezayi state,\cite{read1999} they may appear in the fractional quantum Hall state with filling fraction $\nu=12/5$.\cite{xia2004}
These systems are therefore presently of intense theoretical and experimental interest. 

\citeauthor{feiguin2007} recently initiated the study of interacting non-Abelian anyons with the analysis of nearest neighbour interactions, in \rcite{feiguin2007}, for Fibonacci and more generally, for $\SU(2)_k$ anyons. %
This work was later extended to next-to-nearest neighbour interactions in Refs.~\olcite{trebst2008} and \olcite{trebst2008a}, to higher-spin anyons in \rcite{gils2009}, and to two dimensional systems in Refs.~\onlinecite{poilblanc2011} and \onlinecite{ludwig2011}. These papers identified numerous critical phases, and scaling dimensions of the local scaling operators were extracted using exact diagonalisation by matching the numerically-obtained finite size energy spectra against exact predictions from conformal field theory %
(see Refs.~\onlinecite{cardy1984} and \onlinecite{cardy1986}; also reviewed in \rcite{difrancesco1997}). %
Local scaling operators are of interest as they may appear as perturbations of the critical Hamiltonian, and may be classified by whether or not they respect the topological symmetry of the system.\cite{feiguin2007,gils2009} For Fibonacci anyons undergoing an antiferromagnetic (AFM) interaction, the authors of \rcite{feiguin2007} show that this topological symmetry protects the criticality of the system against all %
translation-invariant perturbations of the Hamiltonian which are ``relevant'' in the renormalisation group sense.

In this paper we characterise %
the %
differences between periodic chains of anyons on the torus and on the disc, and introduce
mappings of these systems 
to equivalent ``spin chains''. %
We show that while the natural definition of translation for a ring of anyons on the torus is closely related to that of a specific spin chain model, 
there are subtleties with
the natural definition of translation on the disc.
As a result, a Hamiltonian which is translation invariant on the disc will only be translation invariant up to a defect on the corresponding spin chain.
The energy spectra of the same local Hamiltonian acting on two periodic chains of anyons, one on a torus and one on a disc, will therefore not in general coincide, highlighting the topological nature of many-body anyon systems. We further show that 
the energy spectrum obtained on the disc always constitutes a subset of the spectrum obtained on the torus, and that for a critical theory, the local scaling operators which may be identified from this subset are precisely those operators which
respect the topological symmetry defined in \rcite{feiguin2007}. 
We also show that similar considerations apply to open chains, where the spectrum of the theory, and for critical theories also the inferred local scaling operator content, is once again affected by the topology of the surface on which the anyons are found.

We will begin in \sref{sec:ASO} by reviewing the influence of manifold topology on the degrees of freedom of an anyonic system, and will show how considerations from topological quantum field theory (TQFT) permit us to extend the popular fusion tree representation for systems of anyons on a disc (see e.g.~Refs.~\onlinecite{bonderson2007,bonderson2008,wang2010}) to surfaces of arbitrary genus, including the derivation of an inner product which is consistent with the chosen vertex normalisation scheme on the sphere. Particular attention is paid to the construction of translation operators for rings of anyons on surfaces of various genera. 
Following this overview of the diagrammatic formalism, 
in \sref{sec:PBC} we then employ the formalism to compare the behaviour of rings of interacting Fibonacci anyons on surfaces of different genus, namely the torus and the disc, specialising to the study of the AFM nearest neighbour interaction for a chain of Fibonacci anyons. We show that at criticality, changing the topology of the manifold on which the anyons are located affects the spectrum of scaling operators having local representations. By explicit construction, we also show that on the disc one may obtain an energy spectrum corresponding to the scaling operator spectrum on the torus by introducing an appropriate set of boundary conditions, and similarly on the torus one may obtain an energy spectrum corresponding to the scaling operator spectrum on the disc. The necessary boundary conditions may always be realised as a local modification to the Hamiltonian on the anyon ring, and thus calculations on the disc may be used to compute the local scaling operator content on the torus, and vice versa. 
Considering the topological symmetry referred to as the ``flux through the torus'' in \rcite{feiguin2007}, we see that the subset of local scaling operators realised on the disc (with trivial boundary charge) constitutes those which carry trivial charge associated with this symmetry.
Related considerations apply for systems of anyons forming open chains, and these are discussed in \sref{sec:OBC}.

An important consequence of these results is that the topological protection of criticality described in \rcite{feiguin2007} for a ring of Fibonacci anyons on the torus is also seen to extend to an equivalent ring of Fibonacci anyons on the disc.

\section{Anyonic states and operators\label{sec:ASO}}

Although many papers have been published which study the behaviour of anyonic systems on surfaces of various topologies,\cite{wen1989,einarsson1990,wen1990,wen1990a,wen1990b,kitaev2003,kitaev2006,feiguin2007,trebst2008,trebst2008a,bais2009,gils2009,gils2009a,buerschaper2010,pfeifer2010,konig2010,pfeifer2011a,ardonne2011,ludwig2011,poilblanc2011,finch2011} little attention has been paid to 
how the diagrammatic formalism may be used to explicitly develop the relationship between states on surfaces of various genera. In this section we address this topic, beginning with a review of the origin and formulation of the diagrammatic representation of states and operators for systems of anyons on surfaces of genus 0 (e.g. sphere, finite disc, infinite disc) in \sref{sec:genus0}. This material may be familiar to many readers. However, we present it here in a manner intended to emphasise the relationship between anyon models and TQFTs,\TQFTcite{} as we will exploit this relationship to generalise the formalism to surfaces of higher genus in \sref{sec:genus1+}. We will also explicitly examine the construction of the translation operator on surfaces of genus 0 and 1, as this will prove important to the study of translation invariant local Hamiltonians on the disc and the torus in Secs.~\ref{sec:PBC}-\ref{sec:OBC}.

Note that in this paper we will only consider anyons on 2D manifolds which are closed, oriented, non-self-intersecting, and embedded in $\mbb{R}^3$. References to anyons on a disc are therefore taken to imply that this is a closed disc (i.e.~that the definition of the disc includes the points on the boundary). Similarly, references to anyons on the infinite disc are taken to imply inclusion of the point at infinity, so that the infinite disc is then seen to be closed by its isomorphism to the sphere $S^2$.

\subsection{Anyons on surfaces of genus 0 (disc, sphere)\label{sec:genus0}}

\subsubsection{Diagrammatic representation of states\label{sec:diagstates_sphere}}

\begin{figure}[tp]
\includegraphics[width=246.0pt]{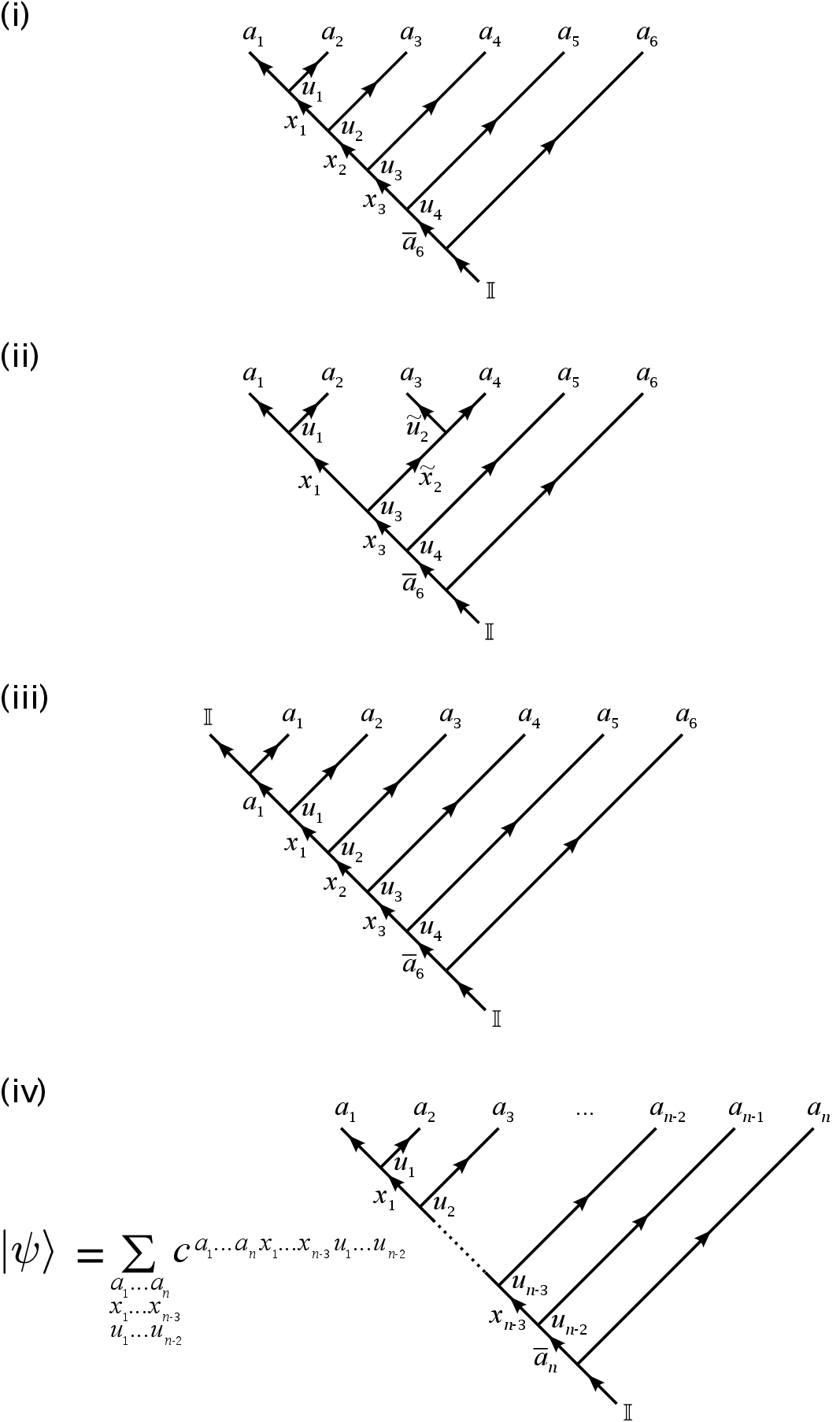}
\caption{Diagrams~(i)-(iii):~Some possible fusion trees for a chain of six anyons with charges $a_1$ to $a_6$ on a surface of genus 0 (disc or sphere). Labels $x_i$ denote intermediate fusion products which may not be uniquely determined by the fusion rules, and labels $u_i$ are associated with vertices and serve to enumerate multiple copies of a given charge for anyon models having some $N^c_{ab}>1$. Note that no vertex index is required for fusion to the vacuum state. A barred label $\bar{a}_i$ denotes the charge dual to $a_i$, such that $\bar{a}_i\times a_i\ni\mathbb{I}$. Tree (ii) is constructed from tree (i) by means of an $F$ move [\protect{\fref{fig:basischange}}(i)], and tree (iii) is constructed from tree (i) by recognising %
that fusion with the vacuum state $\mbb{I}$ is trivial.
Diagram~(iv): Example fusion tree suitable for specifying a state $|\psi\ra$ of $n$ anyons on the disc or sphere.\label{fig:anyonstates_sphere}}
\end{figure}

A system of anyons may be considered to consist of a collection of localised quasiparticle excitations in a two-dimensional medium, for example the topological liquid of a Fractional Quantum Hall (FQH) state.\cite{laughlin1983,arovas1984,girvin1987,read1989a,wilczek1990,read1990,prange1990,moore1991,read1992,das-sarma1997,read1999,xia2004,nayak2008} In general, a system may be considered anyonic if its quasiparticles may be described in terms of a Unitary Braided Tensor Category (UBTC).\footnote{Although non-unitary braided tensor categories may be described using the same formalism, their relevance to the description of physical systems is questionable.\cite{freedman2012} %
In this paper, we will reserve the term ``anyon'' for a quasiparticle excitation of a physical system which may be associated with the leaf of a labelled fusion tree as per e.g.~\protect{\fref{fig:anyonstates_sphere}}(iv), and we therefore restrict ourselves to unitary braided tensor categories.} 
In this paper we will concern ourselves only with anyon models which may be defined on the torus, known as \emph{modular} anyon models (see e.g.~Refs.~\onlinecite{bonderson2007,bonderson2008,wang2010} and \onlinecite{kitaev2006}), for which the properties of the quasiparticles admit description in terms of both a Unitary Braided Modular Tensor Category (UBMTC) and a 2+1D Topological Quantum Field Theory (TQFT)\TQFTcite{} of the Schwarz type.\cite{schwarz1978,kaul2005} Each of the quasiparticle excitations, or anyons, may then be characterised by a label, or charge, which corresponds to a label of the UBMTC. 

However, providing a full description of such a system is in general more complicated than simply cataloguing the value and location of each non-trivial charge. This is because specifying the individual charges of two anyons, $a$ and $b$, does not necessarily uniquely determine the total charge of the pair ($a\times b$). These total charges are constrained by the fusion rules of the UBMTC, which may be written in terms of the multiplicity tensor $N^c_{ab}$ as
\begin{equation}
a\times b\rightarrow \sum_c N_{ab}^c c,\label{eq:fusionrules}
\end{equation}
but when there exist nonzero entries in $N_{ab}^c$ such that multiple terms appear on the right-hand side of this equation, the total charge of $a$ and $b$ may correspond to any of these values $c$ such that $N_{ab}^c\not=0$.
To specify these products, we represent the state of a system of anyons by means of a \emph{fusion tree} (\fref{fig:anyonstates_sphere}).
Labels on the interior edges of the fusion tree graph correspond to the results which would be obtained on measuring the total charge of multiple anyons. For example in \fref{fig:anyonstates_sphere}(i), $x_1$ is the total charge of anyons $a_1$ and $a_2$ together, $x_2$ is the total charge of anyons $a_1$, $a_2$, and $a_3$ together, and so on. The set of valid labellings of a single fusion tree constitutes an orthogonal basis for the Hilbert space of a system of fixed anyons, and a labelling is deemed valid if all fusion vertices correspond to processes associated with non-zero entries in the multiplicity tensor $N_{ab}^c$. In this paper we will normalise all fusion tree bases using factors appropriate to the diagrammatic isotopy convention given in Refs.~\onlinecite{bonderson2007} and \onlinecite{bonderson2008}. For mobile anyons, the co-ordinates of each anyon must be specified in addition to the fusion tree.

\begin{figure}[tp]
\includegraphics[width=246.0pt]{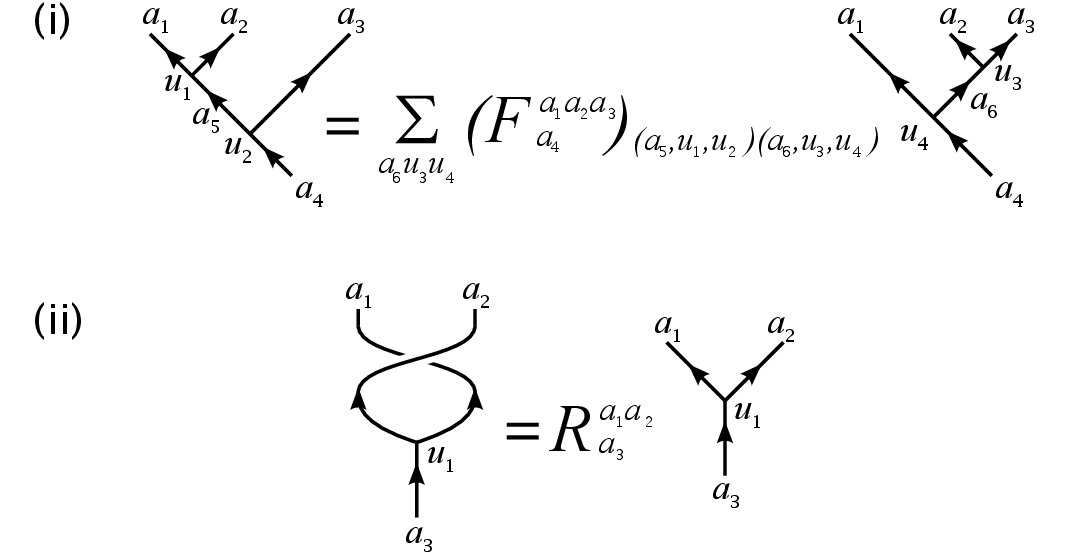}
\caption{Manipulations capable of performing a change of basis on a fusion tree: (i) $F$ move. (ii) Braiding.\label{fig:basischange}}
\end{figure}

Although a single fusion tree does not explicitly state the outcome of all possible measurements, it is possible to convert between different fusion trees using procedures known as $F$ moves and braiding (\fref{fig:basischange}).
In constructing a fusion tree, we have imposed a (possibly arbitrary) linear ordering on the anyons of the system. An $F$ move [\fref{fig:basischange}(i)] alters the structure of the fusion tree while preserving that linear ordering, permitting the computation of additional fusion products [e.g. $\tilde x_2$ in \fref{fig:anyonstates_sphere}(ii) is the combined charge of $a_3$ and $a_4$], while braiding [\fref{fig:basischange}(ii)] permits conversion between different linear orderings.\footnote{In this instance, the process of braiding represents a passive transformation between equivalent bases representing the same physical state. We will subsequently also encounter the use of braiding to denote the active process of particle exchange (\protect{\sref{sec:discoperators}}).} Using these two operations it is possible to determine the probability amplitudes of different outcomes when measuring the total charge of any group of anyons regardless of the fusion tree structure on which the state is initially described.
The tensors $\left(F^{a_1a_2a_3}_{a_4}\right)_{(a_5u_1u_2)(a_6u_3u_4)}$ and $R^{a_1a_2}_{a_3}$ are specified by the UBMTC to which the system of anyons corresponds.

\begin{figure}
\includegraphics[width=246.0pt]{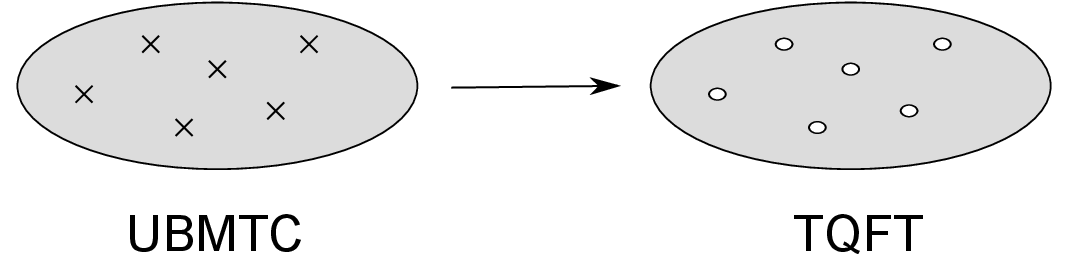}
\caption{Anyonic quasiparticles carrying labels from the UBMTC (denoted $\times$) map to punctures in the manifold when the system is represented by a TQFT.\label{fig:punctures}}
\end{figure}

While the associated UBMTC describes a system of anyons in terms of individual quasiparticles, an equivalent description may also be made in terms of the diffeomorphism-invariant fields of a Schwarz-type 2+1D TQFT. Here, the 2D manifold on which the anyons exist becomes the spatial manifold of the TQFT, with individual anyons corresponding to punctures in this manifold (\fref{fig:punctures}), and the anyon charges corresponding to the individual punctures' boundary charges. 
One may then construct a basis for the system
in terms of the outcomes of %
a complete set of commuting Wilson loop operators, whose expectation values may be identified with the labels of the UBMTC. In a TQFT, a pair of Wilson loop operators which are topologically equivalent 
necessarily constitute a measurement of the same observable. Furthermore, the outcome of a Wilson loop measurement which may be contracted to a point is necessarily trivial. Consequently, we may identify the expectation value of an appropriate Wilson loop operator (of specified orientation, to allow for charges which are not self-dual) with measurement of the total charge on the anyons, or punctures, which it encloses. Where degeneracies exist (i.e. $N_{ab}^c>1$ for some $a,b,c$), the different copies of a particular charge label in the UBMTC can be associated with different expectation values of the Wilson loop operator. %
At a slightly less abstract level, when the Wilson loop operator encircles exactly one puncture, the resulting label from the UBMTC may be understood as an (anyonic) charge associated with the boundary of the puncture.

\begin{figure}[tp]
\includegraphics[width=246.0pt]{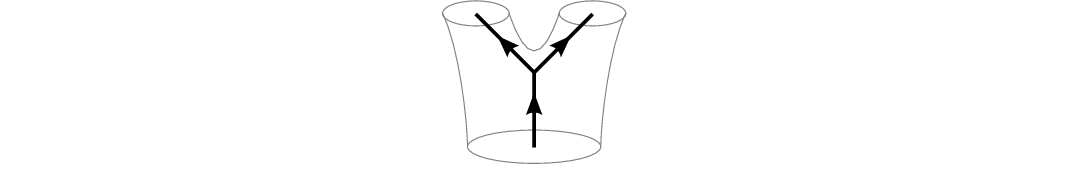}
\caption{Construction of a fusion tree graph from a pair of pants.\label{fig:pantsconstruction}}
\end{figure}

We therefore see that we may map between the TQFT and the UBMTC fusion tree as follows:
First, perform a pairs-of-pants decomposition of the punctured 2D spatial manifold of the TQFT. 
Then, take one specific pair of pants (or 3-punctured 2-sphere), declare that this 2-sphere has an inside and an outside, and specify which is which. Extend this definition of inside and outside consistently over all pairs of pants (this is always possible for a non-self-intersecting, closed, orientable 2-manifold embedded in $\mbb{R}^3$). Having done this, construct the fusion tree by drawing lines inside each pair of pants as shown in \fref{fig:pantsconstruction}.
Now associate a Wilson loop operator which measures charge with each opening of each pair of pants, up to topological equivalence. Specifically, where two pairs of pants connect together, we find two charge measurement operators which are topologically equivalent and so only one of these need be retained. Each line of the fusion tree graph now passes through exactly one Wilson loop, and we label the lines of the graph with the outcome of these charge measurements. 
Finally, if there exist entries in the multiplicity tensor $N^c_{ab}$ 
which are greater than 1, then it is also necessary to associate a degeneracy index with the fusion vertices to enumerate these outcomes, which are also assumed to be specified by the outcomes of %
appropriate measurements.

So far, this fusion tree has been constructed in the space $\mathbb{R}^3$ in which the 2D spatial manifold is embedded. When representing this three-dimensional construction on paper, it is customary to employ a diagrammatic convention whereby the 2D manifold on which the punctures exist is mapped onto a plane perpendicular to the page, and whose projection onto that page forms a horizontal line at the top of the fusion tree diagram (for systems of anyons on the sphere, this is achieved by first identifying that sphere with the Riemann sphere, and then mapping to the infinite plane). Noting that a disc manifold with non-trivial boundary charge is topologically equivalent to a sphere with a large puncture at the south pole, if \emph{all} punctures with non-trivial charges %
are brought to lie on the line at the top of the fusion tree diagram, then the vertical axis of fusion trees drawn in this way (e.g. \fref{fig:anyonstates_sphere}) may then be interpreted as a possible history whereby the present physical state may be obtained from the vacuum (i.e. a state with no punctures),
and corresponds to
the timelike dimension of the TQFT. The lines of the fusion tree correspond to world lines for the quasiparticles presently observed on the manifold, and %
a charge label $\mathbb{I}$ is placed at the bottom of the fusion tree diagram, representing an initial vacuum state.\footnote{We are free to do this because for any UBMTC there exists an identity charge, denoted $\mbb{I}$, such that fusion with this charge does not modify either the space of labels of a given fusion tree, or an individual labelling of this tree. We may therefore, on any diagram, freely insert and delete lines carrying only the trivial charge $\mbb{I}$ without modifying the state which this diagram represents.}%

(Note that for any system of anyons, it is always possible to set the charge at the bottom of the fusion tree to $\mbb{I}$, as described above. This is because there are no fundamental anyonic particles, and thus any system of anyons must always initially be created from a state with trivial anyonic charge. However, when some anyons are then removed by an arbitrarily large distance, leaving behind a subsystem with non-trivial total charge, this may be represented either by explicitly including a branch of the fusion tree corresponding to these distantly removed anyons, or by assigning a fixed, non-trivial total charge to the fusion tree. In this paper, we find it %
appropriate to keep track of the location of all anyons, and therefore consider only systems of anyons with a total charge $\mbb{I}$.)

\begin{figure}
\includegraphics[width=246.0pt]{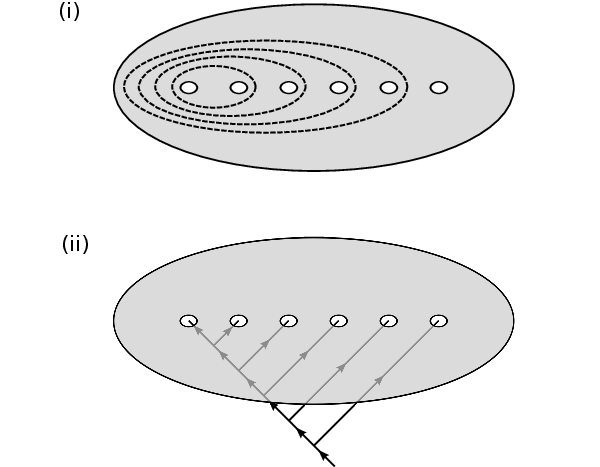}
\caption{(i)~A sample pairs-of-pants decomposition for a 6-punctured finite disc: the manifold is decomposed into pairs of pants by cutting along %
the dotted lines. When the charge associated with the boundary is $\mbb{I}$, the construction described in \protect{\sref{sec:diagstates_sphere}} yields the fusion tree of \protect{\fref{fig:anyonstates_sphere}}(i). (ii)~Relationship of the fusion tree to the manifold of diagram~(i).\label{fig:manifoldandtree}}
\end{figure}

We now observe that by adopting different pairs-of-pants decompositions of the spatial manifold of the TQFT, it is possible to recover all different fusion tree bases of the UBMTC. It is also possible to interchange the definitions of ``inside'' and ``outside'' when constructing the fusion tree from the pairs-of-pants decomposition, but for surfaces of genus 0 this has no effect on the basis obtained.
In \fref{fig:manifoldandtree} we give a simple example of the pairs-of-pants construction, showing the decomposition of a 6-punctured finite disc with trivial charge on the boundary which corresponds to the fusion tree of \fref{fig:anyonstates_sphere}(ii).

We conclude this section with a couple of remarks about specific systems of genus 0:
First, to extend the pairs-of-pants construction to surfaces having less than three punctures, such as the 2-punctured 2-sphere, we note that fusion with the identity label $\mbb{I}$ is trivial. For such a system we may therefore freely introduce additional trivial punctures to obtain a single 3-punctured 2-sphere from which we construct the fusion tree. Similarly, lines carrying trivial charge may be freely added to or removed from any fusion tree diagram [e.g. to obtain \fref{fig:anyonstates_sphere}(iii) from \fref{fig:anyonstates_sphere}(i)].
Second, we note that there exists an important relationship between the sphere and the finite disc. While the infinite disc is topologically equivalent to the Riemann sphere, the finite disc may be treated as the Riemann sphere with a puncture at infinity. The edge of this puncture then constitutes the edge of the disc. The charge associated with the edge of the disc is measured by a Wilson loop of the usual orientation enclosing this puncture on the Riemann sphere, or equivalently, one of reversed orientation enclosing all other punctures on the disc. When the charge associated with this puncture on the Riemann sphere is (and remains) trivial, we may delete the associated line from the fusion tree diagram, and therefore ignore the existence of the boundary when studying anyon behaviour on the finite disc. Third, we note that in the study of lattice models with $n$ sites, we may treat the system as always containing $n$ anyons at fixed locations, even if some of these anyons have trivial charge. The states of these systems can therefore always be represented by a fusion tree with $n$ leaves. The enumeration of the leaves of the fusion tree then corresponds to an enumeration of the lattice sites, and consequently for such a system it is not necessary to separately state the co-ordinates of the individual anyons.

\subsubsection{Inner product\label{sec:discinnerproduct}}

Next, we introduce the diagrammatic representation of the dual space and the inner product. In the diagrammatic representation of the space of states, the conjugation operation $^\dagger$ is implemented by vertically reflecting a fusion tree to obtain a \emph{splitting tree}, taking the complex conjugate of all fusion tree coefficients $c^{a_1\ldots a_{2n}}$, and reversing the direction of all arrows on the tree. In this paper, we will prefer lower indices for the coefficients of a splitting tree, e.g. $c'_{a_1\ldots a_{2n}}$.
The inner product of two diagrams is then performed by connecting the leaves of the fusion and the splitting tree, subject to the requirement that leaves which are connected represent anyons (or punctures) at the same location on the manifold, and that the charges of the connected leaves coincide. Where these conditions do not hold, the inner product of two diagrams is zero.
Recall that the fusion tree is a 2+1-dimensional structure projected onto a two-dimensional page, and thus when performing this connection, both trees must be represented in equivalent projections. Conversion between projections may be achieved by a sequence of appropriately oriented braids.

\begin{figure}[tp]
\includegraphics[width=246.0pt]{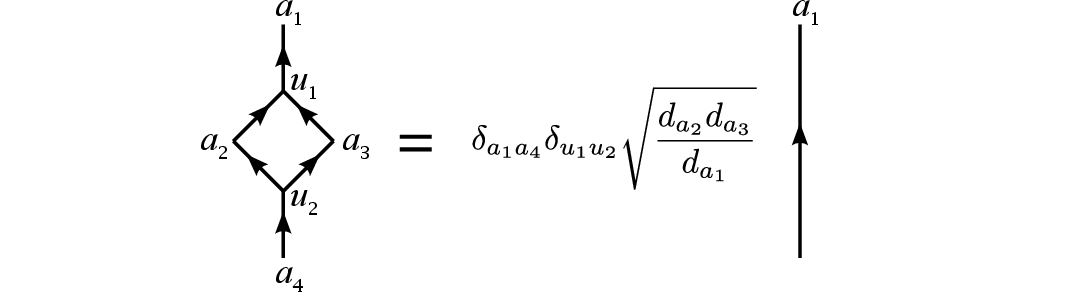}
\caption{Elimination of loops during evaluation of the inner product. The numerical factor given is appropriate to the diagrammatic isotopy convention.\label{fig:innerprod_sphere}}
\end{figure}

Assuming that the inner product has not yet been found to be zero, then once the trees have been connected, $F$ moves are performed, loops are eliminated according to the rule given in \fref{fig:innerprod_sphere}, and trivial punctures are removed, until the resulting diagram has been reduced to a number. This number is then the value of the inner product.
Extension to states represented by a weighted sum over multiple labelled diagrams follows from bilinearity.

\subsubsection{Diagrammatic representation of operators\label{sec:discoperators}}

\begin{figure}[tp]
\includegraphics[width=246.0pt]{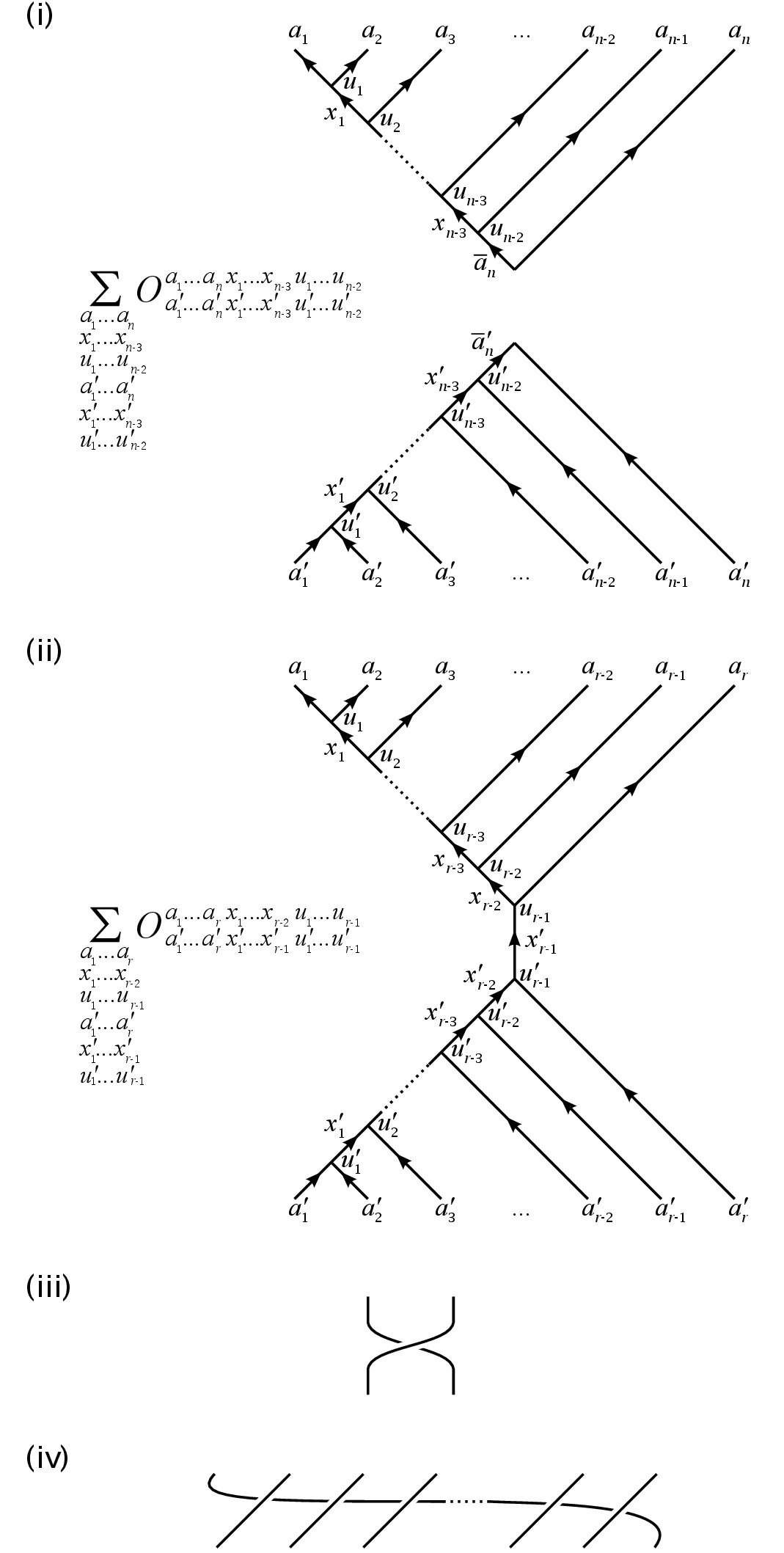}
\caption{Examples of anyonic operators on the disc with $n$ punctures. (i)~A global operator, acting on the total Hilbert space of the system. (ii)~A local operator, acting on $r$ adjacent anyons.
(iii)~The braid operator.
(iv)~The periodic translation operator $\hat T^\mrm{D}$ for a ring of anyons on fixed lattice sites on the disc, closing away from the observer.\label{fig:discoperators}}
\end{figure}

Now that we have presented the diagrammatic formulation for anyonic states and for the inner product, we are in a position to construct anyonic operators. Where these operators act on the entire system, the construction is trivial as an operator is constructed in the usual manner, as a sum over bras and kets:
\begin{equation}
\hat O = \sum_{i,j} O_{ij} |\psi_i\ra\la\psi_j|.\label{eq:braketoperator}
\end{equation}
For anyons the bra is replaced by a splitting tree, the ket by a fusion tree, and the coefficient bears indices corresponding to all labels on the splitting and fusion trees [e.g. \fref{fig:discoperators}(i)].
However, we may also wish to define operators which act only on a finite subregion of the disc. In the same way that the fusion tree specifies how all the anyons in the system may be obtained starting from the vacuum state, in an appropriate basis we may interpret a portion of the fusion tree as specifying how all the anyons within a physically localised subregion may be obtained from a single initial charge. For example, in \fref{fig:anyonstates_sphere}(i), we see that charges $a_1$, $a_2$ and $a_3$ are obtained by splitting an initial charge of $x_2$, and in diagram (ii), charges $a_3$ and $a_4$ are obtained from $\tilde x_2$. We require that our operators respect superselection rules associated with 
the charge labels of the UBMTC, and consequently they cannot change this total charge, but their action within this region is otherwise unconstrained. A completely general local 
operator acting on $r$ sites on the disc may therefore be written in the form of \fref{fig:discoperators}(ii), where the description of the operator as ``local'' means that the $r$ sites are consecutive in the adopted fusion tree basis. As with the state of a system, the choice of fusion and splitting trees employed in this figure merely represent a choice of basis in which to represent the operator, and any alternative choice would have been equally valid. We also note that this construction includes the definition of a global operator on the disc, as the special case $r=n$.

\begin{figure}
\includegraphics[width=246.0pt]{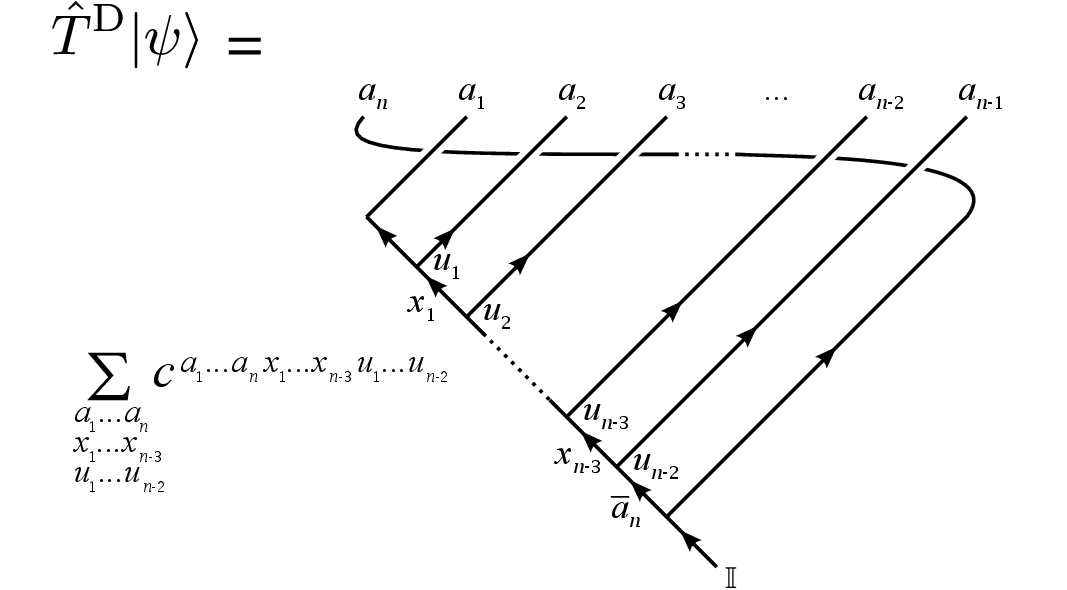}
\caption{Application of the translation operator $\hat T^\Sp$ to a state of $n$ anyons on the disc. The untranslated state $|\psi\ra$ is given in \fref{fig:anyonstates_sphere}(iv).\label{fig:translatedisc}}
\end{figure}

Finally, we note that while any operator on the disc may be represented in the form of \fref{fig:discoperators}(ii), it may frequently be advantageous to represent certain special operators in other forms. Thus, for example, while the braid operator corresponding to the oriented exchange of a pair of anyons may be represented in the form of \fref{fig:discoperators}(ii) for $r=2$, it is usually more convenient to represent it in the form of \fref{fig:discoperators}(iii), from which its unitarity is obvious by diagrammatic isotopy. Similarly, consider a ring of anyons occupying fixed lattice sites on the disc. Exploiting topological invariance, we may construct our fusion tree such that these lattice sites lie in a line at the top of the diagram, and the closure of the ring is implicit, being either towards or away from the observer. If, for definiteness, we assume that the ring closes away from the observer, then we may expediently represent the operator corresponding to periodic translation by one site using the diagram of \fref{fig:discoperators}(iv). Note that this operator may be constructed by composing a series of braids [\fref{fig:discoperators}(iii)], and also that it
respects the interpretation of the vertical axis as a fictional timeline for the creation of the state, as 
the motions of the anyons under the action of this operator are strictly monotonic in time. When this operator is applied to a state, the resulting diagram then describes a process whereby particles are created, migrate to their initial lattice sites, and then %
all move one site periodically around the lattice (\fref{fig:translatedisc}). 

\subsection{Anyons on surfaces of higher genus (\lowercase{e.g.} torus)\label{sec:genus1+}}

Having reviewed the diagrammatic formulation for systems of anyons on the disc or sphere, we will now extend this formulation to surfaces of higher genus, by exploiting the association between modular anyon models and 2+1D TQFTs.

\subsubsection{Diagrammatic representation of states\label{sec:torusfusiontree}}

Extension to surfaces of higher genus is achieved by means of manifold surgery, performed on the punctured manifold inhabited by the 2+1D TQFT. We will be particularly interested in a specific example, the $n$-punctured torus, but the techniques which we will develop are entirely general and thus may be applied to construct diagrammatic representations for states of anyonic systems on surfaces of arbitrary genus.

\begin{figure}
\includegraphics[width=246.0pt]{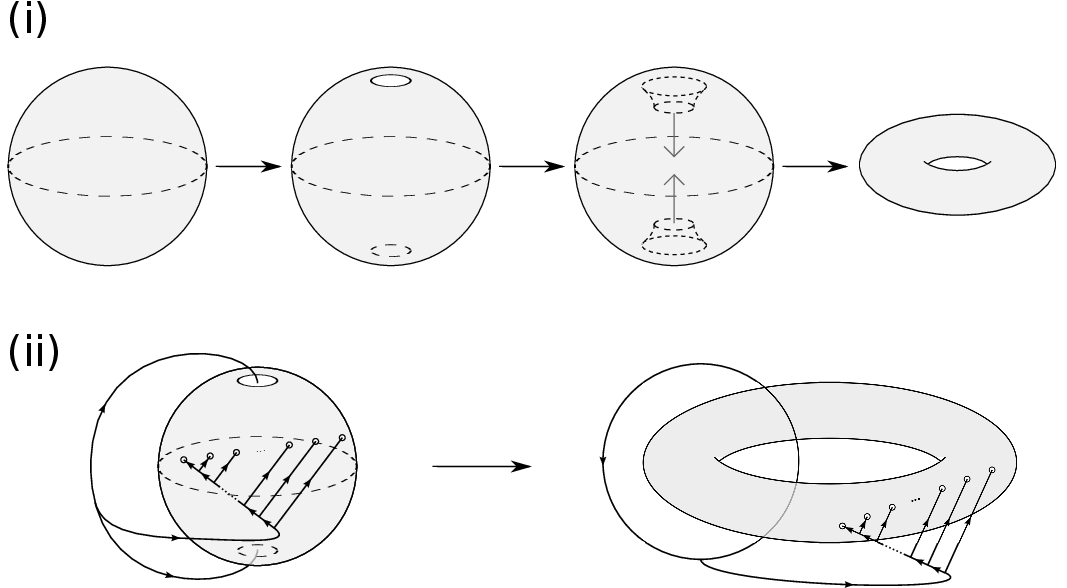}
\caption{(i)~Construction of the torus from the sphere by introducing two punctures, deforming the resulting punctured sphere by migrating the punctures towards the centre, and suturing.\label{fig:maketorus}
(ii)~Construction of an ``outside'' fusion tree for the $(n-2)$-punctured torus, starting from a fusion tree ``outside'' the $n$-punctured sphere.}
\end{figure}

We begin by noting that the %
torus may be constructed from the %
sphere by introducing punctures at the north and south poles, then distorting the sphere so that the puncture at the north pole descends vertically, and the puncture at the south pole rises vertically. When these punctures come into contact, they are sutured [\fref{fig:maketorus}(i)].

\begin{figure}[tp]
\includegraphics[width=246.0pt]{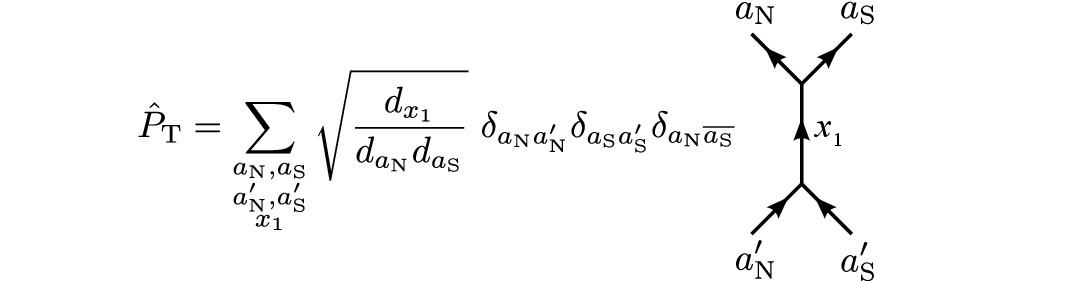}
\caption{Operator $\hat P_\T$ (in the diagrammatic isotopy convention).\label{fig:Pt}}
\end{figure}

Now, we wish to repeat this process for a manifold on which there exists a TQFT. We recognise that 
through the use of Wilson loop operators, charge labels may be
associated with the punctures $a_\mrm{N}$ and $a_\mrm{S}$ at the north and south poles of the sphere respectively. On the sphere, prior to performing the suturing, these observables are independent. On the torus, after suturing, they are topologically equivalent up to a reversal in orientation. Importantly, these observables may be computed purely from the fields on the path of the loop itself, and thus their calculation proceeds identically whether or not the punctures are sutured. From this we infer two important results. First, suturing of these punctures only yields a consistent TQFT on the torus if the values of all Wilson loop observables on the north puncture are the duals of the same observables evaluated on the south puncture. Second, the space of states for the TQFT on the torus is isomorphic to the space of states on the 2-punctured sphere subject to this constraint. In \fref{fig:Pt} we see the operator $\hat P_\T$ which projects from the Hilbert space of the \textit{n}+2-punctured sphere to a reduced Hilbert space isomorphic to the Hilbert space of the $n$-punctured torus.
If we now describe the Hilbert space on the \textit{n}+2-punctured sphere in terms of a fusion tree in the region of $\mbb{R}_3$ colloquially described as ``outside'' the sphere (i.e. extending from the surface of the unpunctured sphere to infinity), then we may use the surgical procedure described to construct a fusion tree for the $n$-punctured torus. Bringing together and suturing the punctures at the north and south poles corresponds to bringing together the equivalent branches of the fusion tree to form a loop [\fref{fig:maketorus}(ii)]. 

\begin{figure}
\includegraphics[width=246.0pt]{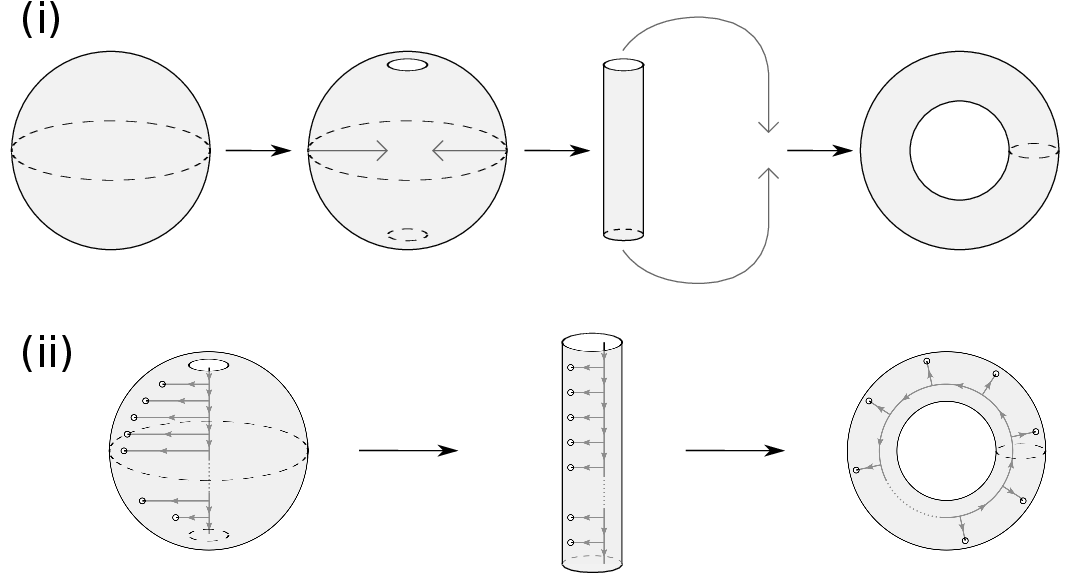}
\caption{(i)~Alternative procedure to construct a torus from the 2-punctured sphere. The procedure presented here is 
compatible with a fusion tree constructed ``inside'' the torus, whereas the procedure presented in \protect{\fref{fig:maketorus}} is %
compatible with a fusion tree constructed ``outside'' the torus.\label{fig:spheretotorus2}
(ii)~Construction of an ``inside'' fusion tree for the $(n-2)$-punctured torus from a fusion tree ``inside'' the $n$-punctured sphere.}
\end{figure}

It is important to note that when constructing the torus from the sphere by means of the surgery procedure described, one necessarily obtains a fusion tree in the region which is again ``outside'' the torus. This is because the branches of the fusion tree on the sphere which terminate in the north pole and south pole punctures must close to form a non-trivial cycle around the torus, and this can only occur if the fusion tree on the sphere inhabits the ``outside'' space. A fusion tree ``inside'' the torus may be obtained by the alternative procedure of first constructing a fusion tree ``inside'' the sphere, lengthening the sphere into a hollow cylinder with the polar punctures at its ends, and then bending this cylinder around into a loop and suturing (\fref{fig:spheretotorus2}). 

Given the existence of this relationship between anyon models on surfaces of higher genus and anyon models on the sphere, we see that on surfaces of higher genus we may employ the pairs-of-pants decomposition approach to construct a fusion tree %
in $\mbb{R}^3$ in precisely the same way as we did for the sphere.
For diagrammatic isotopy conventions to apply, we must now map this fusion tree to the plane of the page in a manner such that the vertical axis corresponds to the time dimension of the 2+1D TQFT and the horizontal axis is a projection of the spatial degrees of freedom, just like we did on the disc.

\begin{figure}
\includegraphics[width=246.0pt]{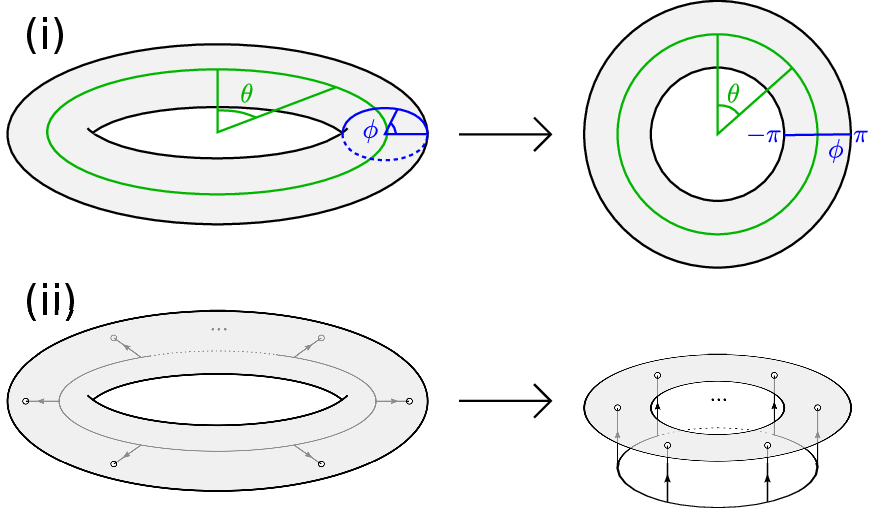}
\caption{COLOUR ONLINE. (i)~Mapping from the torus to an annulus embedded in $\mbb{R}^2$. Note that this map is topology-preserving if the points on the inner and outer edges of the annulus are identified. (ii)~A fusion tree constructed ``inside'' the torus is mapped into a fusion tree for the annulus with identifications.\label{fig:annulus}}
\end{figure}

We begin by mapping the surface of the torus to an annulus on $\mbb{R}^2$, such that the map will be topology-preserving if the inner and outer borders of the annulus are identified (\fref{fig:annulus}).
\begin{figure*}[tp]
\includegraphics[width=492.0pt]{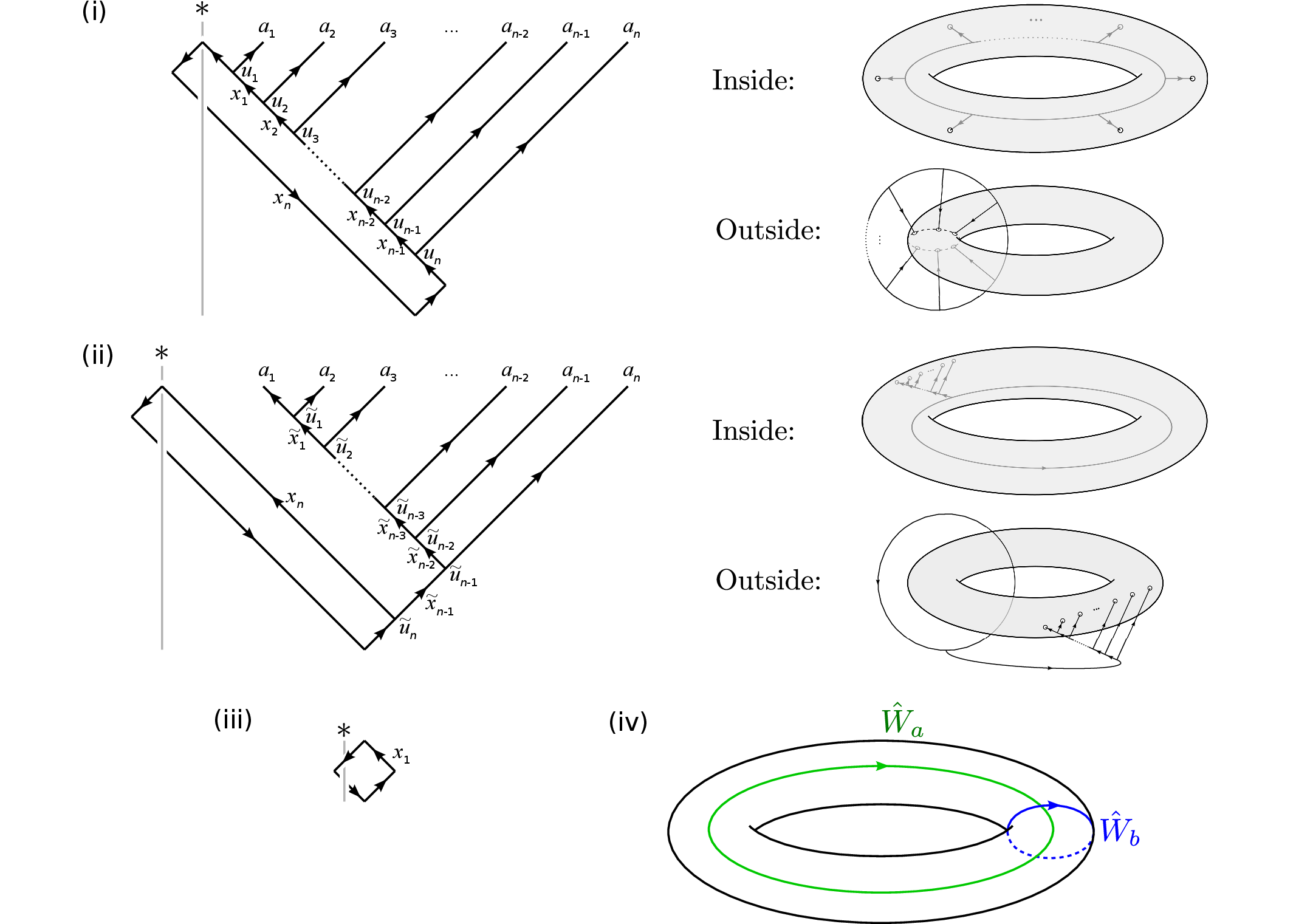}
\caption{COLOUR ONLINE. (i)-(ii)~Fusion trees for systems of $n$ anyons on the torus. The corresponding bases are related by means of a series of $F$ moves, and will be denoted $B_1$ and $B_2$ respectively.
A line marked $*$ constitutes a topological obstruction to the contraction of loops in the fusion tree, indicating that the loop encircles a non-trivial cycle on the torus. Such loops constitute an important part of the description of the state and cannot be eliminated using \protect{\fref{fig:innerprod_sphere}}. %
It is important to recognise that bases of types $B_1$ and $B_2$ may be constructed using either the ``inside'' or the ``outside'' space of the torus, with the relevant constructions being indicated in the accompanying illustrations.
Note that for any given state, the value of $x_n$ is unaffected by changing between a basis of form $B_1$ and one of form $B_2$, and this is reflected in the labelling of the diagrams.
(iii)~Fusion tree for the unpunctured torus. (iv)~Measurements $\hat W_a$ and $\hat W_b$ are associated with non-trivial cycles on the torus.
\label{fig:torusstates}}
\end{figure*}%
We also require that the mapping be chosen so that when the annulus is viewed from above as in \fref{fig:annulus}, the fusion tree may be drawn so that it does not pass under the borders of the annulus. For a fusion tree constructed inside the torus, the borders of the annulus are therefore topologically equivalent to the line of suturing in \fref{fig:maketorus}, and for a fusion tree constructed outside the torus, they are equivalent to the line of suturing in \fref{fig:spheretotorus2}. We now expand the annulus to cover all of the plane $\mbb{R}^2$ except for arbitrarily small discs enclosing the points at the origin and at infinity, marking these points with a star ($*$) and subtending from each of them a line perpendicular to the plane.\footnote{We emphasise that \protect{$*$} denotes an arbitrarily small disc and not a point, as the orientation of the mapping between these discs may potentially become relevant when working with fusion trees which traverse an identification.} Viewing the annulus from the side, so that the copy of $\mbb{R}^2$ into which it is embedded appears as a line, we now project this diagram onto the page to obtain the fusion tree. 

Some example fusion trees for the $n$-punctured torus are shown in \fref{fig:torusstates}(i)-(ii). 
Note the presence of the line labelled $*$, corresponding to the position of the centre of the annulus. This line may be thought of as tracking the location of an obstruction in $\mbb{R}^2$ as a function of time. By virtue of the mapping to the annulus used in the construction of the fusion tree, an arbitrarily small ring encircling this point may be identified with a similarly arbitrarily small ring encircling the point at infinity (which is not shown, but could be brought in to lie on the edge of the diagram by means of an appropriate topology-preserving map, and would then likewise be labelled $*$). Because of this identification, it is impossible to smoothly deform the labelled lines of the fusion tree through the line marked $*$ according to the usual rules of diagrammatic isotopy. Consequently, loops which encircle this obstruction cannot be contracted and eliminated using the identity given in \fref{fig:innerprod_sphere}. More generally, the embedding for a surface of genus $g$ will result in $2g$ such pairs of identifications in $\mbb{R}^2$.

The unpunctured torus has a fusion tree which is given in \fref{fig:torusstates}(iii), and may be obtained using the pairs-of-pants approach by introducing a trivial puncture on the torus, constructing the fusion tree (where the puncture with the inward arrow in \fref{fig:pantsconstruction} is sutured to one of the punctures with an outward arrow), and then deleting the line carrying charge $\mbb{I}$ which is associated with the trivial puncture. 
Again in these examples, each line on the fusion tree diagram may be associated with the outcome of a particular Wilson loop operation in the TQFT. However, on this occasion in addition to the measurements associated with punctures on the surface of the torus, there are also two measurements associated with the non-trivial cycles of the torus [\fref{fig:torusstates}(iv)], and in a given basis, only one of these will encircle a line of the fusion tree. For example, consider the torus with no punctures. The fusion tree may be constructed either ``outside'' the torus (in the region of $\mbb{R}^3$ which extends to infinity), or ``inside'' the torus (in the region of $\mbb{R}_3$ which does not extend to infinity).
The labellings of these two fusion trees both constitute a basis of states, and they are related by means of the topological $S$ matrix, 
\begin{align}
S_{ab}&=\frac{1}{\mc{D}}~\raisebox{-13.5pt}{\includegraphics{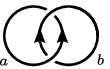}}\label{eq:Smatrix}\\ %
\rule{0pt}{22pt}\mc{D}&=\sqrt{\sum_a d_a^2},\label{eq:totqdim}
\end{align}
according to
\begin{equation}
\raisebox{-12pt}{\includegraphics[width=220.0pt]{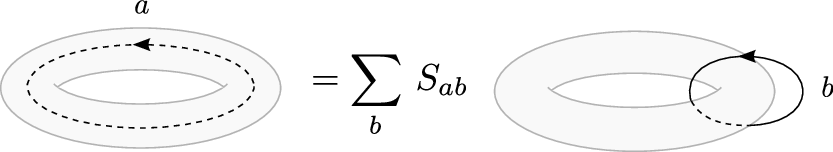}.}\label{eq:Srelate}
\end{equation}
(Note that an anyon model can consequently only be consistently defined on the torus iff the topological $S$ matrix is unitary. This property is the defining characteristic of a modular anyon model.) 

In one of these bases, the fusion tree is encircled by Wilson loop operator $\hat W_A$ of \fref{fig:torusstates}(iv), and in the other basis, by operator $\hat W_B$. Thus by describing a state in one of these bases, we specify the probability amplitudes for the outcomes of measurements around both non-trivial cycles of the torus. %
 
We note that for the torus without punctures, the fusion tree admits as many different labellings as there are species of anyons in the model. 
The Hilbert space of the unpunctured torus is thus $|U|$-dimensional, where $|U|$ represents the number of inequivalent labels in a UBMTC $U$. 
As an example consider Kitaev's toric code, which is commonly understood as exhibiting independent electric and magnetic charges, and has fusion rules corresponding to the quantum double of $\mbb{Z}_2$. 
As the fusion rules of $D(\mbb{Z}_2)$ are Abelian, the elements of $D(\mbb{Z}_2)$ are in 1:1 correspondence with its representations, and 
consequently we may associate the charges of the corresponding UBMTC to the elements of $D(\mbb{Z}_2)$.
Following convention, we may denote these charges $\mbb{I}$, $e$, $m$, and $em$. In the language of electric and magnetic charges, $\mbb{I}$ is the uncharged vacuum state, $e$ corresponds to the presence of an electric charge, $m$ to a magnetic charge, and $em$ corresponds to the presence of both.
For the toric code, all states without punctures are ground states, and thus on the torus the ground state subspace has dimension 4; equivalently, we may say that the ground state on the torus is 4-fold degenerate. Similarly the dimension of the Hilbert space of states on an unpunctured manifold of genus $g$ can easily be seen to be $|U|^g$, and this reproduces the well-known ground state degeneracy of $4^g$ for the toric code on a surface of genus $g$.

Finally, we draw attention to charge $\tilde x_{n-1}$ in \fref{fig:torusstates}(ii). Due to the presence of the topological obstruction denoted by $*$, the loop in this fusion tree is not subject to the delta-function constraints of \fref{fig:innerprod_sphere} which prohibit the existence of tadpole diagrams on the disc. On the torus, charge $\tilde x_{n-1}$ is not constrained to be $\mbb{I}$.

\subsubsection{Inner product\label{sec:torusinnerprod}}

We now introduce a process for computing the inner product on the torus, which is derived from the inner product on the sphere by means of the process of manifold surgery described in \sref{sec:torusfusiontree}. 
This construction generalises immediately to all orientable non-self-intersecting surfaces of higher genus. 

Consider the inner product $\la\psi'^\T|\psi^\T\ra$ between two states $|\psi^\T\ra$ and $|\psi'^\T\ra$ on the torus. For each state in turn we 
reverse 
the construction given in \sref{sec:torusfusiontree}, cutting the 
torus so that it is transformed into a surface isomorphic to the sphere with punctures at north and south poles, and then mapping each state $|\psi^\T\ra$, $|\psi'^\T\ra$ on the torus to an equivalent state $|\psi^\Sp\ra$, $|\psi'^\Sp\ra$ lying within the support of $\hat P_\T$ on the disc.
As a notation convention, superscripts of T and D in this paper will be used to indicate that a particular state or operator lives on the torus or sphere/disc respectively. In contrast the $_\T$ on $\hat P_\T$ is written in subscript, and so is just part of the name we have chosen for this operator and does not denote the topology of the manifold on which the operator exists.
The inner product between two states on the torus is now simply taken to be the inner product between the two equivalent states on the disc,
\begin{equation}
\la\psi'^\T|\psi^\T\ra = \la\psi'^\Sp|\psi^\Sp\ra.\label{eq:torusIP}
\end{equation}

\begin{figure}[tp]
\includegraphics[width=246.0pt]{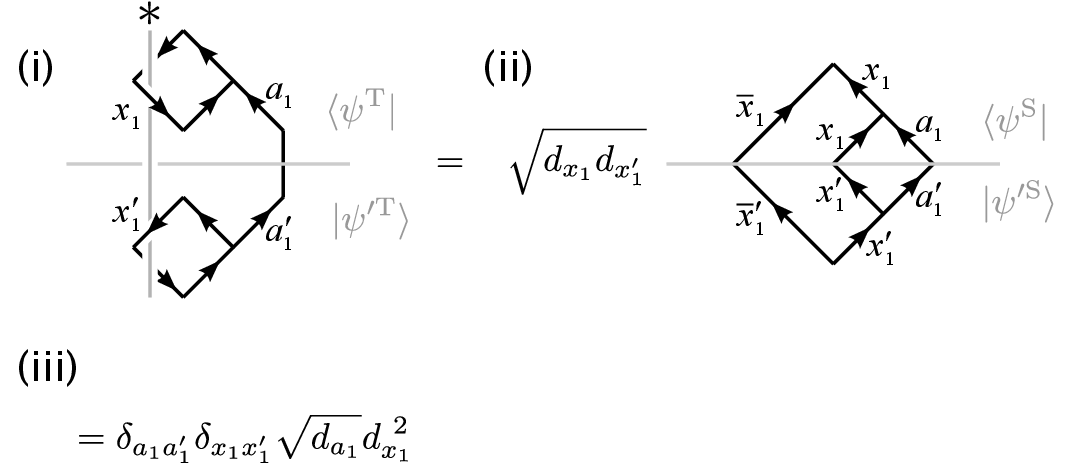}
\caption{Evaluation of the inner product on the singly-punctured torus, in the diagrammatic isotopy convention. (i)~Opposition of torus fusion and splitting trees. (ii)~Equivalent diagram on the sphere. (iii)~Numerical value. Note that in the diagrammatic isotopy convention, mapping states on the torus to states on the sphere effectively amounts to the removal of two vertices, along with their associated numerical factors. This introduces a factor of $\sqrt{d_{x_1}d_{x'_1}}$ in step (ii).
\label{fig:innerprod_torus}}
\end{figure}

We may therefore summarise the computation of the inner product on the torus as follows: First, the fusion and splitting trees are connected at their leaves, as described for the sphere, and any mismatch between charges results in an inner product of zero. If the inner product has not yet been found to be zero, then $F$ moves and \fref{fig:innerprod_sphere} are applied repeatedly until the diagram is reduced to a sum of terms having the form shown in \fref{fig:innerprod_torus}(i). %
\begin{figure}
\includegraphics[width=246.0pt]{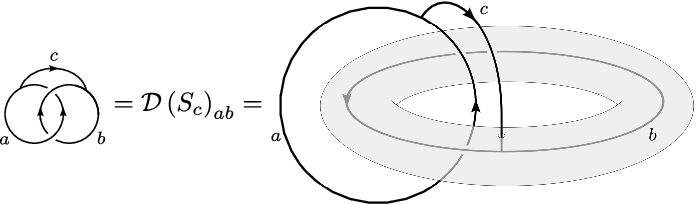}
\caption{The generalisation of the topological $S$ matrix, $\left(S_c\right)_{ab}$, enables us to evaluate the inner product of two states when one of these states is written in a basis constructed ``inside'' the torus, and one is written in a basis constructed ``outside'' the torus. Using this relationship it is possible to convert between bases constructed ``inside'' the torus and bases constructed ``outside'' the torus. $\mc{D}$ is the total quantum dimension, given by \pEref{eq:totqdim}.\label{fig:insideoutside}}
\end{figure}%
These are then evaluated as shown in \fref{fig:innerprod_torus}(ii)--(iii), to obtain the 
value of
the inner product 
(a generalisation of the topological $S$ matrix\cite{kitaev2006} may be used to convert between ``inside'' and ``outside'' bases for the torus, by means of the relationship given in \fref{fig:insideoutside}).

It is instructive to compare this formulation of the inner product with that presented in Appendix~A of \rcite{konig2010}. The formulation of the inner product introduced by \citeauthor{konig2010} similarly guarantees that the physically permissible unique labellings of the fusion tree of the punctured torus yield an orthogonal basis for the Hilbert space, and differs only in the normalisation factors which must be associated with some of the diagrams 
(see \tref{tab:torusnorms}).
\begin{figure}[tp]
\includegraphics[width=246.0pt]{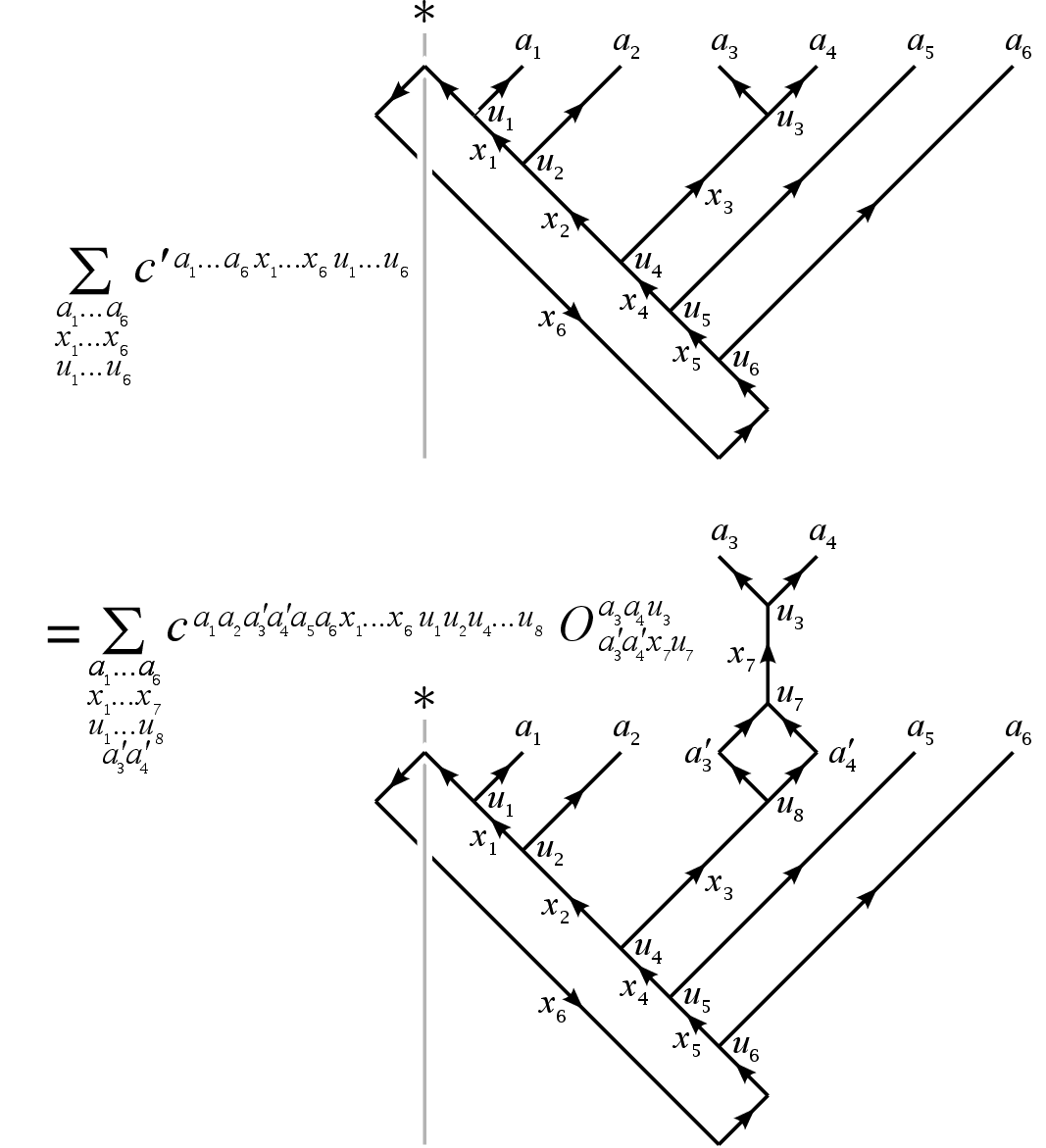}
\caption{An operator $\hat O^\T$ acts on a region of the 6-punctured torus which contains two anyons, and which, if these anyons were not present, would be topologically an unpunctured disc. The above diagrams represent an expression of the form $|\psi^{\prime\mrm{T}}\ra=\hat O^\mrm{T}|\psi^\mrm{T}\ra$, and the basis on the torus has been chosen for convenience.\label{fig:discoperatorontorus}}
\end{figure}%
\begin{table}[bp]
\caption{Inner products of unnormalised diagrams on the 1-punctured torus, for Fibonacci anyon statistics. Labels $a_1$, $x_1$, and $x'_1$ refer to diagram~(i) of \protect{\fref{fig:innerprod_torus}}, and $a'_1$ is set equal to $a_1$.
All inner products not listed below are zero in both conventions.\label{tab:torusnorms}}~\\
\begin{tabular}{|c|c|c|}
\hline
$x_1$, $a_1$, $x'_1$ & Convention of \protect{\sref{sec:torusinnerprod}} & Convention of \protect{\rcite{konig2010}} \\
\hline
$\mbb{I}$, $\mbb{I}$, $\mbb{I}$ & 1 & 1 \\
$\tau$, $\mbb{I}$, $\tau$ & $\phi^2$ & 1 \\
$\tau$, $\tau$, $\tau$ & $\phi^{5/2}$ & $\sqrt{\phi}$\\\hline
\end{tabular}
\end{table}%
\begin{figure*}
\includegraphics[width=492.0pt]{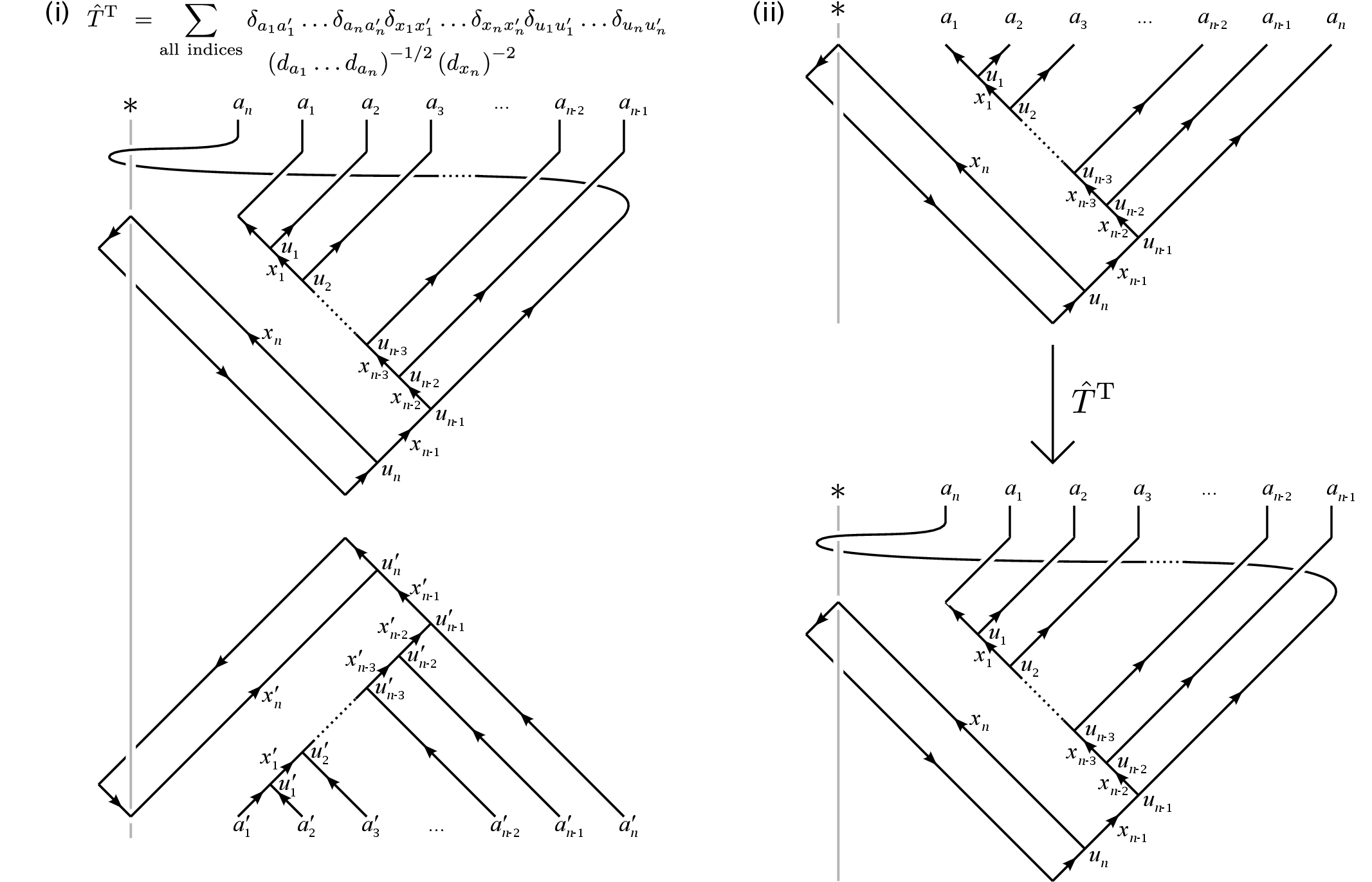}
\caption{The translation operator $\hat T^\T$ on the torus, (i)~represented as an operator in the bra-ket form of \protect{\Eref{eq:braketoperator}}, and (ii)~represented as a mapping between a state $|\psi\ra$ and the translated state $|\psi'\ra=\hat T^\mathrm{T}|\psi\ra$.
\label{fig:torustranslate}}
\end{figure*}%

\subsubsection{Operators on surfaces of higher genus\label{sec:torusoperators}}

As with the sphere, we will now address the diagrammatic representation of operators on surfaces of higher genus. We will begin with a general discussion, and once again will examine explicit examples on the torus.

On a surface of genus $g$ (where $g$ is taken to correspond to the genus of the surface in the absence of punctures), operators may correspond to physical processes acting either on the entire manifold, or on a finite subregion of the manifold. Where operators act on the entire manifold, a completely general construction may once again be achieved by replacing the bras and kets of \Eref{eq:braketoperator} with fusion and splitting tree diagrams for states of the appropriate genus.
However, for operators acting on a finite subregion of the manifold the situation may be simplified somewhat. Momentarily neglecting the existence of punctures on the physical manifold, we examine the topology of the area of support of the operator. If this area of support now lies entirely within a submanifold of genus $g'<g$, 
then 
the operator may be represented using fusion and splitting trees of genus $g'$. 
For example, if we consider an operator on the torus whose support lies within a region which is (again momentarily ignoring any anyons within it) topologically the unpunctured disc, then by locality we need only consider the portion of the fusion tree corresponding to any anyons which do lie within that disc. We then choose a basis where this portion of the tree connects to the rest of the fusion tree via only a single line, such that if we excised this disc from the manifold as a whole, that line would describe the charge on the resulting puncture, and be dual to that on the boundary of the excised disc. We may now represent the operator in the form of an operator on the disc as described in \sref{sec:discoperators}, and apply it by connecting it with the relevant portion of the fusion tree, as shown in the example of \fref{fig:discoperatorontorus}. 
We will, however, make one warning: If the operator shown in \fref{fig:discoperatorontorus} were to be applied to anyons $a_6$ and $a_1$, then transformation from the basis shown into one in which $a_6$ and $a_1$ were adjacent would require the application of the periodic translation operator, which is a non-local operator and will be discussed in \sref{sec:torustranslate}.

Extension of this approach to surfaces and operators of higher genus is straightforward.
\begin{figure*}[tp]
\includegraphics[width=492.0pt]{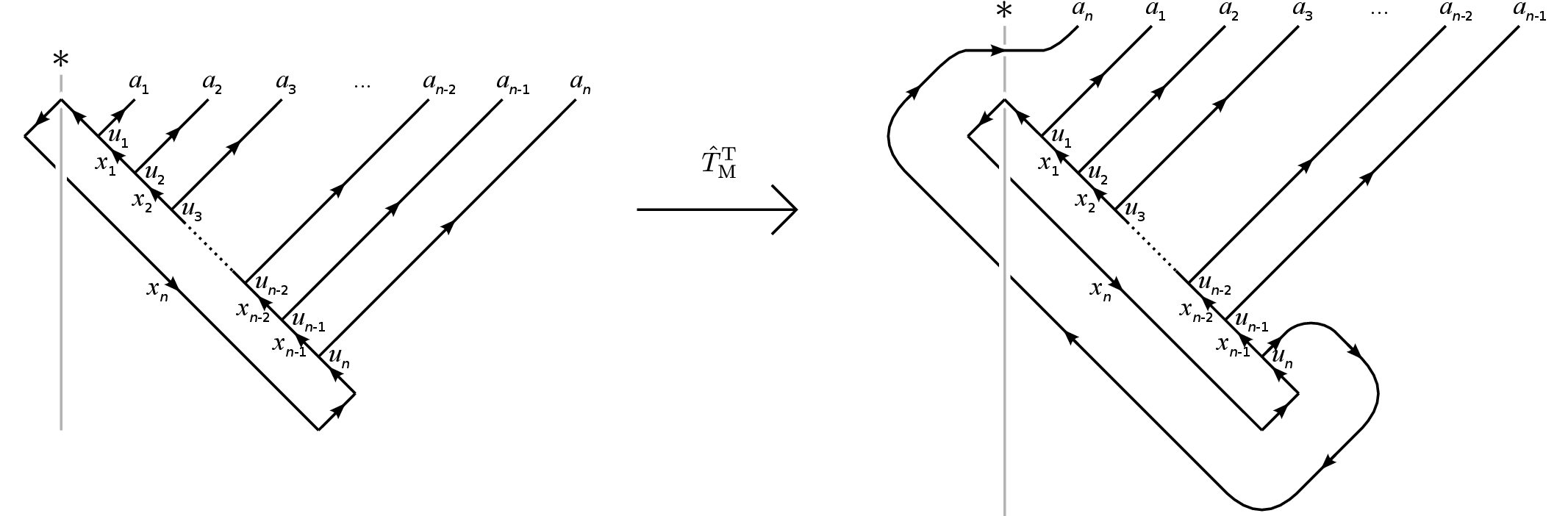}
\caption{Definition of an operator $\hat T^T_\mathrm{M}$ on the torus, which cyclically permutes the degrees of freedom $a_1,\ldots,a_n$ and $x_1,\ldots,x_n$ in basis $B_1$. 
\label{fig:spintranslate}}
\end{figure*}%

There exists one further observation to be made with respect to surfaces of higher genus. Much as the torus admits operators of genus 0 and genus 1, a surface of genus $g$ will admit operators whose support is a region of genus $g'$, for any $g'\leq g$. 
We are not aware of any notation convention for the description of such operators, and on surfaces having genus higher than 1, there is the potential for ambiguity as a given operator diagram of genus $g'$ may conceivably refer to any of $g'\cdot\left(\begin{array}{c}g\\g'\end{array}\right)$ different physical processes, depending on which handles of the manifold are associated with the handles implicit in the operator diagram.
Using a unique label for the pair of obstructions associated with each handle of the manifold serves to alleviate this ambiguity.

\subsubsection{Periodic translation on the torus\label{sec:torustranslate}}

We now consider a specific example system which will be of interest in \sref{sec:PBC}. 
Suppose we have a system of anyons on a torus, arranged on a chain which encircles either the large or the small non-trivial cycle of the torus.\footnote{In describing the non-trivial cycles as ``large'' and ``small'', we assume the torus to have a circular cross-section with respect to a radial cut. More accurately, recall that we are considering only manifolds which are non-self-intersecting and embedded in $\mathbb{R}^3$. We term a non-trivial cycle ``large'' if the torus may be smoothly deformed, without creating self-intersections and without extending to infinity, to be radially symmetric about an axis which is encircled by this non-trivial cycle.}
As these two situations are topologically equivalent, we choose it to encircle specifically the large non-trivial cycle with no loss of generality.
(It is anticipated that we may also evaluate 
rings which twist around both non-trivial cycles of the torus %
using a Dehn twist, but these 
are not considered in the present paper.)

How can we, in the diagrammatic notation, most efficiently represent the process of simultaneously translating each anyon around the torus by one site?

If we construct our fusion tree ``inside'' the torus, advance each anyon one site around the torus, and project onto the page, then the periodic translation operator is seen to act on the torus as shown in \fref{fig:torustranslate}.
We may therefore write the torus translation operator simply as
\begin{equation}
\hat T^\T =~\translopstar{}\,.\label{eq:T^T}
\end{equation}
Note the presence of the star in the diagrammatic representation of the translation operator, reflecting that the periodic translation passes around a non-trivial cycle on the torus, as opposed to merely cyclically permuting the anyons locally on a disc-like region of the torus-shaped manifold.

\begin{figure*}
\includegraphics[width=492.0pt]{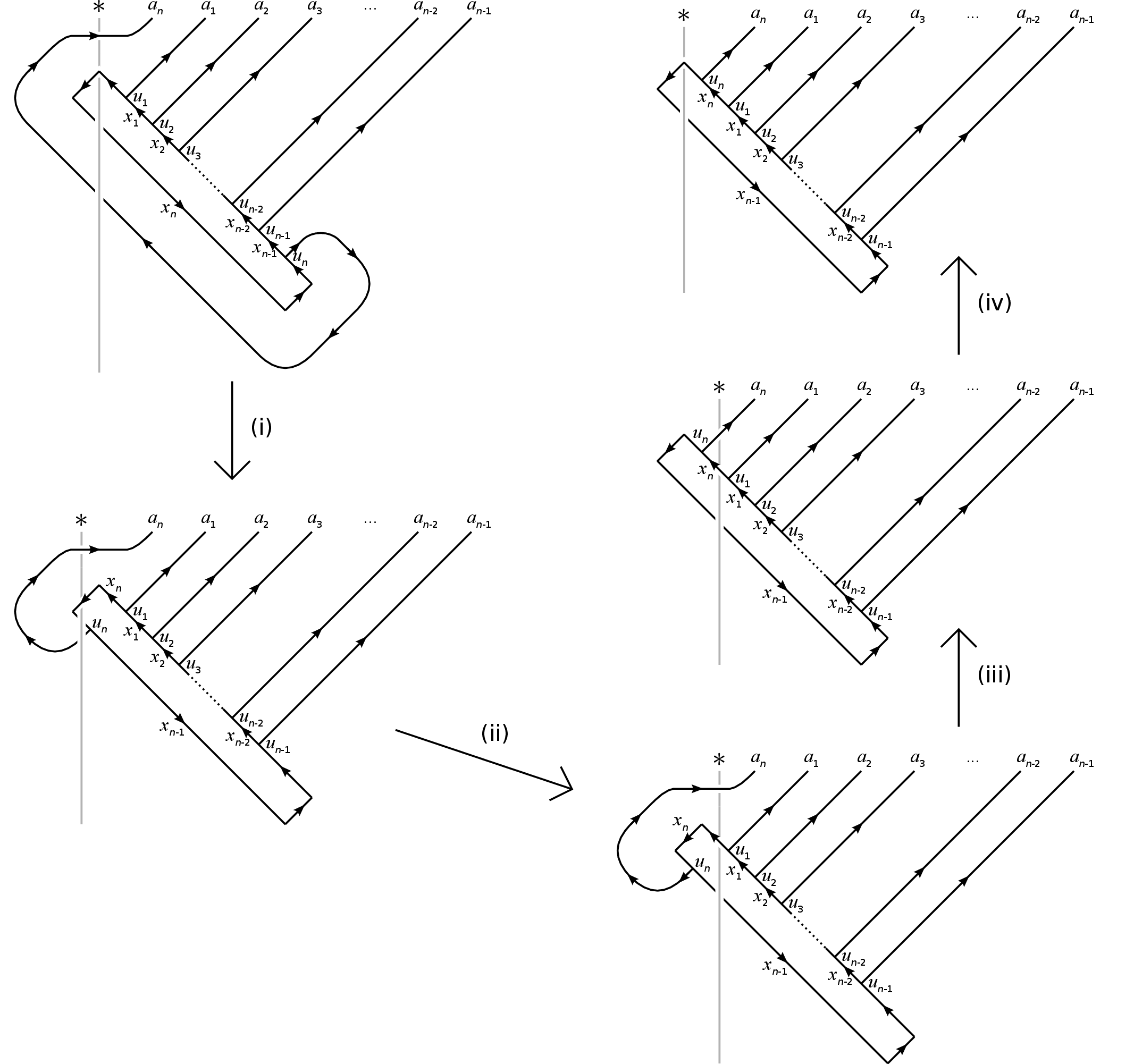}
\caption{Evaluation of the action of $\hat T^T_\mathrm{M}$ as defined in \protect{\fref{fig:spintranslate}}: (i),(iii):~Application of diagrammatic isotopy. (ii),(iv):~Topology-preserving deformation of the manifold. %
Recall that * represents the location of an arbitrarily small ring on the 2D spatial manifold which is identified with a similar ring encircling the point at infinity, and thus shifting the trajectory of * as shown in steps~(ii) and~(iv) just corresponds to performing an entirely local, topologically trivial deformation of the manifold. As both the manifold topology and the fusion tree %
are unaffected by this local deformation, it has no effect on our definition of the state, and no numerical factors are acquired as a result of these actions.
\label{fig:evalspintranslate}}
\end{figure*}

Evaluation of cyclic permutation on the torus in the ``inside'' basis poses an interesting challenge. In contrast with the cyclic translation operator on the disc [$\hat T^\mrm{D}$, \fref{fig:discoperators}(iv)], the operator $\hat T^\T$ can not be constructed from 
local operations by composing a series of braids. Instead we must introduce another new operator, given in \fref{fig:spintranslate}, which we will call the \emph{modified} translation operator, $\hat T^\T_\mrm{M}$. 
By diagrammatic isotopy and topology-preserving deformations of the manifold (\fref{fig:evalspintranslate}), we see that the action of this operator $\hat T^\T_\mrm{M}$ in basis $B_1$ is to cyclically permute the degrees of freedom $a_1,\ldots,a_n$ and $x_1,\ldots,x_n$. 
We may further 
use diagrammatic isotopy to redraw $\hat T^\T_\mrm{M}$ in the form
\begin{equation}
\hat T^\T_\mrm{M}=\,\magtranslopstar{}\,,\label{eq:magtransdiag}
\end{equation}
and this 
may be rewritten using \fref{fig:basischange}(ii) as
\begin{equation}
\hat T^\T_\mrm{M}=\left(R^{a_n\overline{a_n}}_{\mbb{I}}\right)^{-1}~\translopstar{}\,.\label{eq:magtrans}
\end{equation}
Comparing with \Eref{eq:T^T}, we see that $\hat T^\T$ and $\hat T^\T_\mrm{M}$ differ only by the charge-dependent phase $R^{a_n\overline{a_n}}_{\mbb{I}}$. We therefore conclude that the application of the periodic translation operator on the torus to a labelled fusion tree in basis $B_1$ constructed ``inside'' the torus is equivalent to multiplication by $R^{a_n\overline{a_n}}_{\mbb{I}}$ followed by cyclic permutation of all anyon indices $a_1,\ldots,a_n$ and internal indices $x_1,\ldots,x_n$.

\begin{figure}
\includegraphics[width=246.0pt]{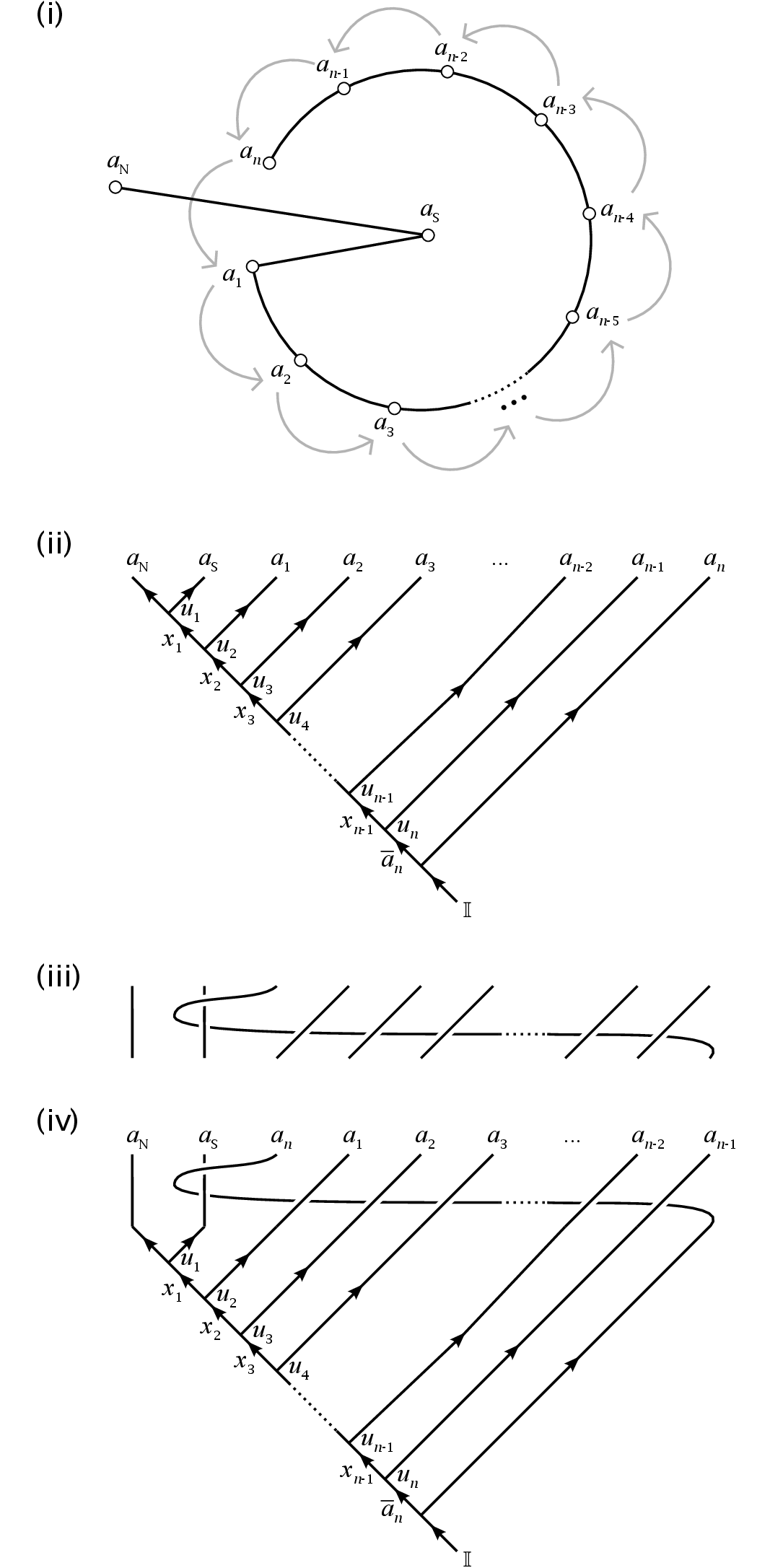}
\caption{(i)~Aerial view of $n+2$ punctures on the infinite plane equivalent to $n$ punctures in a non-trivial ring on the torus. The fusion tree on the plane imposes a linearisation on these punctures, and we may choose this to be as given by the black line. (ii)~The corresponding fusion tree. We may assume this fusion tree to inhabit the curved plane obtained by extending the black line of diagram~(i) into the plane of the page. The grey arrows in (i) indicate the process of periodic translation of the anyons on the ring. Note that during the process of translation, one anyon crosses the plane of the fusion tree while passing between the punctures $a_N$ and $a_S$. This is reflected in the periodic translation operator, labelled (iii). Application of the operator (iii) to states in the form of fusion tree (ii) yields states expressed in the fusion tree basis of diagram (iv). 
\label{fig:planetopview}}
\end{figure}

\begin{figure}
\includegraphics[width=246.0pt]{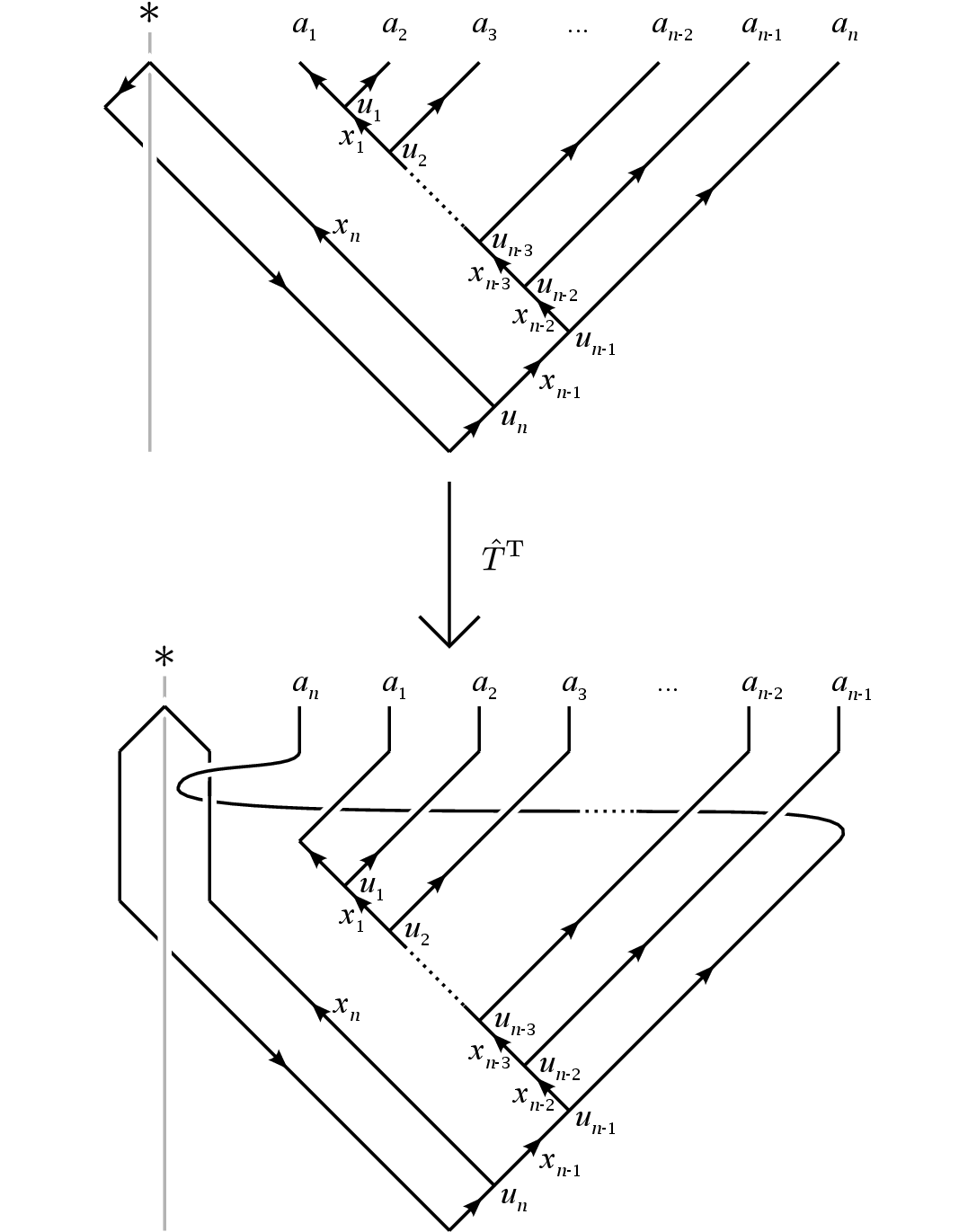}
\caption{Periodic translation operator around the larger non-trivial cycle of the torus, expressed in basis $B_2$ constructed in the ``outside'' space, and represented as a mapping between a state $|\psi\ra$ and the translated state $|\psi'\ra=\hat T^\mathrm{T}|\psi\ra$.
\label{fig:torustranslation_throughloop}}
\end{figure}

To construct a diagrammatic representation of the periodic translation operator in an ``outside'' basis, we will proceed somewhat differently. This time, let us begin with a state written in basis $B_2$ and constructed in the region ``outside'' the torus.
First, we map this state
to the \textit{n}+2-punctured sphere. We then perform a further mapping of this sphere to the infinite plane, to obtain the situation depicted in \fref{fig:planetopview}(i) where the arrangement of punctures is shown on the plane of the page.
Introducing a fusion tree for this arrangement of punctures, 
as shown in \fref{fig:planetopview}(ii), it is easy to construct the appropriate translation operator on the infinite plane [\fref{fig:planetopview}(iii)]. This operator maps states in the basis of \fref{fig:planetopview}(ii) to states in the basis of \fref{fig:planetopview}(iv), in which 
the anyon $a_n$ is explicitly braided around the south polar puncture. The equivalent operator on the torus in the ``outside'' version of basis $B_2$ is given in \fref{fig:torustranslation_throughloop}, where anyon $a_n$ is seen to braid \emph{through} the loop which carries the flux through the torus. 
\begin{figure}[!tp]
\includegraphics[width=246.0pt]{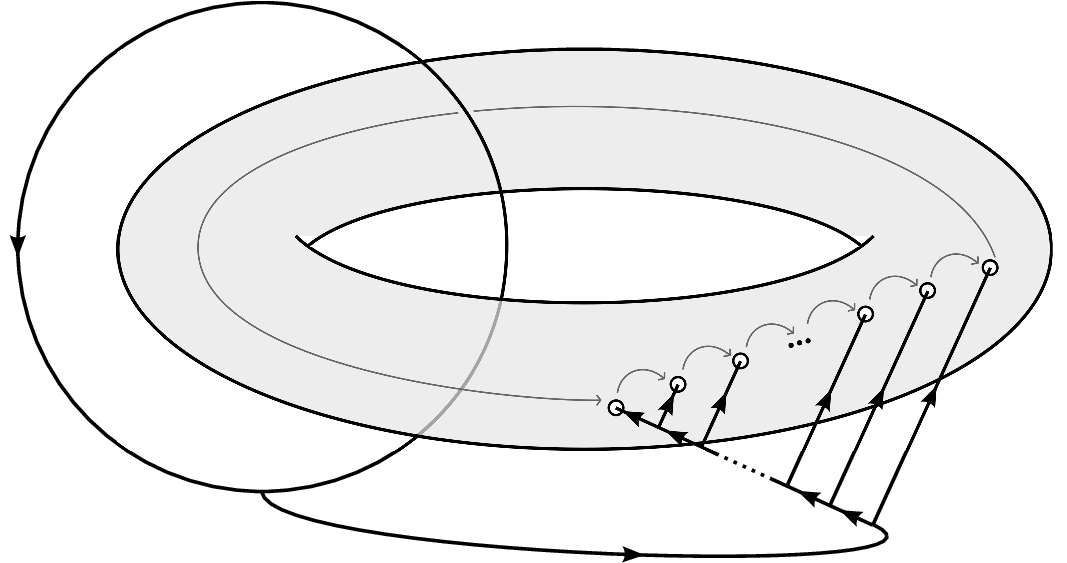}
\caption{In this diagram we see the process of periodic translation represented schematically on the actual toroidal manifold, with the fusion tree visible in the ``outside'' region of $\mbb{R}_3$. It is seen that, in bases constructed in the ``outside'' space, periodic translation on the torus threads an anyon through the non-trivial loop of the fusion tree.\label{fig:torustranslation_intuit}}
\end{figure}%

This may also be intuitively understood by explicitly constructing the fusion tree in the ``outside'' space, as shown in \fref{fig:torustranslation_intuit}, and observing that during periodic translation of the punctures, one anyon is necessarily threaded through the loop of the fusion tree.
Interestingly, and in contrast with bases constructed ``inside'' the torus, the translation operator for a basis ``outside'' the torus can be implemented entirely in terms of local operations once the state has been mapped to the equivalent sphere. This approach can not be applied to ``inside'' bases, as reversing the surgery process given in \fref{fig:spheretotorus2} 
involves cutting the ring of anyons, and this leaves the process of periodic translation on the equivalent sphere undefined.

Once again, notice that in either basis, the translation operator on the torus respects the arrow of time of the associated 2+1D TQFT: In Figs.~\ref{fig:torustranslate} and \ref{fig:torustranslation_throughloop} the trajectories of the punctures during periodic translation are all monotonic in the vertical direction.

\subsubsection{Topological symmetry operators\label{sec:Yoperators}}

\begin{figure*}
\includegraphics[width=492.0pt]{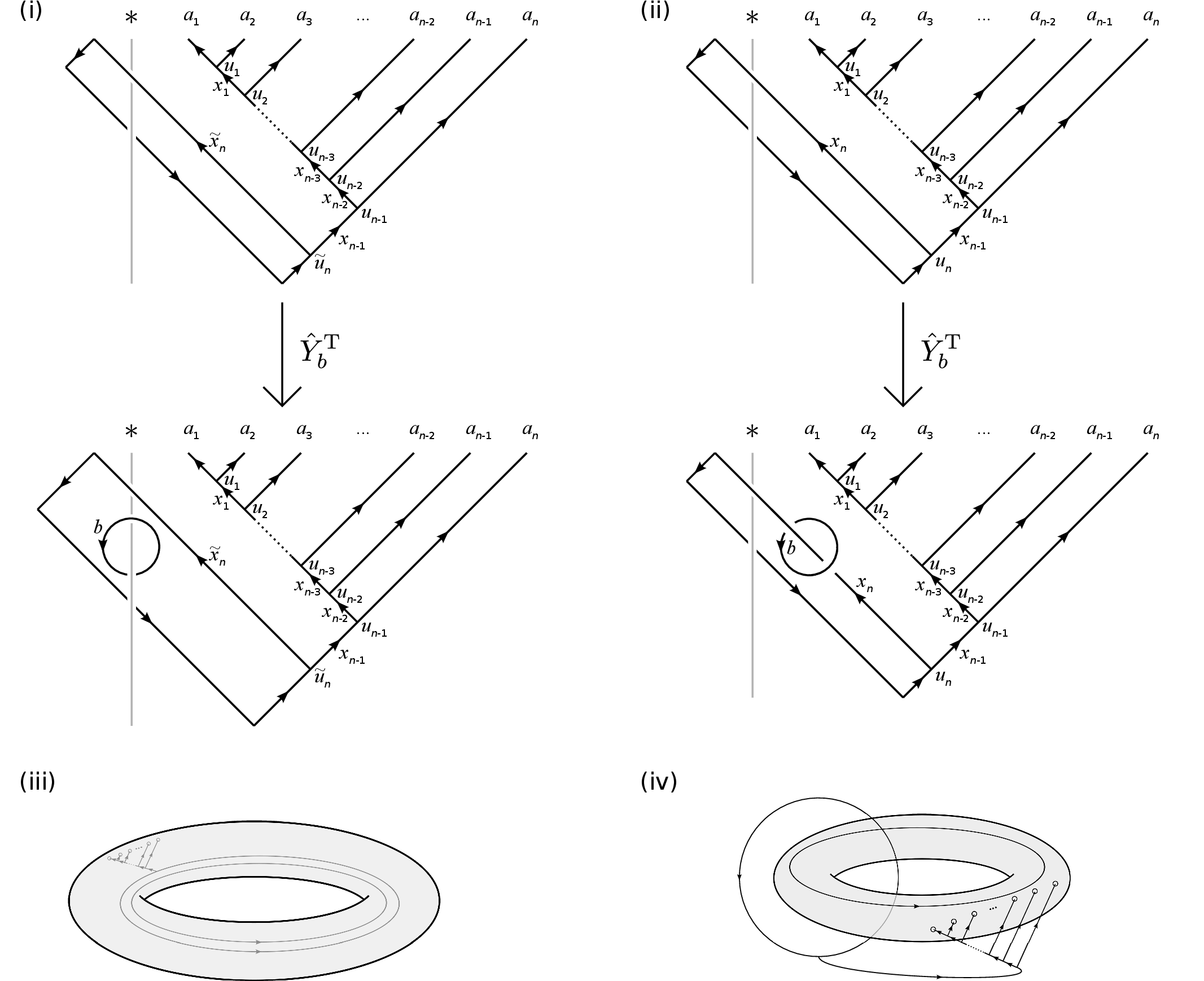}
\caption{Operator $\hat Y^\T_b$ acts on a state on the torus. In diagram~(i), the anyon pair $b$ and $\bar b$ travel around the large non-trivial cycle, and the fusion tree is constructed in the ``inside'' space. In diagram~(ii), the anyon pair still travel around the large non-trivial cycle, but the fusion tree [which may represent the same state as in diagram~(i)] is now constructed in the ``outside'' space. To aid in understanding of the physical process described by $\hat Y^\T_b$, diagrams~(iii)-(iv) illustrate how the resulting fusion trees may be related to the manifold when this is represented explicitly as a torus embedded in $\mbb{R}^3$.
Note that although in each instance operator $\hat Y^\T_b$ admits an interpretation as being equivalent to an anyon pair $b$ and $\bar{b}$ passing around the great cycle of the torus and annihilating, in diagrams~(i) and~(iii) this process is represented in the ``inside'' space, and in diagrams~(ii) and~(iv) it is constructed in the ``outside'' space. This is because, although the anyons themselves live on the surface of the manifold, when we choose to construct a fusion tree in the ``inside'' or ``outside'' space we associate a direction \emph{in that region} with the timelike dimension of the TQFT. As charges $b$ and $\bar b$ are created, pass around the torus, and then annihilate one another again, their histories then correspondingly trace out loops in the ``inside'' or ``outside'' space as appropriate.\label{fig:Yb}}
\end{figure*}

Finally, we will 
find it useful to introduce one more class of operator on the torus which admits a special graphical representation. Consider now a torus with a ring of punctures around the large non-trivial cycle. These operators, which we will denote $\hat Y_b^\T$, describe a process whereby a pair of anyons carrying charges $b$ and $\bar b$ are created from the vacuum, travel around opposite sides of a non-trivial cycle on the torus coplanar with the ring of punctures and without braiding, and then annihilate back to the vacuum. 
Expressed as a map from a state $|\psi\ra$ to a state $|\psi'\ra$ where $|\psi\ra$ and $|\psi'\ra$ are written in a fusion tree basis in the ``inside'' space, an operator 
$\hat Y^\T_b$ may be written as shown in \fref{fig:Yb}(i).
If we now re-express the state $|\psi\ra$ in a basis constructed ``outside'' the torus using \Eref{eq:Srelate}, operator $\hat Y^\T_b$ takes the form shown in \fref{fig:Yb}(ii).
Using the identity 
\begin{equation}
\raisebox{-20pt}{\includegraphics{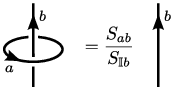}}\label{eq:Sunlink}
\end{equation}
we can see from \fref{fig:Yb}(ii) that $\hat Y^\T_b$ will have eigenvalues $S_{bx_n}/S_{b\mbb{I}}$ in the physical portion of the Hilbert space, and a state will be an eigenvector of $\hat Y^\T_b$ iff it is not in a superposition over label $x_n$.

In the present notation, the $\hat Y$ operator employed in \rcite{feiguin2007} would be denoted $\hat Y^\T_\tau$ and is constructed in the
``inside'' basis, as per \fref{fig:Yb}(i).

\section{Periodic boundary conditions\label{sec:PBC}}

In this section we will consider periodic chains of anyons first on the torus (\sref{sec:PBCtorus}) and then on the disc (\sref{sec:PBCdisc}). We further specialise to models in which every site in the chain carries a fixed, identical charge, and consequently in this section, and also in the next, 
we set $a_1=a_2=\ldots=a_n$. For each topology (torus and disc) we will introduce a translation-invariant local Hamiltonian written as a sum of local terms, and take as a specific example the AFM nearest neighbour interaction for a chain of Fibonacci anyons. 
We assume that an operator which is local acts only on a disc-like subregion of the manifold (i.e. if it is an operator acting on the torus, it does not include any non-trivial cycles). Consequently such an operator may be written in terms of
a fusion tree 
defined on the disc, as per \fref{fig:discoperators}(ii).

To express the states of our system, on the disc we shall use the basis given in \fref{fig:anyonstates_sphere}(iv), and on the torus we shall use a basis of the form given in \fref{fig:torusstates}(i) (basis $B_1$). The ring of punctures is taken as encircling the large non-trivial cycle of the torus, and the fusion tree is constructed in the ``inside'' space.

For the Fibonacci AFM interaction, which is a nearest neighbour interaction, all terms of the Hamiltonian take the form of \fref{fig:discoperators}(ii) for $r=2$.
For clarity, from this point forwards we shall only provide explicit treatments for nearest neighbour Hamiltonians, though most of the arguments and techniques presented readily generalise to $r>2$.

\subsection{Mapping to spins\label{sec:spinmapping}}

In studying one-dimensional systems of anyons using numerical techniques, it may frequently be favorable to map the anyon chain to a spin chain model with constraints.
We will therefore map the degrees of freedom for a one-dimensional system of Fibonacci anyons onto a spin chain. 

In the basis of \fref{fig:anyonstates_sphere}(iv) for the sphere and in basis $B_1$ for the torus [\fref{fig:torusstates}(i)],
for a system of $n$ anyons the Hilbert space of the system is spanned by the $p$ free parameters of the fusion tree, $x_1\ldots x_p$, where $p=n-3$ on the disc and $p=n$ 
on the torus.
We now construct a second, independent quantum system whose Hilbert space $\mc{H}$ admits a tensor product decomposition into $p$ ``sites'', each of local dimension $|U|$, %
\begin{equation}
\mc{H}=\left(\mc{H}_1\right)^{\otimes p},\qquad \mrm{dim}\left(\mc{H}_1\right)=\left|U\right|,
\end{equation}
which we will term a ``spin chain''.
On this spin chain we may now construct a local orthonormal basis on each site, %
identifying the elements of this basis with the charges of the UBMTC. We then map the value of each label $x_i$ on the fusion tree to the state of a corresponding spin (which we will also denote $x_i$) on the spin chain.

The Hilbert space of the resulting spin chain is larger than that of the associated anyon chain, and so is then
restricted to admit only those states which correspond to valid fusion trees under the anyonic fusion rules.

It is important to recognise that under this mapping, the process of translation on the anyon chain does not in general correspond to the natural definition of translation on the spin chain. This is particularly evident when examining the process of periodic translation on anyon rings, which we will consider now in some detail.

\subsection{Periodic translation of anyons, implemented on a chain of spins\label{sec:spintranslate}}

We note that there exists a special relationship between the process of periodic translation on a chain of spins and periodic translation on a ring of anyons encircling a non-trivial cycle of the torus. Under the mapping of \sref{sec:spinmapping}, each degree of freedom $x_1,\ldots,x_n$ on the torus is mapped to a site on the spin chain, and periodic translation on the system of spins, which we will denote $\hat T^\mrm{S}$, cyclically permutes these labels by one place. 
For a state satisfying $a_1=a_2=\ldots=a_n$, 
an equivalent effect may be obtained for the fusion diagram of the torus by applying the operator $\hat T^\T_\mrm{M}$ discussed in \sref{sec:torustranslate}.
Furthermore, for fixed anyon charges $a_1,\ldots,a_n$, 
the factor $\lb(R^{a_n\overline{a_n}}_\mbb{I}\rb)^{-1}$ in \Eref{eq:magtrans} is a constant.
Consequently, 
for a ring of fixed, identical anyons on the torus,
periodic translation 
of the ring of anyons is equivalent up to a phase to periodic translation on the associated spin chain, with this phase given by $R^{a_n\overline{a_n}}_\mbb{I}$, the value of which is specified in the UBMTC describing the anyons. 

Note that for Fibonacci anyons, there are two different possible UBMTCs based on the Fibonacci fusion rules. In one of these the phase $R^{\tau\tau}_\mbb{I}$ takes on the value $e^{4\pi\rmi/5}$, while in the other it is $e^{-4\pi\rmi/5}$. 
The existence of this pair of Fibonacci UBMTCs differing only by complex conjugation of $R^{ab}_c$ will be discussed further 
in \sref{sec:PBCtorus} and Appendix~\ref{apdx:chiral}, but for now it suffices to note that because $R^{a_n\overline{a_n}}_\mbb{I}$ is just a phase,
we have the identity
\begin{equation}
\hat T^\T\hat O\,\hat T^{\T\dagger} = \hat T^\T_\mrm{M}\,\hat O\,\hat T^{\T\dagger}_\mrm{M} = \hat T^\mrm{S}\hat O\,\hat T^{\mrm{S}\dagger}\label{eq:equivtrans}
\end{equation}
for any operator $\hat O$ which acts on a periodic chain of fixed identical anyons. %

Although the translation of operators is therefore relatively simple, some care 
is required when computing the momenta of translation-covariant states: if translation by one site on the torus introduces a phase of $e^{\rmi\theta^\T}$, then translation of the equivalent state by one site on the spin chain will introduce a phase of $e^{\rmi\theta^\mrm{S}}=\left(R^{a_n\overline{a_n}}_\mbb{I}\right)^{-1}e^{\rmi\theta^\T}$.

\subsection{Hamiltonian with periodic boundary conditions on the torus\label{sec:PBCtorus}}

\begin{figure}[tp]
\includegraphics[width=246.0pt]{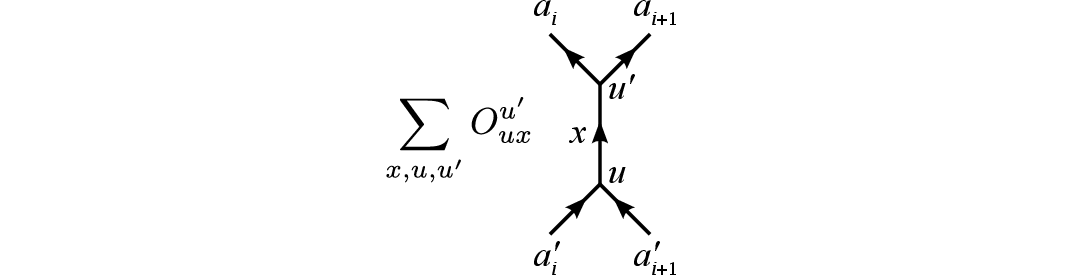}
\caption{Form of the two-site local operator used as a term in the local Hamiltonians $\hat H^\mrm{A,P,T}$ \protect{\eref{eq:H^APT} and $\hat H^\mrm{A,P,D}$ \eref{eq:H^APD}}. Charges $a_i$, $a'_i$, $a_{i+1}$, and $a'_{i+1}$ are assumed to be fixed.
\label{fig:localoperators}}
\end{figure}

We now introduce a translation-invariant anyonic Hamiltonian, $\hat H^\mrm{A,P,T}$. The superscripts $\mrm{A}$, $\mrm{P}$, and $\mrm{T}$ indicate that the Hamiltonian is anyonic, periodic, and constructed on 
the torus respectively.
For $r=2$ (nearest neighbour), we may write
\begin{equation}
\begin{split}
\hat H^\mrm{A,P,T} 
&= \sum_{i=0}^{n-1} \left(\hat T^\T\right)^i\left(\hat h^\mrm{A}_{1,2}\right)\left(\hat T^{\T\dagger}\right)^i\\
&= \sum_{i=1}^n \hat h^\mrm{A}_{i,i+1}
\end{split}\label{eq:H^APT}
\end{equation}
where local operator $\hat h^\mrm{A}_{i,i+1}$ acts on lattice sites $i$ and $i+1$, and takes the form of \fref{fig:localoperators}. Unless otherwise stated, the evaluation of position indices such as $i+1$ is assumed to be periodic in the range $1\ldots n$, so (for example) site $n+1$ is identified with site 1.

As a specific example we will consider the AFM interaction on the golden chain, for which all anyons $a_i$ on the lattice are constrained to have charge $\tau$. 
Because the charges $a_i$, $a_{i+1}$, $a'_i$, and $a'_{i+1}$ 
are fixed and there are no degeneracy indices, we may denote the elements of $\hat h^\mrm{A}_{i,i+1}$ by $\left(h^\mrm{A}_{i,i+1}\right)_x$ where $x$ corresponds to the fusion product of the Fibonacci anyons on sites $i$ and $i+1$ respectively. The AFM Hamiltonian favours the fusion path $\tau\times\tau\rightarrow 1$, and we therefore assign $\lb(h^\mrm{A}_{i,i+1}\rb)_1=-1$ and $\lb(h^\mrm{A}_{i,i+1}\rb)_\tau=0$.

As demonstrated by \citeauthor{feiguin2007},\cite{feiguin2007} a two-body operator $\hat h^\mrm{A}$ acting on two adjacent sites ($a_i$ and $a_{i+1}$) of the golden chain may, in an appropriate basis, also be understood as acting on three of the internal degrees of freedom on the fusion tree [e.g. $x_{i-1}$, $x_i$, $x_{i+1}$ in the basis of \fref{fig:torusstates}(i)]. %
Using this information, we can
construct a three-body operator $\hat h^\mrm{S}$ on the spin chain whose action is locally equivalent to the two-body operator $\hat h^\mrm{A}$ on the system of anyons. %
We first
introduce the spin chain equivalent of applying an $F$ move at sites $i$ and $i+1$ of basis $B_1$, for $i<n$:
\begin{equation}
\begin{split}
&\hat F^\mrm{S}_{i-1,i,i+1}|x_{i-1}x_{i}x_{i+1}\ra\\
&\quad=\sum_{\tilde x_{i}} \lb(F^{a_{i}a_{i+1}x_{i+1}}_{x_{i-1}}\rb)_{x_{i}\tilde x_{i}} |x_{i-1}\tilde x_{i} x_{i+1}\rangle\label{eq:defineF},
\end{split}
\end{equation}
where $^\mrm{S}$ indicates an operator acting on the spin chain. We then write
\begin{equation}
\hat h^\mrm{S}_{i-1,i,i+1} = \lb(\hat F^\mrm{S}_{i-1,i,i+1}\rb)^\dagger\,\hat h^{\prime\mrm{S}}_{\p{\prime}i}\,\hat F^\mrm{S}_{i-1,i,i+1},
\end{equation}
where
\begin{equation}
\hat h^{\prime\mrm{S}}_{\p{\prime}i}|\ldots,x_{i},\ldots\ra = \left(h^\mrm{A}_{i,i+1}\right)_{x_{i}}|\ldots,x_{i},\ldots\ra.
\end{equation}
For a chain of Fibonacci anyons, $a_i=\tau$ for all values of $i$, and the above construction yields $\hat h^\mrm{S}_{i-1,i,i+1}$ corresponding to $\hat h^\mrm{A}_{i,i+1}$ for all $i<n$. 
To construct the final term we will exercise some caution, as $a_n$ and $a_1$ are presently located at opposite ends of the fusion diagram. We therefore begin with the expression
\begin{equation}
\hat h^\mrm{A}_{n,1} = \hat T^\T\hat h^\mrm{A}_{n-1,n}\hat T^{\T\dagger}.
\end{equation}
Since we have fixed the charges of all punctures $a_i$ to be equivalently $\tau$, we may apply \Eref{eq:equivtrans} to obtain
\begin{equation}
\hat h^\mrm{A}_{n,1} = \hat T^\T_\mrm{M}\hat h^\mrm{A}_{n-1,n}\hat T^\T_{\mrm{M}\dagger}.
\end{equation}
As $\hat T^\mrm{M}$ is equivalent to translation on the chain of spins, $\hat T^\mrm{S}$, we see that
\begin{equation}
\hat h^\mrm{S}_{n-1,n,1} = \hat T^\mrm{S}\hat h^\mrm{S}_{n-2,n-1,n}\hat T^{\mrm{S}\dagger}
\end{equation}
as might have been expected.
The total spin chain Hamiltonian may therefore be written
\begin{equation}
\begin{split}
\hat H^\mrm{S,P,T} &= \sum_{i=0}^{n-1} \lb(\hat T^\mrm{S}\rb)^i\lb(\hat h^\mrm{S}_{1,2,3}\rb)\lb(\hat T^{\mrm{S}\dagger}\rb)^i\\
&= \sum_{i=1}^n \hat h^\mrm{S}_{i-1,i,i+1}
\end{split}\label{eq:H^SPT}
\end{equation}
on a periodic spin chain of length $n$.

For Fibonacci anyons with nearest neighbour interactions, Hamiltonian \eref{eq:H^APT} is a quantum critical Hamiltonian, and thus, by their equivalence, so is \eref{eq:H^SPT}.
The low energy properties of such systems are described by a conformal field theory (CFT), and the 
scaling dimensions 
of the local primary fields 
can be extracted from the low energy spectrum of the translation invariant critical model on a finite lattice with periodic boundary conditions (see Refs.~\onlinecite{cardy1984} and \onlinecite{cardy1986}; also reviewed in \rcite{difrancesco1997}).
The results from numerically exactly diagonalising $\hat H^\mrm{S,P,T}$ for AFM Fibonacci chains of lengths 24 and 25 are presented in \tref{tab:torus1}. 
\begin{table}[bp]%
\caption{Energy spectra for rings of (i)~24 and (ii)~25 Fibonacci anyons interacting via an AFM nearest neighbour interaction on the torus, shifted and rescaled to yield scaling dimensions for the associated conformal field theory. For even numbers of anyons, this is the minimal model $\mc{M}(5,4)$, associated with the tricritical Ising model (TIM). For an odd number of anyons, we obtain the spectrum of those operators in $\mc{M}(5,4)$ which %
incorporate a $\mbb{Z}_2$ twist in the boundary conditions of the TIM.\protect{\cite{mossa2008}} As noted in \protect{\cite{feiguin2007}} the scaling dimensions for odd numbers of anyons may be obtained from those for even numbers of anyons by fusing the corresponding scaling fields with the %
holomorphic 
field $\varepsilon''$.%
\label{tab:torus1}}
~\\
\begin{tabular}{|c|cc|c|}
\hline
\multicolumn{4}{|c|}{(i) 24 anyons}\\
\hline
\hline
Numerics&\multicolumn{2}{|c|}{Prediction from CFT}&Flux through torus\\
\hline
0.0000 & 0 & & $\mbb{I}$\\
0.0750 & ~0.0750~ & ($\frac{3}{40}$) & $\tau$\\
0.1989 & 0.2000 & ($\frac{1}{5}$) & $\tau$\\
0.8826 & 0.8750 & ($\frac{7}{8}$) & $\mbb{I}$\\
\p{$^*$}1.0622$^*$ & 1.0750 & ($\frac{3}{40}+1$) & $\tau$\\
\p{$^*$}1.1784$^*$ & 1.2000 & ($\frac{1}{5}+1$) & $\tau$\\
1.1841 & 1.2000 & ($\frac{6}{5}$) & $\tau$\\
\p{$^*$}1.8540$^*$ & 1.8750 & ($\frac{7}{8}+1$) & $\mbb{I}$\\
\p{$^*$}1.9469$^*$ & 2.0000 & ($0+2$) & $\mbb{I}$\\
\p{$^*$}1.9843$^*$ & 2.0750 & ($\frac{3}{40}+2$) & $\tau$\\
\p{$^*$}2.0180$^*$ & 2.0750 & ($\frac{3}{40}+2$) & $\tau$\\
\hline
\end{tabular}\\~\\~\\
\begin{tabular}{|c|cc|c|}
\hline
\multicolumn{4}{|c|}{(ii) 25 anyons}\\
\hline
\hline
Numerics&\multicolumn{2}{|c|}{Prediction from CFT}&Flux through torus\\
\hline
0.0750 & 0.0750 & ($\frac{3}{40}$) & $\tau$\\
\p{$^*$}0.7000$^*$ & ~0.7000~ & ($\frac{7}{10}$) & $\tau$\\
0.8587 & 0.8750 & ($\frac{7}{8}$) & $\mbb{I}$\\
\p{$^*$}1.0662$^*$ & 1.4750 & ($\frac{3}{40}+1$) & $\tau$\\
\p{$^*$}1.4887$^*$ & 1.5000 & ($\frac{3}{2}$) & $\mbb{I}$\\
\p{$^*$}1.6641$^*$ & 1.7000 & ($\frac{7}{10}+1$) & $\tau$\\
\p{$^*$}1.6863$^*$ & 1.7000 & ($\frac{7}{10}+1$) & $\tau$\\
\p{$^*$}1.8163$^*$ & 1.8750 & ($\frac{7}{8}+1$) & $\mbb{I}$\\
\p{$^*$}2.0008$^*$ & 2.0000 & ($0+2$) & $\mbb{I}$\\
\p{$^*$}2.0176$^*$ & 2.0750 & ($\frac{3}{40}+2$) & $\tau$\\
\p{$^*$}2.0516$^*$ & 2.0750 & ($\frac{3}{40}+2$) & $\tau$\\
\hline
\end{tabular}
\\~\\~\\
$0\equiv(\mbb{I},\mbb{I})$, $\frac{3}{40}\equiv(\sigma,\sigma)$, $\frac{1}{5}\equiv(\varepsilon,\varepsilon)$, $\frac{7}{10}\equiv(\varepsilon,\varepsilon')$ or $(\varepsilon',\varepsilon)$, $\frac{7}{8}\equiv(\sigma',\sigma')$, $\frac{6}{5}\equiv(\varepsilon',\varepsilon')$, $\frac{3}{2}\equiv(\varepsilon'',\mbb{I})$ or $(\mbb{I},\varepsilon'')$\\~\\
$^*$ Eigenvalue is twofold degenerate
\end{table}%
The energy eigenvalues have been shifted and rescaled to give the scaling dimensions of the corresponding CFT, which for this Hamiltonian is the minimal model associated with tricritical Ising model, $\mc{M}(5,4)$. 

For each scaling dimension, \tref{tab:torus1} also gives a parameter 
referred to as the ``flux through the torus''. 
If we write our 
states 
on the torus in a basis of type $B_2$ constructed in the ``outside'' region of $\mbb{R}_3$, then this is simply the value of charge label $x_n$. %
It is not difficult to see from \Eref{eq:T^T} and Figs.~\ref{fig:Yb} and \ref{fig:localoperators} that any operator $\hat Y^\T_b$ will commute with a Hamiltonian of local terms $\hat H^\mrm{A,P,T}$ on the torus, and thus for Fibonacci anyons we may simultaneously diagonalise $\hat H^\mrm{A,P,T}$ and $\hat Y^\T_\tau$. It therefore follows that we may associate every energy eigenstate with a corresponding eigenvalue of $\hat Y^\T_\tau$, which in turn corresponds to the measurement of a well-defined charge $x_n$.

For a basis constructed in the ``inside'' space, %
the flux no longer corresponds to any particular label on the diagram. %
Instead, in an ``inside'' basis, to measure the flux one makes a charge measurement around the %
topological obstruction labelled $*$ [as seen in \fref{fig:Yb}(i)]. However, the conclusions drawn from the study of operator $\hat Y^\T_\tau$ in an ``outside'' basis still necessarily hold, and thus we may again simultaneously diagonalise $\hat H^\mrm{A,P,T}$ and $\hat Y^\T_\tau$.
The measurement of the flux through the torus in \rcite{feiguin2007} %
corresponds to the evaluation of the operator $\hat Y^\T_\tau$ (denoted in this paper by $\hat Y$) in an ``inside'' basis of type $B_1$.

\begin{figure}[tp]
\includegraphics[width=246.0pt]{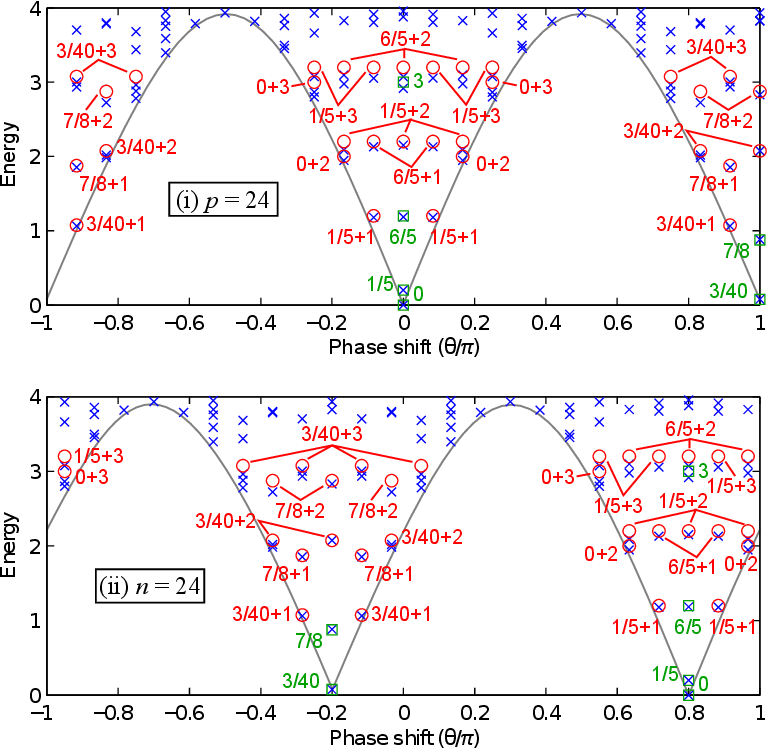} %
\caption{COLOUR ONLINE. Energy vs. phase diagram, where $e^{\mrm{i}\theta}$ is the phase acquired by energy eigenstates on translation by one site, for the Hamiltonian of the golden chain with AFM interaction (i)~made translation-invariant on the corresponding periodic spin chain with $p=24$ spins, and (ii)~naturally translation-invariant on a non-trivial ring of anyons on the torus with $n=24$ anyons. Squares mark theoretical values for primary fields, and circles show selected descendants. Solid lines bound the energies from below. 
The system of spins yields a dispersion diagram identical to that of anyons on the torus, except that the momenta are shifted by an offset of 
$-\rmi\ln\left[\left(R^{\tau\tau}_\mbb{I}\right)^{-1}\right]=-\frac{4\pi}{5}$ as described in \sref{sec:spintranslate}. 
(Note that as described in \protect{\sref{sec:PBCtorus}} and Appendix~\protect{\ref{apdx:chiral}}, the momentum of the ground state is dependent on the choice of Fibonacci UBMTC. In all our numerical calculations we have adopted a UBMTC for which $R^{\tau\tau}_\mbb{I}=\frac{4\pi\rmi}{5}$.)
\label{fig:dispersion}}
\end{figure}

Because the Hamiltonian is translation invariant, %
we may also assign a momentum to each state as shown in the dispersion diagram of \fref{fig:dispersion}. 
This diagram clearly shows the distinction between (i)~periodic translation on sites of the spin chain and (ii)~periodic translation of anyons on the torus, with the difference 
between $\hat T^\mrm{S}$ and $\hat T^\T$
resulting in a relative phase shift of 
$R^{\tau\tau}_\mbb{I}$ %
as described in \sref{sec:spintranslate}. For numerical calculations in this paper, we have adopted a Fibonacci UBMTC for which $R^{\tau\tau}_\mbb{I}=e^{{4\pi\rmi}/{5}}$.

The non-zero momentum of the ground state in \fref{fig:dispersion}(ii) merits some discussion. First, we note that the ground state momentum is precisely $-\rmi\ln\left(R^{\tau\tau}_\mbb{I}\right)=4\pi/5$, and that if we had adopted a UBMTC for which $R^{\tau\tau}_\mbb{I}=e^{-{4\pi\rmi}/{5}}$, we would have similarly obtained a momentum of $-4\pi/5$. 
This is because under
translation, a state will necessarily acquire a phase shift as a result of the braiding structure of the operator $\hat T^\T$.

It is also interesting to note that (as would be expected) the ground state momentum is not reflection-invariant. Spatial reflection maps all braids into their inverses, and is therefore equivalent to exchanging a UBMTC for which $R^{\tau\tau}_\mbb{I}=e^{{4\pi\rmi}/{5}}$ with one for which $R^{\tau\tau}_\mbb{I}=e^{-{4\pi\rmi}/{5}}$, changing the ground state momentum from $4\pi/5$ to $-4\pi/5$. Consequently, we see that the ground state of the anyonic system---and indeed the entire Fibonacci UBMTC---is chiral, and this chirality originates in the chirality of the underlying TQFT. Chirality of anyon models is discussed further in Appendix~\ref{apdx:chiral}.

It is important to recognise that this calculation is specific to systems governed purely by the Fibonacci UBMTC, and the behaviour may differ in models where the Fibonacci UBMTC, %
$\Fib$, %
represents just part of the overall symmetry of the system. For example, the Read--Rezayi state\cite{read1999} exhibits the quantum symmetry $\mbb{Z}_3\otimes \mrm{U}(1)$ which is equivalent, up to an Abelian phase, to $\Fib$, which
is the %
subcategory generated by the even charges of $\SU(2)_3$. In this example, the $\mrm{U}(1)$ sector in the Read--Rezayi state gives rise to a further, model-dependent phase during periodic translation. 
If this phase %
is not fixed %
by the description of the physical system, then it may be considered a gauge freedom. In previous work\cite{feiguin2007,trebst2008,gils2009,ardonne2011,poilblanc2011,poilblanc2012,ludwig2011} this gauge has been chosen to cancel out the factor of $R^{\tau\tau}_\mbb{I}$ which arises from periodic translation, yielding a momentum of zero for the ground state. %
In the present paper, however, we work %
with just the %
Fibonacci UBMTC %
and so, lacking the additional $\mrm{U}(1)$ gauge freedom of the Read--Rezayi state,
we 
necessarily 
obtain nonzero momenta for the ground states of the AFM Hamiltonian on the torus (above) and the disc (\sref{sec:PBCdisc})%
.

\subsection{Hamiltonian with periodic boundary conditions on the disc\label{sec:PBCdisc}}

\begin{figure*}
\includegraphics[width=492.0pt]{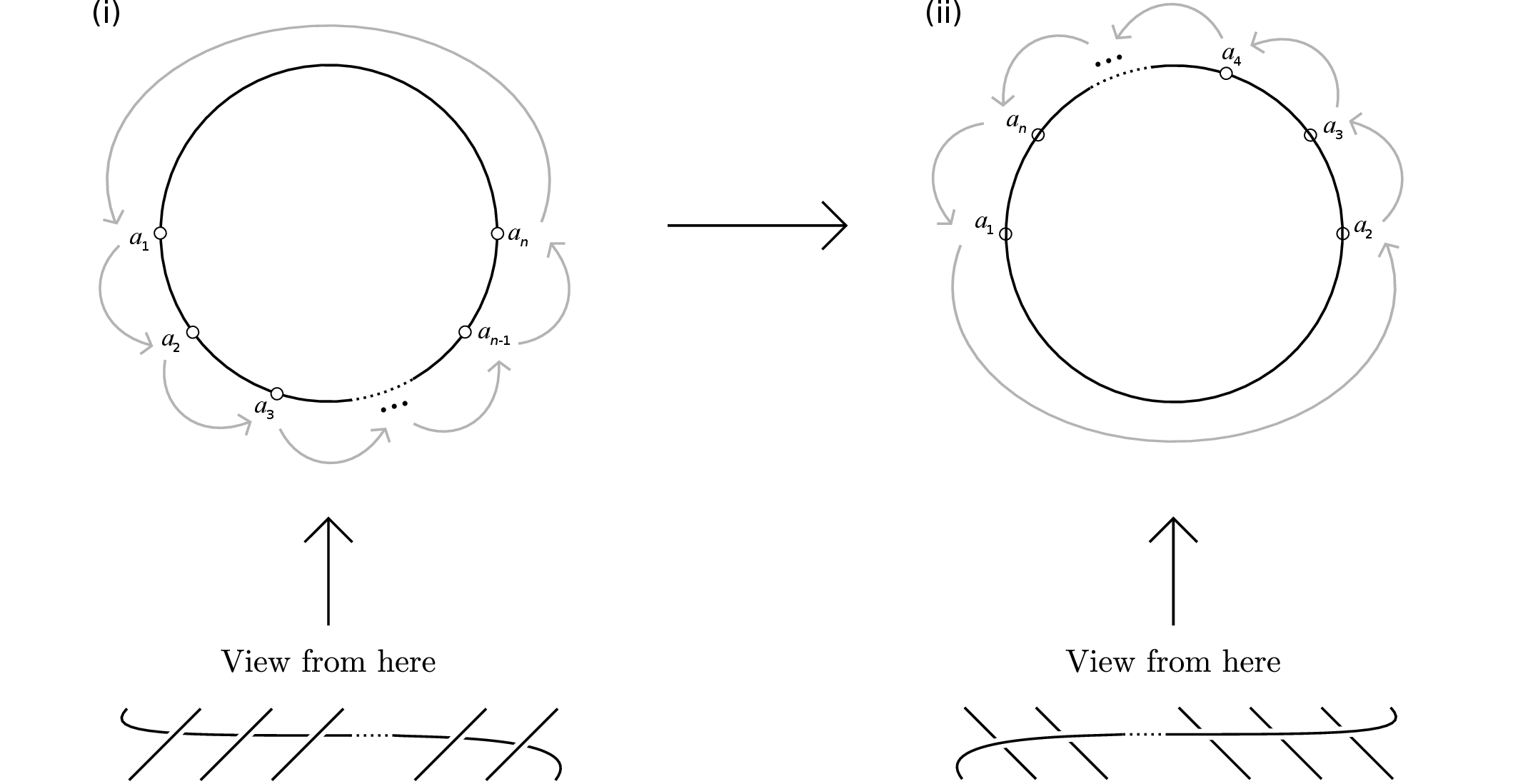}
\caption{The translation operator of \protect{\fref{fig:discoperators}}(iv) may be rewritten in the form of \protect{\Eref{eq:othertranslop}} by means of a topology-preserving change of basis. Diagram~(i) above shows the translation process whose projection yields \protect{\fref{fig:discoperators}}(iv). This anyon ring may be smoothly distorted to obtain that shown in diagram~(ii), for which projection yields the operator of \protect{\Eref{eq:othertranslop}}.\label{fig:pcob}}
\end{figure*}

To construct a periodic translation operator on the disc, we recognise that the anyon sites in \fref{fig:anyonstates_sphere}(iv) lie on a circle, which must be assumed to close either towards or away from the reader. Opting for the latter, we may define the periodic translation operator on the disc according to \fref{fig:discoperators}(iv). By inspection we see that translation may be implemented by means of repeated application of the braiding operator of \fref{fig:discoperators}(iii), which we will denote $\hat B^\mrm{A}_{i,i+1}$. 
The operator in \fref{fig:discoperators}(iv) cyclically permutes all anyons anticlockwise by one lattice site. Up to the passive change of basis shown in \fref{fig:pcob}
it is therefore equivalent to the diagram
\begin{equation}
\raisebox{-12pt}{\includegraphics{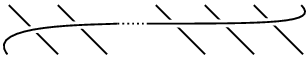}}\,,\label{eq:othertranslop}
\end{equation}
and may consequently be written as
\begin{equation}
\hat T^\mrm{D}=\prod_{i=1}^{n-1}\hat B^\mrm{A}_{i,i+1}.\label{eq:TD}
\end{equation}
We %
now introduce a translation-invariant Hamiltonian
\begin{equation}
\begin{split}
\hat H^\mrm{A,P,D} &= \sum_{i=0}^{n-1} \left(\hat T^\mrm{D}\right)^i\left(\hat h^\mrm{A}_{1,2}\right)\left(\hat T^{\mrm{D}\dagger}\right)^i\\
&= \sum_{i=1}^n \hat h^\mrm{A}_{i,i+1}
\end{split}\label{eq:H^APD}
\end{equation}
where $^\mrm{D}$ denotes that this Hamiltonian acts on the disc.
As before, we also introduce a spin chain whose sites correspond to the degrees of freedom of the anyonic fusion tree. However, on this occasion we use the fusion tree of \fref{fig:anyonstates_sphere}(iv) as a basis of states, and the corresponding spin chain is of length $p=n-3$.
Near the centre of the chain we may construct the spin equivalent of an $F$ move in the same way as for the torus \eref{eq:defineF}, but we see that operators near the end of the chain will act on a reduced number of sites. For example, an $F$ move acting on sites $a_2$ and $a_3$ 
acts only on spin variables $x_1$ and $x_2$, 
mapping $|x_1x_2\ra$ into $\sum_{\tilde x_1} (F^{a_1a_2a_3}_{x_2})_{x_1\tilde x_1} |\tilde x_1 x_2\ra$. 
We will continue to denote the spin chain counterparts of these operators by $\hat F^\mrm{S}_{i,i+1,i+2}$, with the understanding that when evaluating \Eref{eq:defineF} on the disc, any indices $x_0$ or $x_{p+1}$ are to be replaced by charges $a_1$ and $\overline{a_n}$ respectively, and any indices $x_{-1}$ or $x_{p+2}$ are to be replaced by the vacuum charge $\mbb{I}$. We do not modify the spin chain, which continues to run from $x_1$ to $x_p$. This behaviour manifestly breaks translation invariance on the spin chain. 

For values of $i$ sufficiently distant from 1 or $n$ we may also map $\hat h^\mrm{A}_{i,i+1}$ onto a three-site spin operator as before, although this is now denoted $\hat h^\mrm{S}_{i-2,i-1,i}$ as it acts on spin variables $x_{i-2}$, $x_{i-1}$, and $x_i$.
By using the extended definition of $\hat F^\mrm{S}_{i,i+1,i+2}$ we may even write down spin operators equivalent to $\hat h^\mrm{A}_{1,2}$, $\hat h^\mrm{A}_{2,3}$, $\hat h^\mrm{A}_{n-2,n-1}$, and $\hat h^\mrm{A}_{n-1,n}$. However, for $\hat h^\mrm{A}_{n,1}$ we must introduce the spin chain equivalent of the anyonic periodic translation operator on the disc, $\hat T^\mrm{D}$.

To do this, we first construct the spin chain counterpart to the anyonic braiding operator 
given in \fref{fig:discoperators}(iii). This is achieved by introducing a unitary operator $\hat R_i^\mrm{S}$ derived from the tensor $R$ in \fref{fig:basischange}(ii), which operator multiplies a state $|x_i\rangle$ by a phase $R^{a_{i+1}a_{i+2}}_{x_i}$. Using this we may write the 
spin chain equivalent of \fref{fig:discoperators}(iii) as
\begin{equation}
\hat B^\mrm{S}_{i,i+1,i+2} = (\hat F^\mrm{S}_{i,i+1,i+2})^\dagger \hat R^\mrm{S}_{i+1} \hat F^\mrm{S}_{i,i+1,i+2}.
\end{equation}
As with $\hat F^\mrm{S}$, the same special identifications for $x_{-1}$, $x_0$, $x_{p+1}$, and $x_{p+2}$ must be made when applying
either $\hat R^\mrm{S}$ or $\hat B^\mrm{S}$
to a state. Using $\hat{B}^\mrm{S}$ we can define an operator $\hat T^{\prime\mrm{S}}$ on the spin chain which is equivalent to periodic translation on the lattice of anyons,
\begin{equation}
\hat T^{\prime\mrm{S}} = \prod_{i=0}^{p+1} \hat B^\mrm{S}_{\spinsite{i-1},\spinsite{i},\spinsite{i+1}},\label{eq:anyontrans}
\end{equation}
and thus compute the spin chain Hamiltonian which is equivalent to $\hat H^\mrm{A,P,D}$:
\begin{equation}
\hat H^\mrm{S,P,D} = \sum_{i=0}^{p+2} \left(\hat T^{\prime\mrm{S}}\right)^i \left(\hat h^\mrm{S}_{\spinsite{1},\spinsite{2},\spinsite{3}}\right) \left(\hat T^{\prime\mrm{S}\dagger}\right)^i.\label{eq:H^SPD}
\end{equation}
The Hamiltonian $\hat H^\mrm{S,P,D}$ is clearly not translation invariant under the natural definition of translation on a periodic spin chain. However, it does exhibit 
translation invariance under the adjoint action of the anyon-derived translation %
operator %
$\hat T^{\prime\mrm{S}}$,
\begin{equation}
\hat H^\mrm{S,P,D}=\hat T^{\prime\mrm{S}}\hat H^\mrm{S,P,D}\hat T^{\prime\mrm{S}\dagger}. 
\end{equation}

Away from the edges of the fusion tree, the action of 
$\hat T^{\prime\mrm{S}}(\cdot)\hat T^{\prime\mrm{S}\dagger}$ is equivalent to translation on the system of spins, such that (for example) $\hat T^{\prime\mrm{S}}(\hat h^\mrm{S}_{1,2,3})\hat T^{\prime\mrm{S}\dagger}=\hat h^\mrm{S}_{2,3,4}$. However, this does not hold where the translation would yield an operator crossing between sites 1 and $p$. Instead, $\hat T^{\prime\mrm{S}}(\hat h^\mrm{S}_{p-2,p-1,p})\hat T^{\prime\mrm{S}\dagger}$ yields a two-site operator acting on spin sites $p-1$ and $p$, and it is necessary to apply $\hat T^{\prime\mrm{S}}(\cdot)\hat T^{\prime\mrm{S}\dagger}$ 
six times to map $\hat h^\mrm{S}_{p-2,p-1,p}$ into $\hat h^\mrm{S}_{1,2,3}$, with none of the intermediate terms resembling a translation on the spin system of the original operator $\hat h^\mrm{S}_{p-2,p-1,p}$. Nevertheless, the complete Hamiltonian satisfies $\hat T^{\prime\mrm{S}}(\hat H^\mrm{S,P,D})\hat T^{\prime\mrm{S}\dagger} = \hat H^\mrm{S,P,D}$. 

\begin{figure}[tp]
\includegraphics[width=246.0pt]{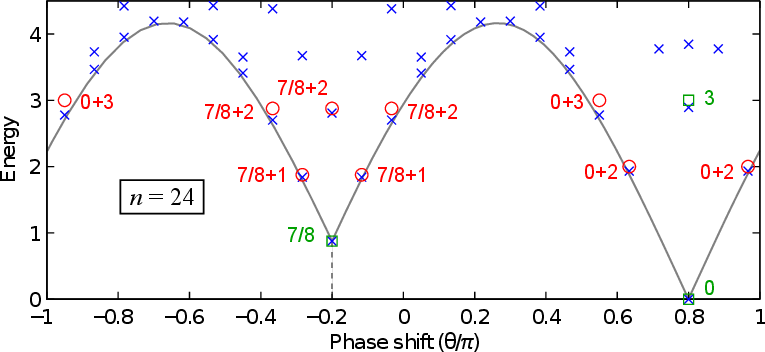}
\caption{COLOUR ONLINE. Energy vs. phase diagram, where $e^{\mrm{i}\theta}$ is the phase acquired by energy eigenstates on translation by one site, for the golden chain with AFM interaction on the disc with $n=24$ anyons. Squares mark theoretical values for primary fields, and circles show selected descendants. Solid lines bound the energies from below. Note that %
phase shifts 
are in agreement with those computed on the torus [\protect{\fref{fig:dispersion}(ii)}], and are offset by $\frac{4\pi}{5}$ relative to the corresponding values on the spin chain [\protect{\fref{fig:dispersion}(i)}]. 
(Note that as described in \protect{\sref{sec:PBCtorus}} and Appendix~\protect{\ref{apdx:chiral}}, the momentum of the ground state is dependent on the choice of Fibonacci UBMTC. In all our numerical calculations we have adopted a UBMTC for which $R^{\tau\tau}_\mbb{I}=\frac{4\pi\rmi}{5}$.)
\label{fig:dispersion_disc}}
\end{figure}

The results of numerically exactly diagonalising $\hat H^\mrm{S,P,D}$ are given in \tref{tab:disc1}, and a dispersion diagram is plotted in \fref{fig:dispersion_disc} for comparison with the torus [\fref{fig:dispersion}(ii)].
\begin{table}[bp]%
\caption{Energy spectra for rings of (i)~24 and (ii)~25 Fibonacci anyons interacting via an AFM nearest neighbour interaction on the disc, shifted and rescaled to yield scaling dimensions for the associated conformal field theory. For even numbers of anyons, this is the minimal model $\mc{M}(5,4)$, associated with the tricritical Ising model (TIM). For an odd number of anyons, we obtain a spectrum of operators in $\mc{M}(5,4)$ which %
incorporate a $\mbb{Z}_2$ twist in the boundary conditions of the TIM.\protect{\cite{mossa2008}} As noted in \protect{\cite{feiguin2007}} the scaling dimensions for odd numbers of anyons may be obtained from those for even numbers of anyons by fusing the corresponding scaling fields with the %
holomorphic 
field $\varepsilon''$. %
\label{tab:disc1}}
~\\
\begin{tabular}{|c|cc|}
\hline
\multicolumn{3}{|c|}{(i) 24 anyons}\\
\hline
\hline
Numerics&\multicolumn{2}{|c|}{CFT prediction}\\
\hline
0.0000 & 0 &\\
0.8750 & ~0.8750~ & ($\frac{7}{8}$)\\
\p{$^*$}1.8380$^*$ & 1.8750 & ($\frac{7}{8}+1$)\\
\p{$^*$}1.9301$^*$ & 2.0000 & ($0+2$)\\
\p{$^*$}2.7012$^*$ & 2.8750 & ($\frac{7}{8}+2$)\\
\p{$^*$}2.7771$^*$ & 2.8750 & ($\frac{7}{8}+2$)\\
\hline
\end{tabular}~~~
\begin{tabular}{|c|cc|}
\hline
\multicolumn{3}{|c|}{(ii) 25 anyons}\\
\hline
\hline
Numerics&\multicolumn{2}{|c|}{CFT prediction}\\
\hline
0.8750 & 0.8750 & ($\frac{7}{8}$)\\
\p{$^*$}1.5000$^*$ & ~1.5000~ & ($\frac{3}{2}$)\\
\p{$^*$}1.8250$^*$ & 1.8750 & ($\frac{7}{8}+1$)\\
\p{$^*$}2.4107$^*$ & 2.5000 & ($\frac{3}{2}+1$)\\
\p{$^*$}2.6939$^*$ & 2.8750 & ($\frac{7}{8}+2$)\\
2.7598 & 2.8750 & ($\frac{7}{8}+2$)\\
\hline
\end{tabular}
\\~\\~\\$0\equiv(\mbb{I},\mbb{I})$, $\frac{7}{8}\equiv(\sigma',\sigma')$, $\frac{3}{2}\equiv(\varepsilon'',\mbb{I})$ or $(\mbb{I},\varepsilon'')$\\~\\$^*$ Eigenvalue is twofold degenerate
\end{table}%
As with the torus the energy spectrum has been shifted and rescaled to give the scaling dimensions of local scaling operators in $\mc{M}(5,4)$. It is noted that the local scaling operators obtained coincide with those calculated using the anyonic scale-invariant MERA in the absence of free charges,\cite{pfeifer2010,pfeifer2011a} and this is to be expected as the local scaling operators employed in the cited %
references are topologically trivial, occupying a region of the manifold which is locally the disc.

This behaviour, where translation invariance exists relative to an operator which is not the natural translation operator on the spin chain,
has previously been observed for certain $\mrm{SU}(2)_k$-invariant spin chain Hamiltonians by \textcite{grosse1994} It is now known that a 
relationship exists between $\SU(2)_k$-invariant spin chains and chains of $\SU(2)_k$ anyons, and although present research has concentrated on anyons on the torus,\cite{feiguin2007,trebst2008} it is nevertheless likely that the models of \citeauthor{grosse1994} may similarly 
be 
mapped into interactions of $\SU(2)_k$ anyons on the disc.
The form of the anyonic translation operator also has practical implications for the restriction of the Hilbert space of the spin chain mentioned in \sref{sec:spinmapping}, and these technical details are discussed in 
Appendix~\ref{apdx:hilbertspace}.

\subsection{Relationship between the torus and the disc}

\subsubsection{Mapping between disc and torus states\label{sec:mapdisctorus}}

By comparing Tables~\ref{tab:torus1} and \ref{tab:disc1} we see that on the disc with trivial boundary charge, we compute scaling dimensions which correspond to those obtained for trivial flux through the torus.\footnote{The situations corresponding to the disc with non-trivial boundary charge are discussed in \protect{\rcite{pfeifer2012}}} The reason for this may be seen by comparing Figs.~\ref{fig:anyonstates_sphere}(iv) and \fref{fig:torusstates}(ii). Without specialising to Fibonacci anyons, we note that for any anyon model, if we
restrict the flux through the torus ${x}_n$ in ``outside'' basis $B_2$ to be $\mbb{I}$, we obtain a fusion tree identical to that of \fref{fig:anyonstates_sphere}(iv). 
The action of a general translation-invariant Hamiltonian
\begin{equation}
\hat H^\mrm{A,P,X} = \sum_{i=0}^{n-1} \left(\hat T^\mrm{X}\right)^i\left(\hat h^\mrm{A}_{1,2}\right)\left(\hat T^{\mrm{X}\dagger}\right)^i\label{eq:neighbour}
\end{equation}
where $\mrm{X}$ stands for $\mrm{T}$ on the torus and $\mrm{D}$ on the disc is therefore equivalent in both cases, and we obtain the observed correspondences between the energy spectrum of the torus and the disc.

For general anyon models with a nearest-neighbour Hamiltonian of the form of \Eref{eq:neighbour}, we may consider both critical and non-critical systems.
In models which are not critical,
the Hamiltonian will be insensitive to non-local properties of the system. In the thermodynamic limit each energy level is therefore $|U|$-fold degenerate on the torus, 
with the degeneracy enumerated by the different values which may be taken by the flux through the torus. As the disc corresponds to the torus with the flux constrained to be $\mbb{I}$, the eigenvalues on the disc in the thermodynamic limit are the same, but nondegenerate.

At criticality, the energy spectrum is no longer necessarily independent of the flux through the torus, and the degeneracy of energy levels on the torus may be broken, with different flux sectors exhibiting different energy spectra.
Nevertheless, the identification between the disc and the torus with trivial flux persists, and for critical local Hamiltonians applied both on the torus and on the disc, 
the disc necessarily exhibits the same energy spectrum 
as obtained on the torus
when the flux through the torus is constrained to be $\mbb{I}$.

Recall now that each scaling dimension obtained from the energy spectrum is associated with a scaling operator (see Refs.~\onlinecite{cardy1984} and \onlinecite{cardy1986}; also reviewed in \rcite{difrancesco1997}). %
Where the ground state is associated with the identity operator of the CFT, these scaling operators are local. In \rcite{feiguin2007}, it is argued that for a critical ring of Fibonacci anyons on the torus, scaling operators may also be classified according to the flux through the torus of their associated energy eigenstates, and it is proposed on topological grounds that only local scaling operators associated with a flux of $\mbb{I}$ may appear as local perturbations of the critical Hamiltonian. In the present paper we see that the spectrum of valid local perturbations on the disc is the same as that on the torus, because only the local scaling operators associated with a flux of $\mbb{I}$ have local counterparts on the disc.

\subsubsection{Conformal field theory with a defect\label{sec:CFTwdefect}}

In Secs.~\ref{sec:PBCtorus} and \ref{sec:PBCdisc} we have seen that a spin chain of length $p$ may be used, via appropriate mappings, to represent states of a system of either $p$ anyons on the torus or $p+3$ anyons on the disc. We also observed that each of these systems comes with its own definition of translation invariance. For the torus this corresponds (up to a state-dependent phase) to the natural definition of translation on a periodic spin chain, whereas for the disc this is given by operator $\hat T^{\prime\mrm{S}}$ of Eq.~\eref{eq:anyontrans} but nevertheless corresponds (when applied to an operator using the adjoint action) to the natural definition of translation invariance on the spin chain on sites sufficiently far from $x_1$ and $x_p$. It is therefore natural to interpret the difference between these two models as being equivalent to the introduction of a defect in translation invariance. 
Furthermore, we will show by construction that %
there exists a second defect which, when introduced manually, will combine with the original defect to restore the full spectrum for a system of anyons on the torus, albeit a torus of length $p-3$. [This length difference arises because the fusion tree for $p$ anyons on the torus has $p$ degrees of freedom, but the fusion tree for $p$ anyonic charges on the disc (assuming trivial boundary charge) has only $p-3$ degrees of freedom, and thus a chain of $p$ anyons on the disc may simulate at most $p-3$ anyons on the torus. A construction achieving precisely this limit is
presented in Appendix~\ref{apdx:anyondefect}.] %

\begin{figure*}
\includegraphics[width=492.0pt]{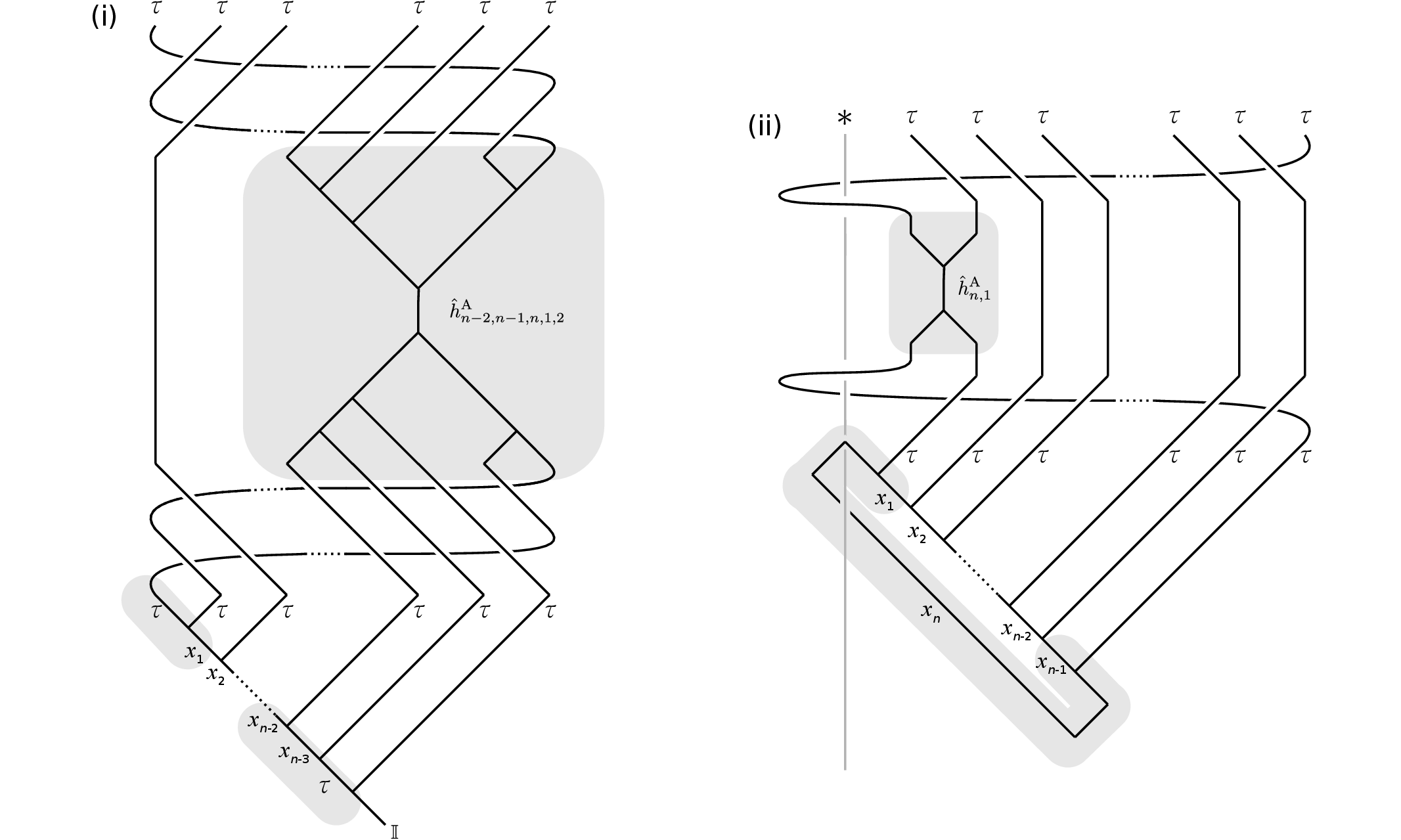}
\caption{(i)~In the basis of \pfref{fig:anyonstates_sphere}(iv), $\hat h^\mrm{A}_{n-2,n-1,n,1,2}$ is seen to depend upon degrees of freedom $x_{n-4}$, $x_{n-3}$ and $x_1$, where the degrees of freedom are labelled $x_1,\ldots,x_{n-3}$. (ii)~Similarly, in basis $B_1$ of \pfref{fig:torusstates}(i), $\hat h^\mrm{A}_{n,1}$ depends upon degrees of freedom $x_{n-1}$, $x_{n}$ and $x_1$, where the degrees of freedom are labelled $x_1,\ldots,x_n$. Thus in each instance, the evaluation of the operator depends on the variables $x_{q-1}$, $x_{q}$, and $x_1$, where $q$ is the total number of labels $x_1,\ldots,x_q$ on the fusion tree, and is $n-3$ or $n$ depending on the topology of the manifold.\label{fig:howitworks}}
\end{figure*}

We now construct a Hamiltonian on a disc of Fibonacci anyons which reproduces the spectrum of the AFM Hamiltonian on the torus. This Hamiltonian satisfies translation invariance on the disc---i.e. it is invariant under the adjoint action of $\hat T^\mrm{D}$ \eref{eq:TD}---except for two local terms.
These terms define a defect $\mc{D}$. Up to some additional degeneracies, which we will ignore, the spectrum of the resulting Hamiltonian for a ring of $n$ anyons on the disc is equivalent to that of $n-3$ anyons on the torus. If $n$ is even, then $n-3$ is odd, and as noted in the caption of \tref{tab:torus1}, the spectrum of a ring of an odd number of anyons on the torus is equivalent to that of an even number of anyons with a $\mbb{Z}_2$ twist.\cite{mossa2008} Thus the fusion of defect $\mc{D}$ with the $\mbb{Z}_2$ twist approximates the inverse of the defect 
identified with the adjoint action of the translation operator $\hat T^{\mrm{D}}$
on the spin chain. 
The Hamiltonian exhibiting defect $\mc{D}$ takes the form
\begin{equation}
\hat H^\mrm{A,P,D\rightarrow T} = \sum_{i=3}^{n-3} \hat h^\mrm{A}_{i,i+1} + \hat h^\mrm{A}_{n-2,n-1,n,1,2} + \hat h^\mrm{A}_{n-1,n,1,2,3}\label{eq:Hdefect_Disc}
\end{equation}
for a ring of $n$ anyons on the disc, 
with invariance under the adjoint action of $\hat T^{\mrm{D}}$ corresponding to
\begin{equation}
\hat H^\mrm{A,P,D\rightarrow T} = \hat T^\mrm{D}(\hat H^\mrm{A,P,D\rightarrow T})\hat T^{\mrm{D}\dagger}.
\end{equation}
Full details of its construction for the AFM or FM Fibonacci chain 
are given in Appendix~\ref{apdx:anyondefect}, but it suffices to note that on the disc, $\hat h^\mrm{A}_{n-2,n-1,n,1,2}$ 
couples the degrees of freedom $x_{n-4}$, $x_{n-3}$ and $x_1$ in \fref{fig:anyonstates_sphere}(iv) in the same way as $\hat h^\mrm{A}_{n,1}$ on the torus couples $x_{n-1}$, $x_n$, and $x_1$ in \fref{fig:torusstates}(i); this is illustrated in \fref{fig:howitworks}. 
Similarly, $\hat h^\mrm{A}_{n-1,n,1,2,3}$ on the disc couples $x_{n-3}$, $x_1$ and $x_2$ [of \fref{fig:anyonstates_sphere}(iv)] in the same way as $\hat h^\mrm{A}_{1,2}$ on the torus couples $x_n$, $x_1$, and $x_2$ [of \fref{fig:torusstates}(i)].

As a consequence of the effective invertibility of the defect in translation, it follows that we may compute the spectrum of a system of anyons on the torus or the disc using a system of anyons also, in each case, either on the torus or on the disc. 
In this section we have shown how to extract the spectrum of a system of anyons on the torus from a system of anyons on the disc, by means of a local modification to the Hamiltonian.
In \sref{sec:mapdisctorus} we noted that one can extract the spectrum of a system of anyons on the disc from a system of anyons on the torus by means of a global operator restricting the flux through the torus to be $\mbb{I}$, and 
for completeness we now note that this may be achieved by adding to the Hamiltonian an appropriate 1-site local operator acting on site $x_n$ of the spin chain. This 1-site operator applies an arbitrarily large energy penalty to states for which $x_n\not=\mbb{I}$.
When this term is introduced, the resulting Hamiltonian has 
a spectrum equivalent to a system of $n$ anyons on the disc. 

Finally, we note that when a defect is invertible, it may also be understood as being equivalent to a set of boundary conditions. A simple example of this is given by the 1D quantum critical Ising model: With periodic boundary conditions, the translation-invariant critical Ising Hamiltonian takes the form
\begin{equation}
\hat H=\left(\sum_{i=1}^{n-1} -\sigma^x_i\sigma^x_{i+1}-\sigma^z_i\right)-\sigma^x_n\sigma^x_1-\sigma^z_n.\label{eq:Ising}
\end{equation}
If we then introduce antiperiodic boundary conditions, which modify the definition of translation, the translation-invariant Hamiltonian then becomes
\begin{equation}
\hat H=\left(\sum_{i=1}^{n-1} -\sigma^x_i\sigma^x_{i+1}-\sigma^z_i\right)+\sigma^x_n\sigma^x_1-\sigma^z_n,
\end{equation}
and this may equivalently be thought of as a Hamiltonian on a 1D chain with periodic boundary conditions and a localised defect. As with the example given for Fibonacci anyons, this defect is invertible: The introduction of a second defect,
\begin{equation}
\hat H=\left(\sum_{i=1}^{n-2} -\sigma^x_i\sigma^x_{i+1}-\sigma^z_i\right)+\sigma^x_{n-1}\sigma^x_n-\sigma^z_{n-1}+\sigma^x_n\sigma^x_1-\sigma^z_n,
\end{equation}
results in a Hamiltonian having the same energy spectrum as the original periodic chain \eref{eq:Ising}.

For Fibonacci anyons, the example which is presented in this paper %
may be
of particular interest as the Hamiltonian \protect{$\hat H^\mrm{A,P,D\rightarrow T}$} is constructed entirely on the anyons lying on the ring. 
An alternative construction is presented in \rcite{pfeifer2012}, in which free charges on the manifold are admitted in addition to the charges of the anyon ring.

\section{Open boundary conditions\label{sec:OBC}}

On 1D systems with open boundary conditions the situation is somewhat simpler, but %
again some care must be taken as the fusion tree basis will again depend upon the topology of the quantum liquid. 
For example, as in \rcite{feiguin2007}, one might choose to study the Hamiltonian corresponding to free boundary conditions on the torus,
\begin{equation}
\hat H^{\mrm{A,F,T}} = \sum_{i=1}^{n-1} \hat h^\mrm{A}_{i,i+1},
\end{equation}
where F denotes free boundary conditions, which maps to the spin chain as
\begin{equation}
\hat H^\mrm{S,F,T} = \sum_{i=1}^{n-1} \hat h^\mrm{S}_{\spinsite{i-1},\spinsite{i},\spinsite{i+1}}.\label{eq:HF_torus}
\end{equation}
[This may be contrasted with \Eref{eq:H^SPT}.]
Similarly, one could place the same Hamiltonian on the disc:
\begin{align}
\hat H^{\mrm{A,F,D}} &= \sum_{i=1}^{n-1} \hat h^\mrm{A}_{i,i+1},\\
\hat H^\mrm{S,F,D} &= \sum_{i=1}^{n-1} \hat h^\mrm{S}_{\spinsite{i-2},\spinsite{i-1},\spinsite{i}}.\label{eq:HF_disc}\end{align}

Once again the spectrum for the Hamiltonian on the disc is seen to be a subset of that on the torus (\tref{tab:OBC_fixed}), and once again by means of appropriate modifications of the Hamiltonians, corresponding to alternative choices of boundary conditions, 
we may obtain either set of scaling dimensions on either topology. %
\begin{table}[bp]
\caption{Numerical results and CFT assignments for the smallest scaling dimensions on open chains of Fibonacci anyons of length $n$ with AFM coupling and free boundary conditions: (i)-(ii)~On the torus, Hamiltonian $\hat H^\mrm{S,F,T}$ \protect{\eref{eq:HF_torus}}. (iii)-(iv) On the disc, Hamiltonian $\hat H^\mrm{S,F,D}$ \protect{\eref{eq:HF_disc}}.
\label{tab:OBC_fixed}}
~\\
\begin{tabular}{|c|cc|}
\hline
\multicolumn{3}{|c|}{(i) 24 anyons, torus}\\
\hline
\hline
Numerics&\multicolumn{2}{|c|}{CFT prediction}\\
\hline
\p{$^*$}0.0000$^*$ & 0 &\\
0.6000 & ~0.6000~ & ($\frac{3}{5}$)\\
1.6009 & 1.6000 & ($\frac{3}{5}+1$)\\
\p{$^*$}2.0186$^*$ & 2.0000 & ($0+2$)\\
2.5765 & 2.6000 & ($\frac{3}{5}+2$)\\
2.5808 & 2.6000 & ($\frac{3}{5}+2$)\\
\hline
\end{tabular}~~~
\begin{tabular}{|c|cc|}
\hline
\multicolumn{3}{|c|}{(ii) 25 anyons, torus}\\
\hline
\hline
Numerics&\multicolumn{2}{|c|}{CFT prediction}\\
\hline
0.1000 & 0.1000 & ($\frac{1}{10}$)\\
1.1000 & ~1.1000~ & ($\frac{1}{10}+1$)\\
\p{$^*$}1.4845$^*$ & 1.5000 & ($\frac{3}{2}$)\\
2.0901 & 2.1000 & ($\frac{1}{10}+2$)\\
\p{$^*$}2.4670$^*$ & 2.5000 & ($\frac{3}{2}+1$)\\
3.0524 & 3.1000 & ($\frac{1}{10}+3$)\\
\hline
\end{tabular}
\\~\\~\\
\begin{tabular}{|c|cc|}
\hline
\multicolumn{3}{|c|}{(iii) 24 anyons, disc}\\
\hline
\hline
Numerics&\multicolumn{2}{|c|}{CFT prediction}\\
\hline
0.0000 & 0 &\\
2.0000 & ~2.0000~ & ($0+2$)\\
2.9762 & 3.0000 & ($0+3$)\\
3.9137 & 4.0000 & ($0+4$)\\
3.9820 & 4.0000 & ($0+4$)\\
4.7976 & 5.0000 & ($0+5$)\\
\hline
\end{tabular}~~~
\begin{tabular}{|c|cc|}
\hline
\multicolumn{3}{|c|}{(iv) 25 anyons, disc}\\
\hline
\hline
Numerics&\multicolumn{2}{|c|}{CFT prediction}\\
\hline
1.5000 & 1.5000 & ($\frac{3}{2}$)\\
2.5000 & ~2.5000~ & ($\frac{3}{2}+1$)\\
3.4726 & 3.5000 & ($\frac{3}{2}+2$)\\
3.4956 & 3.5000 & ($\frac{3}{2}+2$)\\
4.4025 & 4.5000 & ($\frac{3}{2}+3$)\\
4.4621 & 4.5000 & ($\frac{3}{2}+3$)\\
\hline
\end{tabular}
\\~\\~\\
$0\equiv\mbb{I}$, $\frac{1}{10}\equiv\varepsilon$, $\frac{3}{5}\equiv\varepsilon'$, $\frac{3}{2}\equiv\varepsilon''$
\\~\\
$^*$ Eigenvalue is twofold degenerate 
\end{table}

\section{Conclusion}

In summary, this paper may be divided into two parts. In the first (\sref{sec:ASO}) we have drawn attention to the importance of manifold topology in the study of anyonic systems. By reference to the underlying TQFT, we have explained how to construct diagrammatic representations of anyonic states, operators, and the inner product, for surfaces of arbitrary genus, and we have done so explicitly for the torus, sphere, and disc. 

In the second part of the paper (Secs.~\ref{sec:PBC}--\ref{sec:OBC}) we used these results to study the behaviour of an example system, consisting of a ring or chain of interacting Fibonacci anyons on either the torus or the disc. 
It has previously been shown that this chain is described by the same CFT as the tricritical Ising model, and that on the torus its criticality is topologically protected.\cite{feiguin2007} 
We have shown that 
criticality is similarly protected on the disc, using exact diagonalisation to calculate the scaling dimensions of the local scaling operators which may appear as perturbations to the critical Hamiltonian. %
We also see that the %
low-energy properties of rings of anyons on the torus and on the disc may be considered equivalent up to 
the appropriate choice of %
boundary conditions.

As a whole, this paper therefore presents the means to relate systems of interacting anyons on manifolds of differing topology, and applies this to examples using Fibonacci anyons. In particular, insight is gained into the topological protection of criticality of these systems and into the robustness of this protection across surfaces of different genus, and equivalent protection 
is seen to be exhibited on both the torus and the disc.

\begin{acknowledgments}
The authors would like to thank Miguel Aguado for insightful discussions. The authors acknowledge the support of the Australian Research Council (FF0668731, DP0878830, DP1092513, APA). This research was supported in part by the Perimeter Institute for Theoretical Physics, as well as by the National Science Foundation under grant DMR-0706140 (AWWL). We also thank the Ontario Ministry of Research and Innovation ERA for financial support.
\end{acknowledgments}

\appendix

\section{Alternative fusion tree and translation operator for anyons on the torus\label{apdx:alttrees}}

\begin{figure}
\includegraphics[width=246.0pt]{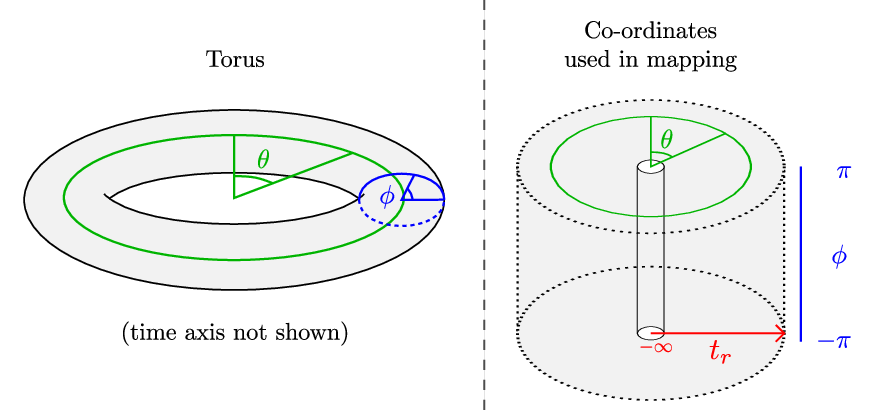}
\caption{COLOUR ONLINE. Co-ordinate system employed in step one of the mapping from the basis of~\pEref{eq:flatbasis} to the basis of \pfref{fig:torusstates}(i).\label{fig:pita}}
\end{figure}

In Eq.~(14) of \rcite{konig2010}, states on the torus are described using a fusion tree substantially different to the ones presented in \fref{fig:torusstates}, namely
\begin{equation}
\raisebox{-22pt}{\includegraphics[width=209.0pt]{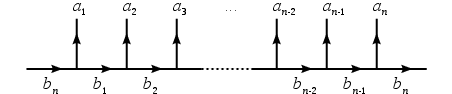}}.
\end{equation}
Note that the ends of the fusion tree are labelled with the same charge (here, $b_n$). In order to graphically represent the translation operator on this basis, we must make this identification between the ends of the tree a little more explicit, drawing the basis as
\begin{equation}
\raisebox{-30pt}{\includegraphics[width=209.0pt]{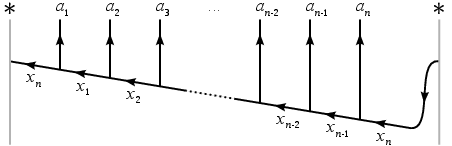}}
\label{eq:flatbasis}
\end{equation}
and identifying the two lines labelled $*$. We have also tilted the basis from the horizontal in order to assist in formally identifying the vertical axis with time, and replaced all $b_i$ with $x_i=\overline{b_i}$ to emphasise the (deceptive) visual similarity of this basis to \fref{fig:torusstates}(i).
We observe the following similarities between the basis of \Eref{eq:flatbasis} and the bases of \fref{fig:torusstates}:
\begin{enumerate}
\item In both \fref{fig:torusstates} and \Eref{eq:flatbasis}, the arrow of time runs up the page.
\item In both \fref{fig:torusstates} and \Eref{eq:flatbasis}, there are additional topological features which distinguish the fusion trees of the torus from those of the disc.
\end{enumerate}
However, in \fref{fig:torusstates}, these topological features consist of pairs of identified arbitrarily small rings which act as obstructions, but do not make contact with the fusion tree. In \Eref{eq:flatbasis}, on the other hand, there is
an identification between opposite ends of the anyon chain. 
Conversion between the basis of \Eref{eq:flatbasis} and that of %
\fref{fig:torusstates}(i), %
can be achieved as follows:
\begin{enumerate}
\item Close the identified ends of \Eref{eq:flatbasis} into a circle with the $a_i$ pointing radially out. This fusion tree diagram now has a radial arrow of time. Introduce a radial co-ordinate system ($t_r$,$\theta$). The second spatial co-ordinate $\phi$ on the surface of the torus is assumed to lie perpendicular to the plane of the page, on an interval [$-\pi$,$\pi$] with ends identified (\fref{fig:pita}). Note that $t_r=0$ corresponds to the infinite past, and obstructs any attempt to remove the loop from the diagram using \fref{fig:innerprod_sphere}.
\item Map the ($t_r$,$\theta$) plane onto an infinitely long hollow cylinder parameterised by ($t_z$,$\theta$) where $t_z = \ln (t_r)$. The $\phi$ co-ordinate, being perpendicular to this plane, now corresponds to a thickening of the hollow cylinder to have a cross-section which is the annulus (with outer and inner edges identified). Co-ordinates $\theta$ and $\phi$ now parameterise this cross-section as shown in \fref{fig:annulus}.
\item As described in the main text, expand the annulus so that $\theta$ and $\phi$ now parameterise almost the entire plane, lacking only arbitrarily small discs enclosing the two points at the origin and infinity. Label these points $*$.
\item Project the hollow cylinder into the plane of the page, oriented so that the axis $t_z$ corresponds to the vertical axis lying on the plane of the page.
\end{enumerate}
Using this process to map between the bases of \Eref{eq:flatbasis} and \fref{fig:torusstates}(i) (basis $B_1$), it is easy to rewrite the operators $\hat T^\T$ \eref{eq:T^T} and $\hat T^\T_\mrm{M}$ \eref{eq:magtransdiag} in the basis of \Eref{eq:flatbasis}. Explicitly, we find
\begin{align}
\hat T^\T&=\raisebox{-12pt}{\includegraphics[width=209.0pt]{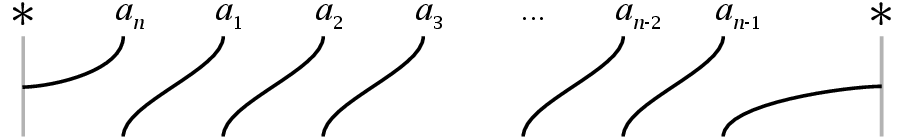}}\label{eq:'T^T}\\
\hat T^\T_\mrm{M} &= \raisebox{-12pt}{\includegraphics[width=209.0pt]{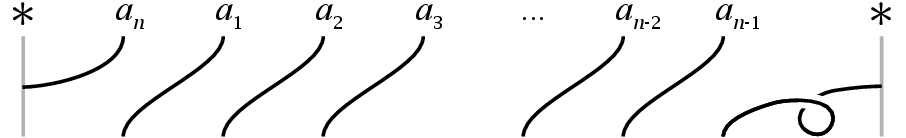}}.\label{eq:'T^T_M}
\end{align}

\begin{figure}
\includegraphics[width=246.0pt]{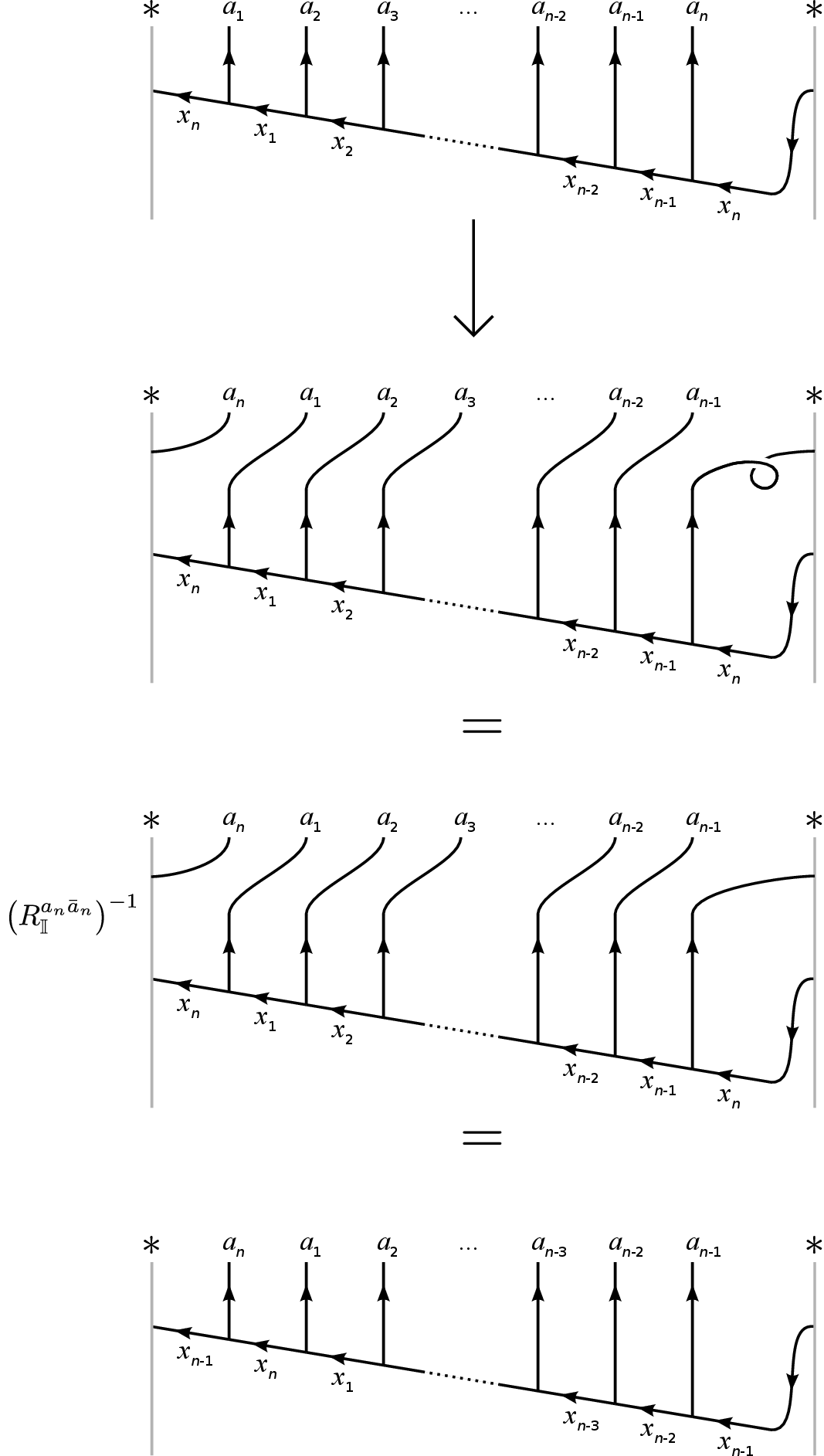}
\caption{Application of the operator $\hat T^\T_\mrm{M}$ in the basis of \protect{\Eref{eq:flatbasis}}. We know from \protect{\fref{fig:evalspintranslate}} and the mapping between basis $B_1$ [\protect{\fref{fig:torusstates}}(i)] and basis \eref{eq:flatbasis} that the action of $\hat T^\T_\mrm{M}$ is to cyclically permute all charge labels $a_i$ and $x_i$, from which follows the existence of a factor of $R^{a_n\bar{a}_n}_\mbb{I}$ associated with translating the vertex involving charges $x_{n-1}$, $x_n$, and $a_n$ through the two identified points $*$. %
\label{fig:flatbasisTM}}
\end{figure}

\begin{figure}
\includegraphics[width=246.0pt]{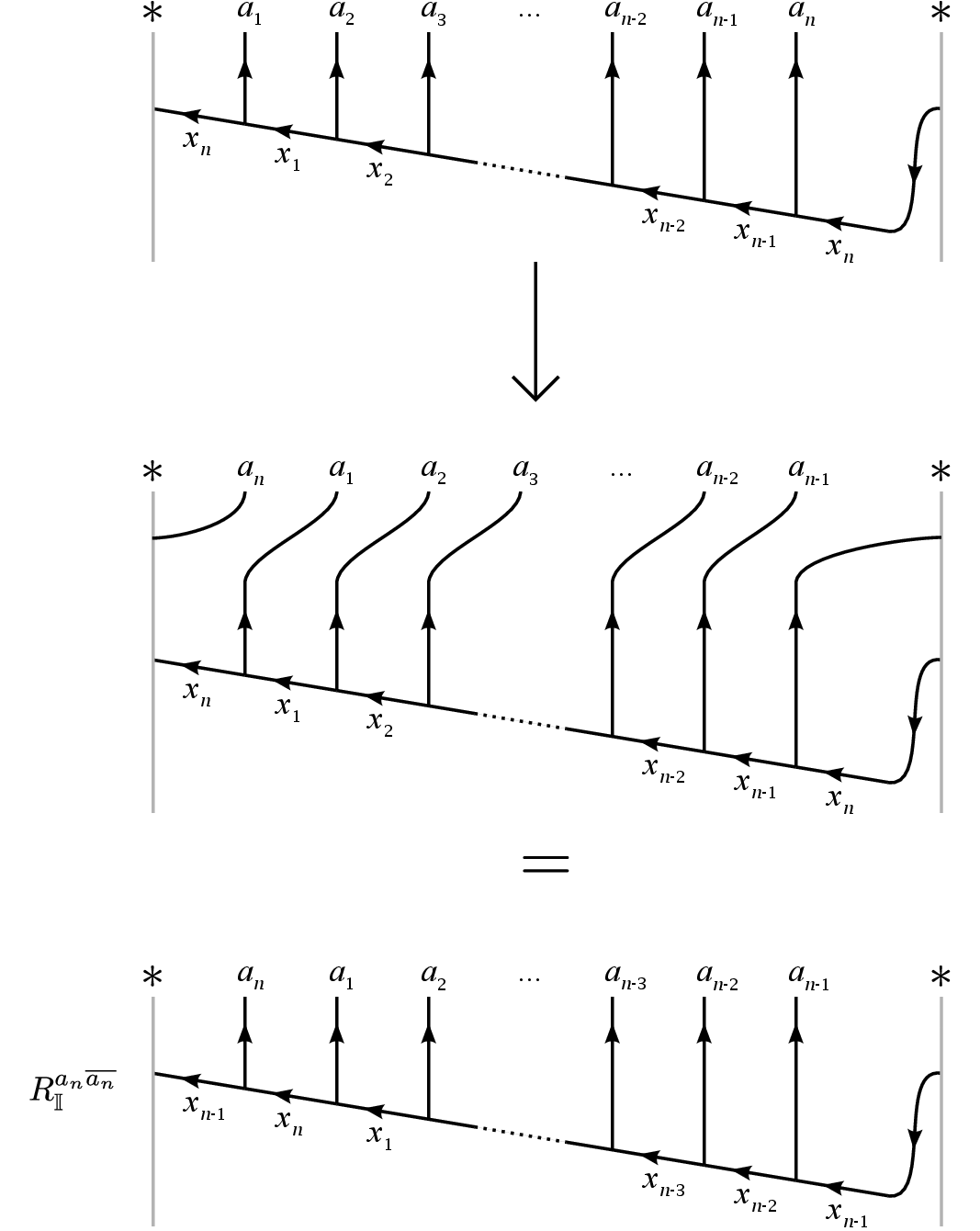}
\caption{Periodic translation in the basis of \protect{\Eref{eq:flatbasis}}. The factor of $R^{a_n\bar{a}_n}_\mbb{I}$ in this expression originates from translating the vertex involving charges $x_{n-1}$, $x_n$, and $a_n$ through the two identified points $*$. %
\label{fig:flatbasistranslation}}
\end{figure}

Evaluation of these operators in basis \eref{eq:flatbasis} is non-trivial. The rules of diagrammatic isotopy, commonly used in the evaluation or simplification of fusion tree diagrams, are formulated for regions which represent a projection of the 2+1-disc, and they therefore do not necessarily hold when the fusion tree traverses an identification. We therefore begin by transforming $\hat T^\T_\mrm{M}$ back into basis $B_1$ [\fref{fig:torusstates}(i)], because (as shown in \fref{fig:evalspintranslate}) in this basis the action of $\hat T^\T_\mrm{M}$ can be evaluated purely by means of diagrammatic isotopy applied to regions which are locally projections of the 2+1-disc, and local topology-preserving deformations which relocate the obstruction but leave the fusion tree unchanged. From this we see that in basis $B_1$, $\hat T^\T_\mrm{M}$ has the sole effect of cyclically permuting all $a_i$ and $x_i$. As the mapping between basis $B_1$ and basis \eref{eq:flatbasis} does not affect any of the charge labels, its action must be identical in basis \eref{eq:flatbasis}. On applying $\hat T^\T_\mrm{M}$ in basis \eref{eq:flatbasis}, as shown in \fref{fig:flatbasisTM}, we see that in this particular basis, and on this specific topology, pushing the vertex involving charges $x_{n-1}$, $x_n$, and $a_n$ through the identification attracts a factor of $R^{a_n\bar{a}_n}_\mrm{I}$. 
As $\hat T^\T$ and $\hat T^\T_\mrm{M}$ differ only by a phase, this knowledge suffices to permit us to likewise evaluate the action of $\hat T^\T$ in basis \eref{eq:flatbasis}, as shown in \fref{fig:flatbasistranslation}.

It is important to note that there exist two choices of mapping from the $(t_r,\theta)$ plane to the hollow cylinder, equivalent up to spatial reflection. 
Care must therefore be taken to keep track of parity when working with chiral Hamiltonians or UBTCs.
Spatial reflection is discussed further in Appendix~\ref{apdx:chiral}.
 
\section{Chiral symmetry and Fibonacci anyons\label{apdx:chiral}}

In \sref{sec:PBCtorus} we noted that %
the ground state of the AFM golden chain has a non-zero momentum,
implying that %
it is chiral%
. %
We will now explain this observation in more detail.

\begin{figure}
\includegraphics[width=246.0pt]{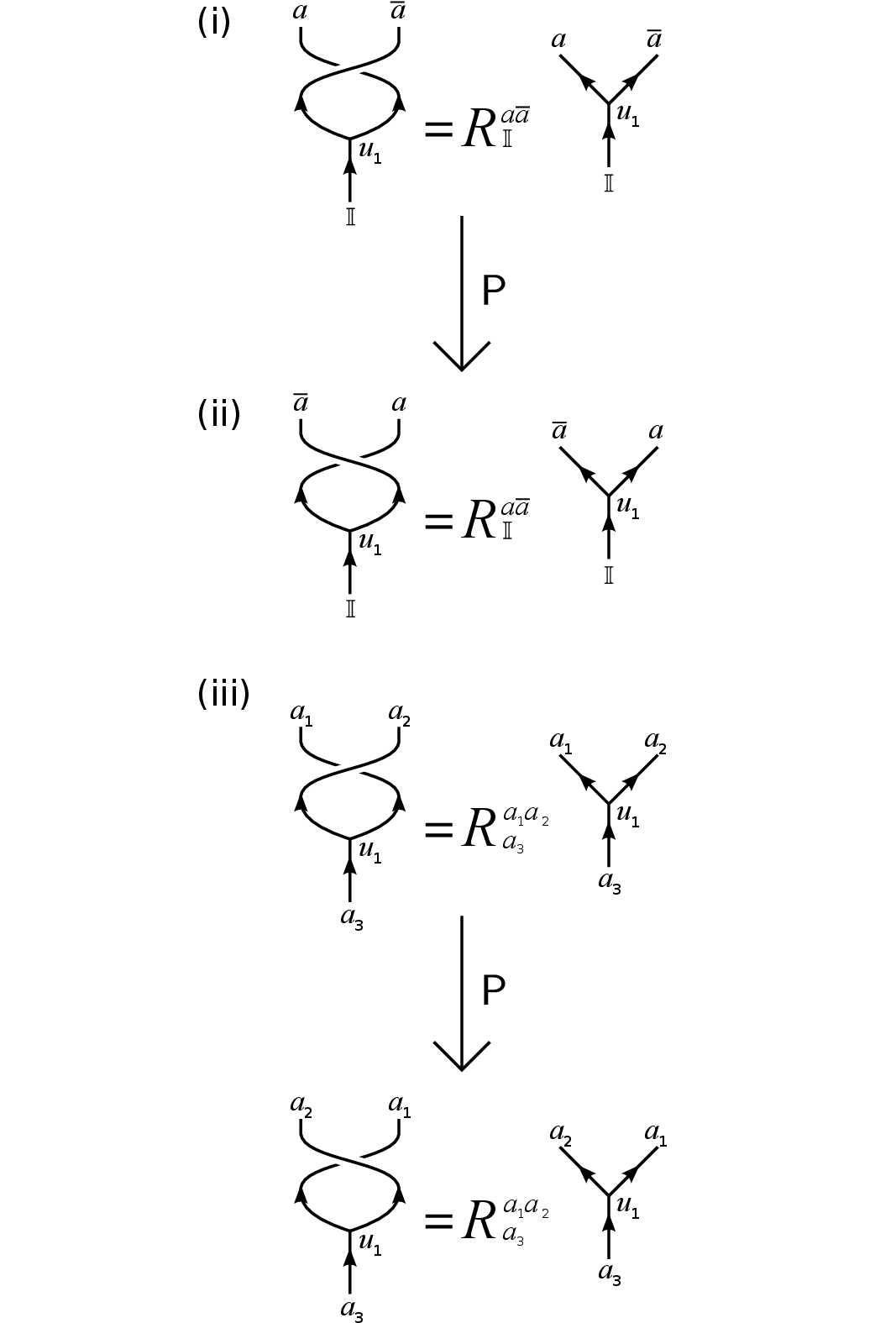}
\caption{(i)~A state $|\psi,U\ra$ having two quasiparticle excitations, and being described by a UBTC $U$, acquires a phase $R^{a\bar{a}}_\mbb{I}$ on braiding. (ii)~Application of the parity mapping $\textsf{P}$ to diagram~(i) yields an equivalent relationship for the reflected system, in which clockwise braiding of $a$ with $\bar a$ has been replaced by anticlockwise braiding of $\bar a$ with $a$. %
(iii)~More generally, reflection of the diagram which defines the braiding operation shows that all $R^{a_1a_2}_{a_3}$ in $\textsf{P}(U)$ are given by %
$(R^{a_2a_1}_{a_3})^*$ in $U$.\label{fig:braidreflection}}
\end{figure}

\begin{figure}
\includegraphics[width=246.0pt]{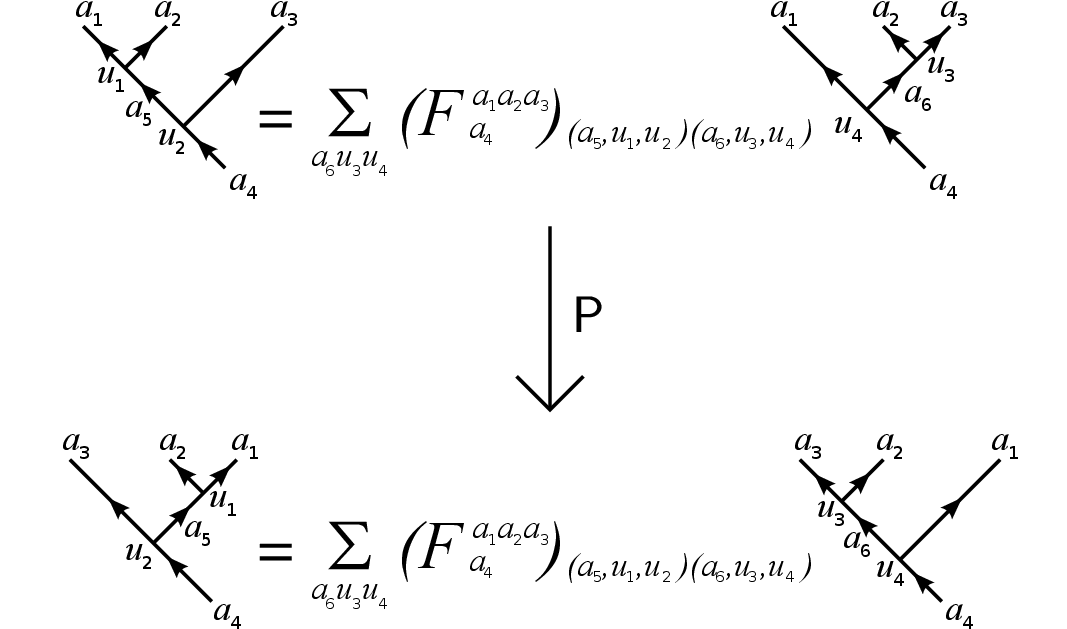}
\caption{As with braiding, we may examine the effect of the parity mapping \textsf{P} on $F$-moves. We find that each matrix $F^{a_1a_2a_3}_{a_4}$ in $\textsf{P}(U)$ is given by $(F^{a_3a_2a_1}_{a_4})^{-1}$ in $U$.\label{fig:Freflection}}
\end{figure}

We take as our starting point the definition of chirality, namely that a state or system is chiral if it is not invariant under a process of spatial reflection such as the mapping $x\rightarrow -x$. 
In the fusion tree representation for a system of anyons, the spatial dimensions are collapsed by projection onto the horizontal axis of the fusion tree diagram, and thus horizontal reflection on the plane of the page constitutes a natural implementation of this process.
As shown in \fref{fig:braidreflection}, it then follows that in the description of the reflected system every braiding coefficient $R^{a_1a_2}_{a_3}$ is replaced by the complex conjugate of $R^{a_2a_1}_{a_3}$, and similarly one can also show that each $F$ matrix $F^{a_1a_2a_3}_{a_4}$ is replaced by the inverse of $F^{a_3a_2a_1}_{a_4}$ (\fref{fig:Freflection}).
If we denote the parity mapping by $\textsf{P}$, then its action on a UBTC $U$, is given by %
\begin{equation}
\textsf{P}:\left\{\begin{array}{ll}R^{a_1a_2}_{a_3}&\!\!\longrightarrow (R^{a_2a_1}_{a_3})^*\quad\forall\quad\{a_1,a_2,a_3\}\in U\\
F^{a_1a_2a_3}_{a_4}&\!\!\longrightarrow (F^{a_3a_2a_1}_{a_4})^{-1}\quad\forall\quad\{a_1,a_2,a_3,a_4\}\in U.\end{array}\right.
\label{eq:mappingP}
\end{equation}
We may describe a UBTC $U$ as invariant under the action of \tsf{P} iff $\tsf{P}(U)=U$.
We will further distinguish between UBTCs for which there exists a permutation of charge labels \tsf{Perm} such that $\tsf{Perm}[\tsf{P}(U)]=U$, and those for which no such permutation exists. The set of UBTCs for which $\tsf{Perm}[\tsf{P}(U)]=U$ includes some non-\tsf{P}-invariant UBTCs, and also all \tsf{P}-invariant UBTCs, for which \tsf{Perm} is trivial.

Having established the action of the parity transformation \tsf{P} on UBTCs, we may now consider the effects of reflection on states and operators described by a UBTC $U$. We begin by recognising that
specification of the relevant UBTC $U$ constitutes an essential part of the description of the state. Even on the most trivial topology, it would be erroneous, for example, to identify the ``vacuum'' state (no quasiparticle excitations) of a system %
of Fibonacci anyons as being physically equivalent to the ``vacuum'' of a system described by, say, $D(\mbb{Z}_2)$. If we have a description of the underlying physical systems from which the anyonic quasiparticles emerge, then we can observe that the two systems are distinct as their manifolds are covered by different quantum spin liquids, and thus are capable of supporting different quasiparticle excitations. If we are working entirely at the level of the coarse-grained description of anyons and fusion trees, then in lieu of explicitly describing this quantum spin liquid, we
specify the relevant UBTCs by fixing $F^{a_1a_2a_3}_{a_4}$ and $R^{a_1a_2}_{a_3}$, and this therefore constitutes an essential part of the description of the state. With this in mind, it follows that for a system described by a UBTC $U$, spatial reflection will always map a state $|\psi,U\ra$ into a state $|\psi',\textsf{P}(U)\ra$. Consequently, for all models described by a UBTC satisfying $\tsf{Perm}[\tsf{P}(U)]=U$, \tsf{P} acts as an intertwinor on the Hilbert space $\mc{H}$ of the model, i.e. $\tsf{P}(\mc{H})=\mc{H}$. Conversely, if no such permutation \tsf{Perm} exists, then $\tsf{P}(\mc{H})$ and $\mc{H}$ are necessarily disjoint.\footnote{Recall that any state of a UBTC, even the ``vacuum'' state, is characterised by the excitations which the quantum spin liquid is \emph{capable} of supporting. Thus when \tsf{Perm} does not exist, it follows that $\mc{H}$ and $\tsf{P}(\mc{H})$ have no states in common.} %
The existence of a permutation of charge labels \tsf{Perm} such that $\tsf{Perm}[\tsf{P}(U)]=U$ is therefore seen as a necessary precondition for the construction of a state which is invariant under reflection (i.e. non-chiral).

As an example, consider UBMTCs having the Fibonacci fusion rules. 
In constructing such a UBMTC there is a phase degree of freedom in the definition of $F^{a_1a_2a_3}_{a_4}$, corresponding to the value of $\theta$ in
\begin{equation}
F^{\tau\tau\tau}_\tau=\left(\begin{array}{cc}\phi^{-1}&\mrm{e}^{\rmi\theta}\phi^{-1/2}\\\mrm{e}^{-\rmi\theta}\phi^{-1/2}&-\phi^{-1}\end{array}\right).
\end{equation} 
Once this degree of freedom has been fixed, solving the hexagon equations yields
two distinct %
UBMTCs having Fibonacci anyon fusion rules and the same set of $F$ matrices. One of these UBMTCs has $R^{\tau\tau}_\mbb{I}=e^{-4\pi\rmi/5}$ and $R^{\tau\tau}_\tau=e^{3\pi\rmi/5}$, and the other has $R^{\tau\tau}_\mbb{I}=e^{4\pi\rmi/5}$ and $R^{\tau\tau}_\tau=e^{-3\pi\rmi/5}$. 
We will denote these UBMTCs by $\Fib_\pm$, where 
$\pm$ refers to the sign on the exponent in $R^{\tau\tau}_\mbb{I}$. 
For a single copy of $\Fib_+$ or $\Fib_-$, no permutation \tsf{Perm} exists such that $\tsf{Perm}[\tsf{P}(\Fib_\pm)]=\Fib_\pm$, and thus any state described by a single Fibonacci UBMTC %
is necessarily chiral%
.\footnote{Note that this includes even the ``vacuum'' state of $\Fib_\pm$. If we had a vacuum state from which we could generate excitations in either of $\Fib_+$ and $\Fib_-$, it would follow that this ``vacuum'' was not described by a single copy of $\Fib$, but rather by the UBMTC $\Fib_+\otimes\Fib_-$. This is consistent with our earlier assertion that the Hilbert spaces of two UBTCs related by parity, such as $\Fib_+$ and $\Fib_-$, are disjoint.}

\begin{figure}
\includegraphics[width=246.0pt]{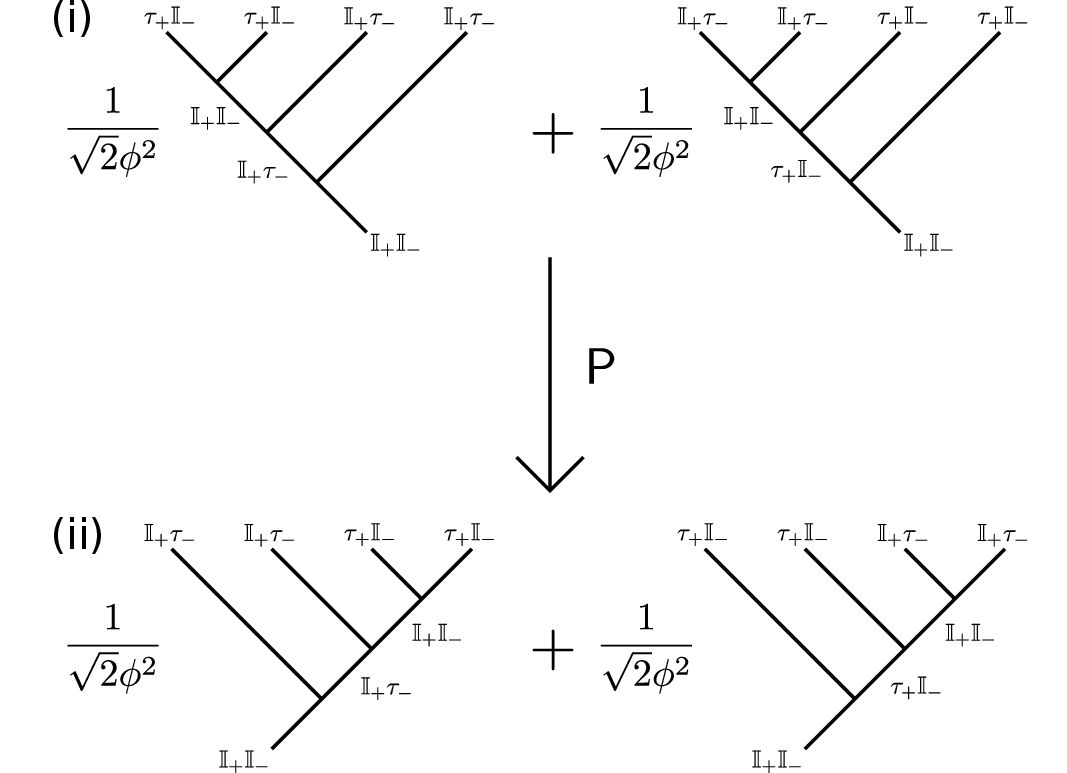}
\caption{(i)~Fusion tree description of a state \mbox{$|\psi,\Fib_+\otimes\Fib_-\ra$.} 
Horizontal reflection yields diagram~(ii). On restoring the original fusion tree basis using $F$-moves, this is seen to be equal to~(i), and thus \mbox{$|\psi,\Fib_+\otimes\Fib_-\ra$} is invariant under reflection. Arrows are not required for $\Fib_+\otimes\Fib_-$ as all charges are self-dual.\label{fig:reflinvstate}}
\end{figure}

In contrast, now consider the product $\Fib_+\otimes\Fib_-$, which is an example of a quantum double model and known to be non-chiral.\cite{gils2009a,levin2005} In this UBMTC, every branch of a fusion tree carries a compound charge consisting of one label from $\Fib_+$ and one from $\Fib_-$.
Recognising that the action of \tsf{P} is to map a UBMTC $\Fib_\pm$ into $\Fib_\mp$, we see that the UBMTC $\Fib_+\otimes\Fib_-$ is invariant under $\textsf{Perm}\circ\textsf{P}$ for
\begin{equation}
\textsf{Perm}:a_+b_-\longrightarrow b_+a_-\quad\forall\quad\{a,b\}\in\{\mbb{I},\tau\}.\label{eq:perm}
\end{equation}
For a model described by this UBMTC, the parity mapping \tsf{P} therefore acts on the Hilbert space as an intertwinor, and we may construct states which are reflection-invariant, i.e. invariant under the action of \tsf{P}. An example of such a state is given in \fref{fig:reflinvstate}.

We conclude this Appendix with three further observations.
First, in this Appendix we have emphasised the importance of specifying the relevant UBTC when describing a state, operator, or physical system, and have used the notation $|\psi,U\ra$ to highlight that the specification of $U$ constitutes a necessary part of the description of a state.
However, when working with multiple objects or systems all described by a single UBTC $U$, it is frequently unnecessary to repeatedly specify the UBTC in this manner,
and it is more usual to simply write e.g. $|\psi\ra$ for brevity. 
Nevertheless, even when specification of the UBTC is suppressed %
(as is the norm in the literature, and as has been done in all of this paper except for the present Appendix), it is always implicit that the fusion tree description of a state or operator is associated with a specific UBTC. Consequently, a necessary precondition for reflection invariance of any state or operator is the existence of a permutation \tsf{Perm} such that $\tsf{Perm}[\tsf{P}(U)]=U$.

Second, we note that none of the UBMTCs obtained from the quantum groups $\mrm{SU(2)}_k$, where $k$ is a finite integer, admit the definition of a charge permutation \tsf{Perm} such that $\tsf{Perm}[\tsf{P}(U)]=U$. Consequently, the state of any system described entirely in terms of one of these UBMTCs is always necessarily chiral.
This result includes the UBMTCs which describe
both Ising anyons and Fibonacci anyons, as the former corresponds to $\mrm{SU(2)}_2$, and the latter to
the restriction to integer charges of $\mrm{SU(2)}_3$.

Finally, we observe that although the chiral nature of the golden chain 
is made explicitly manifest in terms of the ground state momentum of Figs.~\ref{fig:dispersion} and \ref{fig:dispersion_disc}, this particular manifestation of chirality
may be concealed by the use of a modified translation operator which eliminates all non-trivial braids, for example
\begin{align}
\hat T^\mrm{D}_\mrm{M}&=(R^{\tau\tau}_\mbb{I})^{-1}\,\translop\label{eq:TDM}\\
&=\,\magtranslop\label{eq:TDM2}
\end{align}
on the disc. [Applying this operator to the fusion tree of \fref{fig:anyonstates_sphere}(iv), one readily sees that its action may be evaluated without requiring knowledge of $R^{a_1a_2}_{a_3}$.] However, without physical motivation for the loop in \Eref{eq:TDM2} this %
elimination of the ground state momentum is artificial, %
and %
may be compared to
arbitrarily
introducing
extra factors of $-1$ %
into the
translation operator %
on a chain of fermions, %
purely 
to suppress %
any 
minus signs (and non-zero momenta) %
arising 
from %
fermionic statistics. %
As a mathematical transformation it may sometimes be useful, but the model no longer corresponds directly to the original physical system.
To understand why the operators $\hat T^\Sp$ and $\hat T^\T$ [\fref{fig:discoperators}(iv), \Eref{eq:T^T}, and \fref{fig:torustranslation_throughloop}] represent the natural definitions of translation on the disc and torus respectively, the reader is directed to \sref{sec:ASO}, where these operators are derived, %
and also to
their application to %
the Heisenberg spin chain in Appendix~\ref{sec:heisenberg}.

\section{Restricting the Hilbert space\label{apdx:hilbertspace}}

To restrict the Hilbert space of the spin chain so that it corresponds to a valid fusion tree, we must exclude states forbidden by the fusion rules. For example, for a chain of Fibonacci anyons we prohibit all states in which $x_i=x_{i+1}=1$. On the torus this condition is applied for all $x_i$, $i$ ranging from 1 to $p$ inclusive, and identifying $i=p+1$ with $i=1$, but on the disc the constraint applies only for $1\leq i<p$. 

If this restriction is enforced by including terms in $\hat h^\mrm{S}_{i-1,i,i+1}$ which apply an arbitrarily large energy penalty to invalid states,
then the behaviour of $\hat T^\mrm{D}$ on the disc is such that this restriction is appropriately applied to all pairs $\{i,i+1\},\ 1\leq i<p$ and not to the pair $\{p,1\}$. Thus the structure of $\hat T^\mrm{D}$ makes it possible to easily enforce the restriction on the Hilbert space via Eqs.~\eref{eq:H^SPT} and \eref{eq:H^SPD}, just by modifying $\hat h^\mrm{S}_{i-1,i,i+1}$ to impose large energy penalties on the unphysical states.

\section{Construction of the five-body operators in Hamiltonian $\hat H^\mrm{A,P,D\rightarrow T}$\label{apdx:anyondefect}}

\begin{figure}[!tp]
\includegraphics[width=246.0pt]{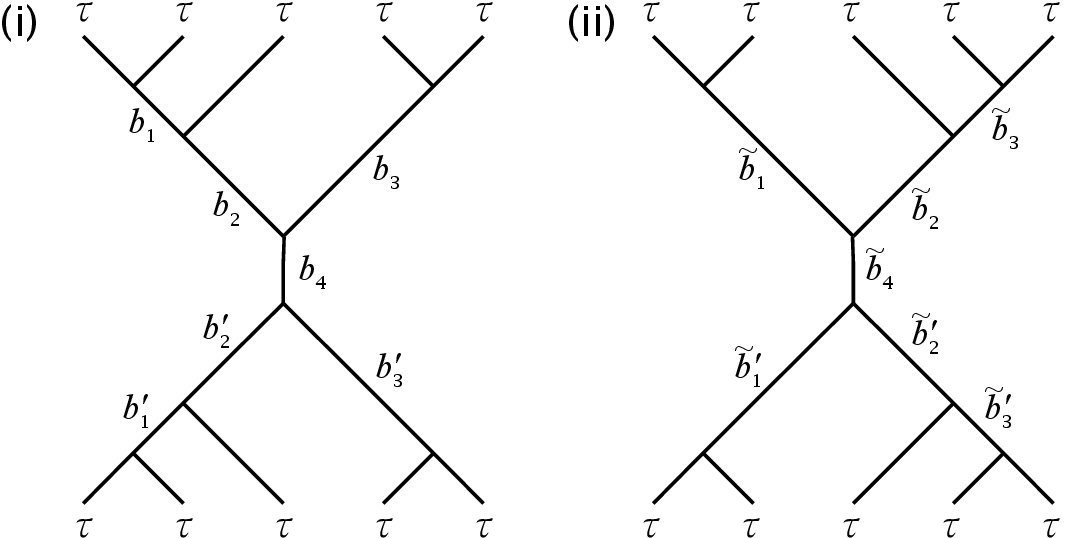} %
\caption{Structures of operators used in the definition of $\hat H^{\mrm{A,P,D\rightarrow T}}$ \eref{eq:Hdefect_Disc} for a chain of FM or AFM interacting Fibonacci anyons. Arrows are not required for Fibonacci anyons as all charges are self-dual.
\label{fig:apdxBoperators}}
\end{figure}

This appendix presents the construction of the terms $\hat h^\mrm{A}_{n-2,n-1,n,1,2}$ and $\hat h^\mrm{A}_{n-1,n,1,2,3}$ in Hamiltonian $\hat H^\mrm{A,P,D\rightarrow T}$ \eref{eq:Hdefect_Disc} for a chain of AFM or FM interacting Fibonacci anyons. These operators will be defined using the fusion trees of \fref{fig:apdxBoperators}(i) and (ii) respectively.
The rationale behind this construction is that behaviour on the disc may be achieved which is analogous to that on the torus, if we introduce operators which couple the outermost degrees of freedom on the disc fusion tree [$x_1$ and $x_{n-3}$ in \fref{fig:anyonstates_sphere}(iv)] as if they were adjacent degrees of freedom (e.g. $x_1$ and $x_n$) in the ``inside'' torus basis of \fref{fig:torusstates}(i).

For $\hat h^\mrm{A}_{n-2,n-1,n,1,2}$ in the basis of the first diagram of \fref{fig:apdxBoperators}(i), we may identify $b_1\equiv x_{p-1}$, $b_2\equiv x_p$, $b_3\equiv x_1$, $b_4\equiv z_1$, where we have labelled by $z_1$ the total charge of the five anyons on sites $\{n-2,n-1,n,1,2\}$. The 
coefficients $\left(\hat h^\mrm{A}_{n-2,n-1,n,1,2}\right)^{b_1b_2b_3}_{b'_1b'_2b'_3b_4}$ of this operator therefore describe a mapping from states labelled by $x_{p-1}$, $x_p$, $x_1$, and $z_1$ into states labelled by $x'_{p-1}$, $x'_p$, $x'_1$, and $z_1$. 
To reproduce the behaviour of the operator $\hat h^\mrm{S}_{\spinsite{p-1},\spinsite{p},\spinsite{1}}$ corresponding to $\hat h^\mrm{A}_{p,1}$ on the $p$-site torus, we want this operator to be independent of $z_1$, and to correspond to $\hat h^\mrm{S}_{\spinsite{p-1},\spinsite{p},\spinsite{1}}$ on the subspace labelled by $\{x_{p-1},x_p,x_1,x'_{p-1},x'_p,x'_1\}$. Its coefficients are therefore given by 
\begin{equation}
\forall\ z_1:\left(\hat h^\mrm{A}_{n-2,n-1,n,1,2}\right)^{x_{p-1}x_px_1}_{x'_{p-1}x'_px'_1z_1}=\left(\hat h^\mrm{S}_{\spinsite{p-1},\spinsite{p},\spinsite{1}}\right)^{x_{p-1}x_px_1}_{x'_{p-1}x'_px'_1}.
\end{equation}
Similarly, $\hat h^\mrm{A}_{n-1,n,1,2,3}$ in the basis of the third diagram of \fref{fig:apdxBoperators}(ii) takes its coefficients from $\hat h^\mrm{S}_{\spinsite{p},\spinsite{1},\spinsite{2}}$ for all $z_2$, where $z_2$ represents the total charge of the five anyons on sites $\{n-1,n,1,2,3\}$ and we make the identifications $\tilde b_1\equiv x_{p}$, $\tilde b_2\equiv x_1$, $\tilde b_3\equiv x_2$, $\tilde b_4\equiv z_2$.

Finally, on the torus the Hilbert space is restricted to states which satisfy
the constraints $x_{i+1} \in x_i\times \tau$ for all values of $i$, and site $p+1$ is identified with site $1$. 
On the disc this restriction is enforced by the fusion rules for $1\leq i \leq p-1$, but for $i=p$ it must be applied manually by inserting an arbitrarily large energy penalty into the 5-site Hamiltonian terms for states satisfying $x_p=x_1=1$.

\section{Application of the diagrammatic formalism to the Heisenberg spin chain\label{sec:heisenberg}}

\begin{figure}
\includegraphics[width=246.0pt]{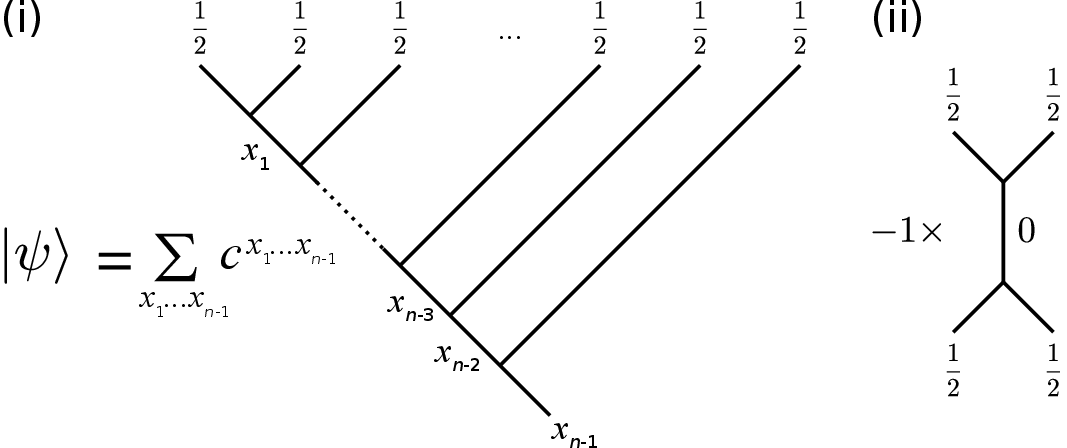}
\caption{(i)~Fusion tree used as a basis of states for the $n$-site Heisenberg spin chain. Note that in contrast to \protect{\fref{fig:anyonstates_sphere}}(iv), the total charge (or spin) is not constrained to be zero, and
a tree with total charge $x_{n-1}$ therefore represents a vector space of 
dimension $d_{x_{n-1}}$. 
(ii)~Diagrammatic representation of nearest neighbour interaction in the AFM spin-$\frac{1}{2}$ Heisenberg model. 
The leading ``$-1$'' in this diagram is a numerical multiplier. When this operator acts on a pair of sites $i$ and $i+1$, it yields a negative energy contribution if and only if the combined spin of these two sites is zero.
\label{fig:heisenberg}}
\end{figure}

In this paper we have primarily dealt with models such as the AFM golden chain, which are defined on a ring of anyons. It is instructive to compare these with models such as the Heisenberg model, which is defined on a spin chain but which can also be expressed in the diagrammatic notation used in this paper.\cite{pfeifer2011a} As an example, we will now consider the spin-$\frac{1}{2}$ AFM Heisenberg model with periodic boundary conditions, assumed to be constructed on a manifold which is topologically the disc. Because this model possesses SU(2) symmetry, we may represent it in the diagrammatic notation using a UBMTC based on the fusion rules for representations of SU(2). States can be represented in the form of \fref{fig:heisenberg}(i), and the nearest-neighbour interaction takes the form of \fref{fig:heisenberg}(ii).

We may now %
analyse this system in two different ways. Either we may obtain the energy spectrum by exactly diagonalising the original spin chain, and compute momenta using the natural definition of translation on that spin chain, or we may write it in the diagrammatic notation, and map this to a different spin chain as
described in \sref{sec:spinmapping}. We would then compute the energy on the chain of fusion tree variables $x_1,\ldots,x_{n-1}$, and the momenta using translation operator $\hat T^\mrm{D}$ from \fref{fig:anyonstates_sphere}(i).
As is to be expected, the results obtained using these different methods agree to the limits of numerical precision.

This %
observation %
has bearing upon the definition of the translation operator. If we use the %
definitions for $\hat T^\Sp$ and $\hat T^\T$ given in \fref{fig:anyonstates_sphere}(i), \Eref{eq:T^T}, and \fref{fig:torustranslation_throughloop}, then we find that the ground state of the AFM golden chain has non-trivial momentum. Suppose that, instead, we assume %
a momentum of zero for the ground state of the golden chain, and adopt
\begin{equation}
\hat T^\Sp_\mrm{M}=\left(R^{a_n\overline{a_n}}_\mbb{I}\right)^{-1}\,\translop{}\,\label{eq:magtransdisc}
\end{equation}
as the translation operator
on the disc and
\begin{equation}
\hat T^\T_\mrm{M}=\left(R^{a_n\overline{a_n}}_\mbb{I}\right)^{-1}\,\translopstar{}\,\tag{\ref{eq:magtrans}}
\end{equation}
on the torus (noting that $\hat T^\T_\mrm{M}$ is the modified translation operator originally introduced in \sref{sec:torustranslate}, corresponding to cycling of the fusion tree variables $x_1,\ldots,x_{n-1}$). 
For consistency we would then also have to use \Eref{eq:magtransdisc} when working with the diagrammatic representation of the AFM Heisenberg chain, with $R^{a_n\overline{a_n}}_\mbb{I}
=R^{\frac{1}{2}\frac{1}{2}}_0=-1$. However, the momenta obtained using this operator are %
inconsistent with results obtained by exactly diagonalising the original spin chain, indicating that $\hat T^\Sp$, and not $\hat T^\Sp_\mrm{M}$, is the correct definition for the periodic translation operator on the disc. 
As we require that
the torus with trivial flux be consistent with the disc, we also %
obtain that $\hat T^\T$, and not $\hat T^\T_\mrm{M}$, is the correct periodic translation operator on the torus. Thus %
study of the AFM Heisenberg spin chain supports our claim that the ground state of the AFM golden chain has non-zero momentum, as observed in \fref{fig:dispersion}(ii) for the torus and \fref{fig:dispersion_disc} for the disc.

\bibliography{Anyonpaper}

\end{document}